\documentclass[aps,twocolumn,prd,showpacs,nofootinbib]{revtex4}

\usepackage{bm}
\usepackage{latexsym}
\usepackage{amsfonts,amssymb,amsmath}
\usepackage{graphicx,epsfig}
\usepackage{psfrag}
\usepackage{subfigure}
\usepackage{ulem}
\usepackage{hyperref}
\usepackage{color}
\usepackage{appendix}

\hypersetup{
    bookmarks=true,         % show bookmarks bar?
    unicode=false,          % non-Latin characters in Acrobat’s bookmarks
    pdftoolbar=true,        % show Acrobat’s toolbar?
    pdfmenubar=true,        % show Acrobat’s menu?
    pdffitwindow=false,     % window fit to page when opened
    pdfstartview={FitH},    % fits the width of the page to the window
    pdftitle={My title},    % title
    pdfauthor={Author},     % author
    pdfsubject={Subject},   % subject of the document
    pdfcreator={Creator},   % creator of the document
    pdfproducer={Producer}, % producer of the document
    pdfkeywords={keyword1} {key2} {key3}, % list of keywords
    pdfnewwindow=true,      % links in new window
    colorlinks=true,       % false: boxed links; true: colored links
    linkcolor=red,          % color of internal links
    citecolor=cyan,        % color of links to bibliography
    filecolor=magenta,      % color of file links
    urlcolor=green,           % color of external links
    linktocpage=true
}

\newcommand{\dd}{\mathrm{d}}
\newcommand{\gsim}{\gtrsim}

\def\spose#1{\hbox to 0pt{#1\hss}}

\def\lta{\mathrel{\spose{\lower 3pt\hbox{$\mathchar"218$}}
     \raise 2.0pt\hbox{$\mathchar"13C$}}}
\def\gta{\mathrel{\spose{\lower 3pt\hbox{$\mathchar"218$}}
     \raise 2.0pt\hbox{$\mathchar"13E$}}}

\newcommand{\ie}{\textsl{i.e.~}}

\newcommand{\eg}{\textsl{e.g.~}}

\newcommand{\Hu}{\mathcal{H}}

\newcommand{\GN}{G_{_\mathrm{N}}}

\newcommand{\Mp}{M_{_\mathrm{Pl}}}

\def\beq{\begin{equation}}
\def\eeq{\end{equation}}
\def\bea{\begin{eqnarray}}
\def\eea{\end{eqnarray}}
\def\eqref{\ref}

\newcommand{\mean}[1]{\left\langle #1 \right\rangle}

\newcommand{\ue}{\mathrm{e}}

\newcommand{\Rea}{\Re \mathrm{e}\,}
\newcommand{\Ima}{\Im \mathrm{m}\,}

\begin{document}

\title{Cosmological Inflation and the Quantum Measurement Problem}

\author{J\'er\^ome Martin} \email{jmartin@iap.fr}
\affiliation{Institut d'Astrophysique de Paris, UMR 7095-CNRS,
Universit\'e Pierre et Marie Curie, 98 bis boulevard Arago, 75014
Paris, France}

\author{Vincent Vennin} \email{vennin@iap.fr}
\affiliation{Institut d'Astrophysique de Paris, UMR 7095-CNRS,
Universit\'e Pierre et Marie Curie, 98 bis boulevard Arago, 75014
Paris, France}

\author{Patrick Peter} \email{peter@iap.fr}
\affiliation{Institut d'Astrophysique de Paris, UMR 7095-CNRS,
Universit\'e Pierre et Marie Curie, 98 bis boulevard Arago, 75014
Paris, France}

\date{\today}

\begin{abstract}
  According to cosmological inflation, the inhomogeneities in our
  universe are of quantum mechanical origin. This scenario is
  phenomenologically very appealing as it solves the puzzles of the
  standard hot big bang model and naturally explains why the spectrum
  of cosmological perturbations is almost scale invariant. It is also
  an ideal playground to discuss deep questions among which is the
  quantum measurement problem in a cosmological context. Although the
  large squeezing of the quantum state of the perturbations and the
  phenomenon of decoherence explain many aspects of the quantum to
  classical transition, it remains to understand how a specific
  outcome can be produced in the early universe, in the absence of any
  observer. The Continuous Spontaneous Localization (CSL) approach to
  quantum mechanics attempts to solve the quantum measurement question
  in a general context. In this framework, the wavefunction collapse
  is caused by adding new non linear and stochastic terms to the
  Schr\"odinger equation. In this paper, we apply this theory to
  inflation, which amounts to solving the CSL parametric oscillator
  case. We choose the wavefunction collapse to occur on an eigenstate
  of the Mukhanov-Sasaki variable and discuss the corresponding
  modified Schr\"odinger equation. Then, we compute the power spectrum
  of the perturbations and show that it acquires a universal shape
  with two branches, one which remains scale invariant and one with
  $n_{_{\rm S}}=4$, a spectral index in obvious contradiction with the
  Cosmic Microwave Background (CMB) anisotropy observations. The
  requirement that the non-scale invariant part be outside the
  observational window puts stringent constraints on the parameter
  controlling the deviations from ordinary quantum mechanics. Due to
  the absence of a CSL amplification mechanism in field theory, this
  has also for consequence that the collapse mechanism of the
  inflationary fluctuations is not efficient. Then, we determine the
  collapse time. On small scales the collapse is almost instantaneous,
  and we recover exactly the behavior of the CSL harmonic oscillator
  (a case for which we present new results), whereas, on large scales,
  we find that the collapse is delayed and can take several e-folds to
  happen. We conclude that recovering the observational successes of
  inflation and, at the same time, reaching a satisfactory resolution
  of the inflationary ``macro-objectification'' issue seems
  problematic in the framework considered here. This work also
  provides a complete solution to the CSL parametric oscillator
  system, a topic we suggest could play a very important r\^ole to
  further constrain the CSL parameters. Our results illustrate the
  remarkable power of inflation and cosmology to constrain new
  physics.
\end{abstract}

\pacs{98.80.Cq, 98.80.Qc, 03.65.Ta, 03.65.Yz, 05.40.-a}
\maketitle

\section{Introduction}
\label{sec:intro}

Inflation is currently the leading paradigm for explaining the
physical conditions that prevailed in the very early
universe~\cite{Starobinsky:1980te,Guth:1980zm,Linde:1981mu,Albrecht:1982wi,Linde:1983gd}. It
solves the puzzles of the standard hot big bang phase and it explains
the origin of the inhomogeneities in our
universe~\cite{Mukhanov:1981xt,Mukhanov:1982nu,Starobinsky:1982ee,Guth:1982ec,Hawking:1982cz,Bardeen:1983qw}
(for reviews, see
Refs.~\cite{Mukhanov:1990me,Martin:2003bt,Martin:2004um,Martin:2007bw,Linde:2007fr,Sriramkumar:2009kg,Peter:2009}). According
to the inflationary scenario, these inhomogeneities result from the
amplification of the unavoidable vacuum quantum fluctuations of the
gravitational and inflaton fields during a phase of accelerated
expansion. In particular, inflation predicts an almost scale invariant
power spectrum for the cosmological
fluctuations~\cite{Stewart:1993bc}, a prediction which fits very well
the high accuracy astrophysical data now at our
disposal~\cite{Larson:2010gs,Komatsu:2010fb,Martin:2006rs,Lorenz:2007ze,Lorenz:2008je,Martin:2010kz,Martin:2010hh}.

\par

Often less emphasized is the fact that inflation is also particularly
remarkable from the theoretical point of view. Indeed, the
inflationary mechanism for the production of cosmological
perturbations makes use of general relativity and quantum mechanics,
two theories that are notoriously difficult to combine. Moreover, this
mechanism leads to theoretical predictions that are possible to study
observationally with great accuracy. In fact, inflation is probably
the only case in physics where an effect based on general relativity
and quantum mechanics leads to predictions that, given our present day
technological capabilities, can be tested experimentally.

\par

The situation described above can be used to investigate deep
questions. Among these deep questions is how the quantum measurement
problem looks in a cosmological context. According to inflation, the
Cosmic Microwave Background (CMB) radiation
anisotropy~\cite{Smoot:1992td} is an observable and is therefore
described by a quantum operator. As a consequence, when one looks at a
CMB map, one observes the result of a measurement of that
observable. According to the postulates of quantum mechanics in the
Copenhagen interpretation, this means that the wavefunction of the
inflationary perturbations has collapsed to an eigenvector of this
operator and that the CMB map corresponds to one of its
eigenvalue. The problem with this approach is that the collapse is
supposed to occur only when an observer performs a measurement on the
system. Clearly, there was no observer before or when the CMB was
emitted. This seems to contradict the phenomenological fact that
large-scale structure formation started early in the history of the
universe since these structures are seeded by the same early physics
which led to CMB fluctuations. As a matter of fact, CMB fluctuations
can also be understood as the earliest hint that primordial
inhomogeneities had already started to grow at that time.
Furthermore, in some sense, the observers are actually the end product
of the structure formation process! Of course, this measurement
problem is already present in conventional laboratory situations but
it seems to be exacerbated (to use the words of
Ref.~\cite{Sudarsky:2009za}) in a cosmological context.

\par

Important steps towards a better understanding of these issues have
already been accomplished. In particular, it was shown that the
inflationary accelerated expansion transforms a coherent vacuum state
into a strongly squeezed state~\cite{Grishchuk:1990bj}, the
corresponding squeezing being much more important than whatever can be
realized in the laboratory~\cite{Grishchuk:1992tw}. In this limit, the
predictions of the quantum formalism are indistinguishable from that
of a theory where the fluctuations are just assumed to be realizations
of a classical stochastic
process~\cite{Polarski:1995jg,Lesgourgues:1996jc,Grishchuk:1997pk}. The
classical limit is a subtle concept in quantum mechanics but, in this
sense (and in this sense only!), the system can be characterized as
being classical~\cite{Kiefer:2008ku}. Moreover, the large-scale
cosmological perturbations are not isolated and, as a consequence, the
phenomenon of decoherence~\cite{Zurek:1981xq,Zurek:1982ii,Joos:1984uk}
is relevant for them. This has for consequence that their density
matrix becomes diagonal before recombination, a criterion which is
also considered as necessary in order to understand the
quantum-to-classical
transition~\cite{Polarski:1995jg,Lesgourgues:1996jc,Kiefer:2006je,Kiefer:1998qe,Egusquiza:1997ez,Anderson:2005hi,Burgess:2006jn,Martineau:2006ki}. However,
it is known that decoherence {\it per se} does not solve the
measurement problem~\cite{Adler:2001us,Schlosshauer:2003zy}. Indeed,
it remains to understand how a single outcome can be produced. This
point is particularly important given that we only have one CMB map,
that is to say only one measurement of the corresponding
observable. In other words, even if the cosmological fluctuations can
be viewed as a classical stochastic problem, this does not explain how
a given realization of this process becomes an actual perception. This
``macro-objectivation'' problem is already present in a conventional
situation but, as already mentioned before, it becomes particularly
embarrassing in the context of inflation where the collapse of the
wavefunction cannot be due to the presence of a conscious
observer. Facing this situation, the common attitude is to postulate
that decoherence should be combined with a new interpretational
scheme, different from the Copenhagen interpretation. Typically, in
cosmology, the many world approach is often implicitly
assumed~\cite{mukhanov2005physical,Kiefer:2008ku,Nomura:2011dt,Nomura:2011rb,Bousso:2011up,weinberg2008cosmology}. Another
frequently mentioned possibility, which seems to be particularly well
suited to the cosmological context, is to consider that the
wavefunction only represents the information that we have on the
system~\cite{2002quant.ph..5039F}.  In this case, the issue of the
wavefunction collapse becomes irrelevant since it just corresponds to
a situation where the observer updates its knowledge (in the Bayesian
sense) about the physical properties of the system. Other attempts,
such as the non-local hidden variable theories, have also been
tried~\cite{Holland:1993ee,Peter:2005hm,Peter:2006id,Peter:2006hx,Peter:2008qz,PintoNeto:2011ui}. In
all these cases, the cosmological situation does not differ much from
a conventional laboratory situation and, moreover, does not lead to
new, falsifiable, predictions\footnote{In the case of the Bohm-de
  Broglie approach, there could be a transitory regime, before
  ``quantum equilibrium'' is reached, where the predictions differ
  from conventional quantum
  mechanics~\cite{Valentini:2006yj}. Cosmology is also precisely
  considered as a situation where this regime could be
  relevant~\cite{Valentini:2008rg,Pearle:2005rc}.}. Then, it becomes a
question of taste which approach best fits one's own prejudices.

\par

However, there exists an exception to the conclusion of the previous
discussion, namely the case of the collapse
models~\cite{Pearle:1976ka,Ghirardi:1985mt,Pearle:1988uh,Diosi:1988vj,Ghirardi:1989cn,Weinberg:2011jg}
(for reviews, see Refs.~\cite{Bassi:2003gd,Bassi:2012bg}). In this
approach, the Schr\"odinger equation is modified by adding non linear
and stochastic terms which renders dynamical the collapse of the
wavefunction. The model has nice features: firstly, the approach seems
to follow a conservative strategy since, in physics, it is standard to
first consider a linear theory and then, in order to have a more
accurate description, to consider non linear corrections; in some
sense, the collapse theories follow this line of argument. Secondly,
there is now a single law of evolution for the state vector and,
thirdly, the Born laws can be derived instead of postulated. There are
also disadvantages such as the property that energy is not conserved
or the fact that the relativistic formulation of the theory appears to
be technically and conceptually difficult to develop (however, see
Ref.~\cite{Bedingham:2010hz}). But, clearly, the main advantage in
comparison to the possibilities discussed above is that this approach
is falsifiable since it leads to predictions different from that of
conventional quantum mechanics. This fact has been widely used in
order to constrain collapse theories in the
laboratory~\cite{Bassi:2003vf,Adler:2004un,Adler:2004rf,Bassi:2005fp,Bassi:2012bg}
but, clearly, it is also important to see whether this could be done
in a cosmological
context~\cite{Pearle:2007rw,Pearle:2010uu,Lochan:2012di}. It is
therefore interesting to investigate what the collapse theories have
to say about the inflationary mechanism. Notice that, regardless of
one's opinion about collapse theories, the subject is worth studying:
a supporter would argue that the cosmological measurement problem can
possibly find a natural solution within this theory and an opponent
would hope that the constraints obtained in a cosmological context can
rule out the theory. In fact, this last question turns out to be very
important. Indeed, as already mentioned, the constraints that exist on
collapse theories are usually obtained from physical phenomena that
can be observed in the laboratory. Therefore, by studying collapse
theories in the context of cosmology and inflation, one can hope to
derive very relevant new constraints since one now deals with
characteristic scales (energy, length etc \dots) which typically
differ by many orders of magnitude from those used to in a down to
earth context. This illustrates again the conceptual relevance of
inflation when it comes to very fundamental questions and its power to
constrain alternatives to gravity but also to quantum mechanics. In
some sense, inflation represents an ideal playground to test new
theories. Notice in passing that the very same strategy was used in
the case of the so-called trans-Planckian problem of
inflation~\cite{Brandenberger:2000wr,Martin:2000xs,Martin:2000bv}
where it was shown that the inflationary observables could possibly
contain an imprint (although probably small) of string theory.

\par

We are using (modified) Schr\"odinger-type of equation to describe the
behavior of cosmological perturbations. This is justified because each
Fourier mode of those effectively evolves in an independant way and
cosmological expansion permits to define a priviledged time. This
allows for a sensible treatment of cosmological perturbations even
though a fully relativistic CSL model, which could be naively expected
to be required, is still lacking. At this moment, surprizingly, it is
easier to treat inflationary perturbations than ordinary particle
physics.

\par

It should also be emphasized that the idea of applying collapse
theories to inflationary perturbations of quantum-mechanical origin
was first considered in
Refs.~\cite{Perez:2005gh,DeUnanue:2008fw,DiezTejedor:2011jq}. In these
articles, a phenomenological model for the collapse process was
assumed and the corresponding physical properties were derived. In
particular, the power spectrum of the perturbations was calculated and
was shown to deviate from the standard predictions. Therefore,
Refs.~\cite{Perez:2005gh,DeUnanue:2008fw,DiezTejedor:2011jq} have
demonstrated that, in principle, it is possible to observationally test
collapse theories in a cosmological context. Our approach differs from
that of Refs.~\cite{Perez:2005gh,DeUnanue:2008fw,DiezTejedor:2011jq}
in the fact that we use the Continuous Spontaneous Localization (CSL)
model to implement the collapse dynamics. This has the advantage that
our calculations can be directly confronted and compared to other
results obtained in other branches of physics.

\par

This paper is organized as follows. In the next section,
Sec.~\ref{sec:infpert}, we present a brief review of the theory of
inflationary cosmological perturbations of quantum mechanical
origin. We especially focus on the calculation of the power spectrum
since this quantity is the tool that allows us to relate the
inflationary theory with the CMB observations. Then, in
Sec.~\ref{sec:measureproblem}, we discuss the cosmological measurement
problem and we explain how high accuracy CMB measurements can
constrain inflation. In Sec.~\ref{sec:grw}, we consider collapse
theories, in particular, its CSL version, which is, as already
mentioned, the case we use in this article.  These sections aim at
rendering the present work self-contained for readers with different
expertises. Then, we show how the harmonic oscillator can be treated
in this context. This case is particularly relevant for cosmological
fluctuations since it corresponds to the small-scale limit (in
comparison to the Hubble radius) of the theory of cosmological
perturbations. In Sec.~\ref{sec:infcsl}, we apply the CSL theory to
inflation and to the calculation of the power spectrum.  We use this
result to constrain the parameter that controls the deviations from
ordinary quantum mechanics. In Sec.~\ref{sec:collapsepert}, we study
in more details the collapse phenomenon and explicitly compute the
collapse time on small and large scales. In Sec.~\ref{sec:conclusion},
we summarize our results and present our conclusions. We end the paper
with an Appendix~\ref{sec:appendix} where it is shown that changing
the ``temporal gauge'' in which the modified Schr\"odinger equation is
written does not affect the shape of the power spectrum. This
calculation reinforces the generic character of the results obtained
in this work. 

\section{Inflationary Cosmological Perturbations}
\label{sec:infpert}

\subsection{Basic Formalism}
\label{subsec:basic}

By definition, inflation is a phase of accelerated expansion that took
place in the very early Universe, prior to the standard hot Big-Bang
phase
~\cite{Starobinsky:1980te,Guth:1980zm,Linde:1981mu,Albrecht:1982wi,Linde:1983gd}
(for reviews, see
Refs.~\cite{Martin:2003bt,Martin:2004um,Martin:2007bw}).  As is
well known, postulating such a phase of evolution allows us to solve
the standard problems of the hot Big-Bang model. Given that, at very
high energies, field theory is the relevant framework to describe
matter, a natural way to realize inflation is to consider that a real
scalar field (the ``inflaton'' field) dominated the energy density
budget of matter in the early Universe. Moreover, this assumption is
compatible with the observed homogeneity, isotropy and flatness of the
early Universe. Technically, the above-mentioned situation can be
described by the metric tensor ${\rm d}s^2=-{\rm d}t^2+a^2(t)\delta
_{ij}{\rm d}x^i{\rm d}x^j$, where $a(t)$ is the
Friedman-Lema\^{\i}tre-Robertson-Walker (FLRW) scale factor and $t$
the cosmic time\footnote{Unless explicit mention of the contrary, we
  shall in what follows assume natural units in which $\hbar=c=1$ so
  that the Newton constant $\GN$ is related with the Planck mass $\Mp$
  through $8\pi\GN=\Mp^{-2}$}. The Einstein equations imply that
$\ddot{a}/a=-(\rho+3p)/(6\Mp^2)$, $\rho$ and $p$ being the energy
density and pressure of the matter sourcing the gravitational field
and $\Mp$ the Planck mass (a dot denotes a derivative with respect to
the cosmic time $t$). For a scalar field, this reduces to
$\ddot{a}/a=V(\varphi)(1-\dot{\varphi}^2/V)/(3\Mp^2)$, where
$V(\varphi)$ is the scalar field potential. This means that inflation
(\ie $\ddot{a}>0$) can be obtained provided the inflaton slowly rolls
down its potential so that its potential energy dominates over its
kinetic energy. This also shows that the inflaton potential must be
sufficiently flat, a requirement which is not always easy to obtain in
realistic situations and makes the inflationary model building problem
a difficult issue \cite{Lyth:1998xn}. The physical nature of the
inflaton field has not been identified (there are many candidates)
and, as a consequence, the shape of $V(\varphi)$ is not known. Of
course, different $V(\varphi)$ lead to different inflationary
expansions but, since these different potentials must all be
sufficiently flat, the corresponding scale factors are all
approximately given by de Sitter solution. This solution is described
by the scale factor $a(t)\simeq {\rm e}^{Ht}$, where $H\equiv
\dot{a}/a$ is the Hubble parameter, a slowly-varying quantity directly
related to the energy scale of inflation. Observationally, this last
quantity is not known but is constrained ~\cite{Martin:2006rs} to be
between the Grand Unified Theory (GUT) scale, that is to say $\sim
10^{15}\, \mbox{GeV}$, and $\sim 1\, \mbox{TeV}$. The previous
considerations show that inflation can also be viewed as a phase of
quasi-exponential expansion.

\par

A concrete illustration of the above discussion consists in
considering power-law inflation~\cite{Lucchin:1984yf}. Although it is
based on a specific model with potential $V(\varphi)=M^4{\rm
  e}^{-\alpha \varphi/\Mp}$ (with $\alpha$ constant), it captures, in
a simple way, all the essential properties of inflation and, moreover,
is the only scenario which permits an exact integration of the
equations of motion (at the background level but also at the
perturbative level, see below). The corresponding scale factor is
given by
\begin{equation}
a(\eta)=\ell_0 \left(-\eta\right)^{1+\beta},
\end{equation}
where $\ell_0$ is a length the value of which is fixed once the energy
scale of inflation is known and $\eta $ in the conformal time defined
by ${\rm d}t=a{\rm d}\eta$, see Eq.~(\ref{eq:metric}).
The quantity $\beta $ is a free parameter
such that $\beta \le -2$ and is related to $\alpha$ through
$\alpha^2/2=(\beta+2)/(\beta+1)$. The case $\beta =-2$ represents the
de Sitter solution since it implies $\alpha =0$, \ie a flat potential
(and, of course, in cosmic time, the solution $a\propto 1/\eta $ is
given by an exponential). Therefore, different $\beta $ represents
different inflationary solutions and $\beta $ must always be close to
$-2$ in order for the potential to be sufficiently flat. As announced,
power-law inflation illustrates well the discussion of the previous
paragraph.

\par

The above arguments can be considered as strong hints in favor of
inflation. However, soon after its advent, it was realized that
inflation, combined with quantum mechanics, leads to an even more
impressive result, namely it naturally explains the origin of the
Cosmic Microwave Background (CMB) anisotropies and of the large-scale
structures. According to the inflationary paradigm, these deviations
from homogeneity and isotropy originate from the unavoidable
zero-point quantum fluctuations of the coupled inflaton and
gravitational fields. Statistically, the fluctuations are
characterized by their two-point correlation function or power
spectrum. The observations
~\cite{Larson:2010gs,Komatsu:2010fb,Martin:2006rs,Lorenz:2007ze,Lorenz:2008je,Martin:2010kz,Martin:2010hh}
indicate that the corresponding power spectrum is close to the
Harrison-Zel'dovich, scale invariant, power spectrum with equal power
on all scales. That this power spectrum represents a good fit to the
astrophysical data was in fact realized before the advent of inflation
but no convincing fundamental theory was known to explain this result.

\par

The main success of inflation is that it precisely predicts an almost
scale invariant power spectrum, the small deviations from scale
invariance being connected with the micro-physics of inflation
~\cite{Mukhanov:1981xt,Mukhanov:1982nu,Hawking:1982cz,Starobinsky:1982ee,Guth:1982ec,Bardeen:1983qw}. The
fact that different types of inflationary scenarios lead to a power
spectrum which is, at leading order, always close to scale invariance
is connected with the fact that the inflationary scale factor is
always close to the de Sitter solution (see above) or, equivalently,
with the fact that the inflaton potential is always almost flat. The
deviations from scale invariance are related to the deviations from a
flat potential and, therefore, depend on the detailed shape of the
potential. As a consequence, measuring them allows us to say something
about $V(\varphi)$ and there is currently an important effort in this
direction using the high accuracy CMB data that have been released in
the past years.

\par

Let us now see how the results reviewed before can be
derived. Clearly, in order to model the cosmological fluctuations, one
needs to go beyond homogeneity and isotropy. The most general metric
describing small fluctuations of the scalar type on top of a FLRW
Universe can be written as \cite{Mukhanov:1990me}
\begin{eqnarray}
\label{eq:metric}
\dd s^2=a^2\left(\eta\right)\Bigl\lbrace
-\left(1-2\phi\right)\dd\eta^2
+2\left(\partial_iB\right)\dd x^i\dd \eta\nonumber\\
+\left[\left(1-2\psi\right)\delta_{ij}
+2\partial_i\partial_jE
\right]\dd x^i\dd x^j\Bigr\rbrace .
\end{eqnarray}
A similar approach could be used to take into account tensor
perturbations (\ie gravity waves). Here, we do not include them since
they are subdominant in the CMB, representing less than $\sim 20~\%$
at $2 \sigma$ confidence level~\cite{Martin:2006rs} and, in addition,
doing it would not bring any new aspects to the question we want to
investigate in this article. In Eq.~(\ref{eq:metric}), the four
functions $\phi$, $B$, $\psi$ and $E$ are of course functions of time
and space since we consider an inhomogeneous and anisotropic
situation. As is well known, the above approach is redundant because
of gauge freedom
\cite{Mukhanov:1990me,Bardeen:1980kt,Martin:1997zd}. A careful study
of this question shows that the gravitational sector can in fact be
described by a single, gauge-invariant, quantity, the Bardeen
potential $\Phi_{_{\rm B}}$ defined by \cite{Bardeen:1980kt}
\begin{equation}
\label{eq:defbardeen}
\Phi_{_{\rm B}}\left(\eta,\bm{x}\right)=
\phi+\frac{1}{a}
\left[a\left(B-E^\prime\right)\right]^\prime,
\end{equation}
where a prime denotes a derivative with respect to the conformal time $\eta$.
In the same manner, the matter sector can be modeled by the gauge
invariant fluctuation of the scalar field
\begin{equation}
\delta\varphi^{\left(\mathrm{gi}\right)}\left(\eta,\bm{x}\right)=
\delta\varphi+\varphi^\prime\left(B-E^\prime\right)\, .
\end{equation}
The two quantities $\Phi_{_{\rm B}}$ and
$\delta\varphi^{\left(\mathrm{gi}\right)}$ are related by a perturbed
Einstein constraint. This implies that the scalar sector can in fact
be described by a single quantity. For this reason, we now introduce
the so-called Mukhanov-Sasaki variable
\cite{Mukhanov:1981xt,Kodama:1985bj} which is a combination of the
Bardeen potential and of the gauge invariant field
\begin{equation}
v\left(\eta,\bm{x}\right)=a
\left[\delta\varphi^{\left(\mathrm{gi}\right)}
+\varphi^\prime\frac{\Phi_{_{\rm B}}}{\Hu}\right]\, ,
\end{equation}
where ${\cal H}\equiv a'/a$. All the other relevant quantities can be
expressed in terms of $v(\eta,{\bm x})$ which, therefore, fully
characterizes the scalar sector. 

\par

The next step consists in deriving an equation of motion for
$v(\eta,{\bm x})$. This can be done directly from the perturbed
Einstein equations but, here, we first establish the action for the
quantity $v(\eta,{\bm x})$. Expanding the action of the system (\ie
Einstein-Hilbert action plus the action of a scalar field) up to
second order in the perturbations, one obtains~\cite{Mukhanov:1990me}
\begin{equation}
\label{eq:action}
{}^{\left(2\right)}\delta S=\frac{1}{2}
\int{\mathrm{d}^4x
\left[\left(v^\prime\right)^2
-\delta^{ij}\partial_iv\partial_jv
+\frac{\left(a\sqrt{\epsilon_1}\right)^{\prime\prime}}{a\sqrt{\epsilon_1}}
v^2\right]}\, ,
\end{equation}
where $\epsilon_1=1-\Hu^\prime/\Hu^2$ is the first slow-roll parameter
\cite{Schwarz:2001vv,Leach:2002ar}. As the formula
$\ddot{a}/a=H^2(1-\epsilon_1)$ shows, the condition $\epsilon_1<1$ is
in fact sufficient to have inflation. Moreover, we have slow-roll
inflation
\cite{Stewart:1993bc,Martin:1999wa,Martin:2000ak,Leach:2002ar,Schwarz:2001vv}
if $\epsilon_1\ll 1$. In this case, it is easy to show that
$\epsilon_1\simeq (\Mp^2/2V^2)({\rm d}V/{\rm d}\varphi)^2$, \ie
$\epsilon_1$ is in fact a measure of how much the inflaton potential
deviates from a flat potential. Equivalently, according to the
previous considerations, this is also a measure of how much the
inflationary expansion deviates from a pure de Sitter solution. In the
case of power-law inflation, one has $\epsilon_1=(2+\beta)/(1+\beta)$
and, of course, $\epsilon_1=0$ when $\beta =-2$ (de Sitter
solution). The scale factor can also be rewritten as $a(\eta)\simeq
\ell_0(-\eta)^{-1-\epsilon_1}$ and this formula is in fact valid for
any slow-roll model of inflation, \ie for arbitrary shaped
potentials, not necessarily of the exponential type.
In this sense, power-law inflation
with $\beta \lesssim -2$ is a simple representative of all the
slow-roll scenarios. Therefore, the fact that, in this paper, we focus
on this particular model for technical reasons (again, because this
model allows an easy integration of the equations of motion at the
background and perturbative level) does not restrict in any way the
generality of our considerations.

\par

Our next move consists in Fourier transforming the quantity
$v(\eta,{\bm x})$. This is of course justified by the fact that we
work with a linear theory and, hence, all the modes evolve
independently. We have 
\begin{equation}
\label{eq:tfv}
v\left(\eta,\bm{x}\right)=
\frac{1}{\left(2\pi\right)^{3/2}}
\int_{\mathbb{R}^3}{\dd^3\bm{k}\,
v_{\bm{k}}\left(\eta\right)}
\ue^{i\bm{k}\cdot \bm{x}}\, ,
\end{equation}
with $v_{-\bm{k}}=v_{\bm{k}}^*$ because $v(\eta,{\bm x})$ is
real. Then inserting this expansion into Eq.~(\ref{eq:action}), one
arrives at~\cite{Mukhanov:1990me}
\begin{eqnarray}
\label{eq:actionfourier}
{}^{\left(2\right)}\delta S&=&
\int\dd \eta\int \dd^3\bm{k}
\left\lbrace v_{\bm{k}}^\prime{v_{\bm{k}}^*}^\prime
+v_{\bm{k}}v_{\bm{k}}^*\left[
\frac{\left(a\sqrt{\epsilon_1}\right)^{\prime\prime}}{a\sqrt{\epsilon_1}}
-k^ 2\right]\right\rbrace ,
\nonumber\\
\end{eqnarray}
where the integral over ${\bm k}$ is taken over half the Fourier space
only. Next, we define $p_{\bm{k}}$, the variable canonically
conjugate to $v_{\bm{k}}$ 
\begin{equation}
p_{\bm{k}}=\frac{\delta\mathcal{L}}{\delta{v_{\bm{k}}^*}^\prime}
=v_{\bm{k}}^\prime\, ,
\end{equation}
where ${\cal L}$ is the Lagrangian density in Fourier space that can be
derived from Eq.~(\ref{eq:actionfourier}). This allows us to calculate
the Hamiltonian which reads
\begin{equation}
\label{eq:hamilton}
H=\int \dd^3\bm{k}
\left\lbrace p_{\bm{k}}p_{\bm{k}}^*
+v_{\bm{k}}v_{\bm{k}}^*
\left[k^2-
\frac{\left(a\sqrt{\epsilon_1}\right)^{\prime\prime}}
{a\sqrt{\epsilon_1}}\right]\right\rbrace\, .
\end{equation}
This Hamiltonian represents a collection of parametric oscillators
(\ie one oscillator per mode), the time-dependent frequency of which
can be expressed as
\begin{equation}
\label{eq:defomega}
\omega^2\left(\eta, \bm{k}\right)=k^2-
\frac{\left(a\sqrt{\epsilon_1}\right)^{\prime\prime}}
{a\sqrt{\epsilon_1}}\, .
\end{equation}
We see that the frequency depends on the scale factors and its
derivatives (up to the fourth). This means that different
inflationary backgrounds (\ie different inflaton potentials) lead to
different $\omega(\eta,{\bm k})$ and, therefore, to different behaviors
for $v_{\bm k}(\eta)$. From Eq.~(\ref{eq:hamilton}) or
Eq.~(\ref{eq:actionfourier}), it is easy to derive the equation of
motion for the Mukhanov-Sasaki variable. One obtains
\begin{equation}
\label{eq:eomv}
v_{\bm k}''+\omega^2\left(\eta, \bm{k}\right)v_{\bm k}=0,
\end{equation}
which confirms that each mode behaves as a parametric oscillator. Once
a model of inflation has been chosen, the potential $V(\varphi)$ is
known and, hence, the corresponding scale factor can be
calculated. This, in turn, allows us to determine $\omega^2(\eta,{\bm
  k})$ and, then, one can solve the equation of
motion~(\ref{eq:eomv}). However, in order to find the solution for the
Fourier component of the Mukhanov-Sasaki variable, one also needs to
specify the initial conditions. Classically, there does not seem to
exist a natural criterion to choose them. However, when quantization
has been performed, the requirement that it be initially in the vacuum
state of the theory leads to well-defined initial conditions. We now
turn to these questions.

\subsection{Quantization in the Schr\"odinger Picture}
\label{subsec:quantification}

In this section, we review how the cosmological perturbations are
quantized. Very often in the literature, this is done in the
Heisenberg picture. Here, we carry out the quantization in the
Schr\"odinger picture~\cite{Martin:2007bw} because this is more
convenient for the problem we want to investigate in this article. In
order to quantize the system, it is also more convenient to work with
real variables. Therefore, we introduce the following definitions
\begin{equation}
\label{eq:defvRI}
v_{\bm{k}}\equiv \frac{1}{\sqrt{2}}
\left(v_{\bm{k}}^\mathrm{{\rm R}}+
iv_{\bm{k}}^\mathrm{{\rm I}}\right)\, ,\quad
p_{\bm{k}}\equiv \frac{1}{\sqrt{2}}
\left(p_{\bm{k}}^\mathrm{{\rm R}}+
ip_{\bm{k}}^\mathrm{{\rm I}}\right)\, .
\end{equation}
In the Schr\"odinger approach, the quantum state of the system is
described by a wavefunctional, $\Psi\left[v(\eta,{\bm
    x})\right]$. Since we work in Fourier space (and since the theory
is still free in the sense that it does not contain terms with power
higher than two in the Lagrangian), the wavefunctional can also be
factorized into mode components as
\begin{equation}
\Psi\left[v(\eta,{\bm x})\right]=\prod _{\bm k}
\Psi_{\bm k}\left(v_{\bm{k}}^\mathrm{R},
v_{\bm{k}}^\mathrm{I}\right)
=\prod_{\bm k}\Psi^{\rm R}_{\bm k}\left(v_{\bm{k}}^\mathrm{R}\right)
\Psi ^{\rm I}_{\bm k}\left(v_{\bm{k}}^\mathrm{I}\right).
\end{equation}
Quantization is achieved by promoting $v_{\bm k}$ and $p_{\bm k}$ to
quantum operators, $\hat{v}_{\bm k}$ and $\hat{p}_{\bm k}$, and by
requiring the canonical commutation relations
\begin{equation}
\left[\hat{v}_{\bm k}^{\rm R},\hat{p}_{\bm q}^{\rm R}\right]
=i\delta\left({\bm k}-{\bm q}\right), \quad 
\left[\hat{v}_{\bm k}^{\rm I},\hat{p}_{\bm q}^{\rm I}\right]
=i\delta\left({\bm k}-{\bm q}\right).
\end{equation}
These relations admit the following representation
\begin{eqnarray}
\label{eq:ElemActions1}
\hat{v}_{\bm{k}}^\mathrm{R,I}\Psi&=&
v_{\bm{k}}^\mathrm{R,I}\Psi\, ,\quad
\hat{p}_{\bm{k}}^\mathrm{R,I}\Psi=
-i\frac{\partial\Psi}{\partial v_{\bm{k}}^\mathrm{R,I}}\, .
\label{eq:ElemActions2}
\end{eqnarray}

The wavefunctional $\Psi\left[v(\eta,{\bm x})\right]$ obeys the
Schr\"odinger equation which, in this context, is a functional
differential equation. However, since each mode evolves independently,
this functional differential equation can be reduced to an infinite
number of differential equations for each $\Psi_{\bm k}$. Concretely,
we have
\begin{equation}
\label{eq:schrodinger}
i\frac{\Psi_{\bm k}^{\rm R,I}}{\partial \eta}
=\hat{\mathcal{H}}_{\bm{k}}^\mathrm{R,I}\Psi_{\bm k}^{\rm R,I},
\end{equation}
where the Hamiltonian densities
$\hat{\mathcal{H}}_{\bm{k}}^\mathrm{R,I}$, are related to the
Hamiltonian by $\hat{H}=\int \dd^3\bm{k}
\left(\hat{\mathcal{H}}_{\bm{k}}^\mathrm{R}
  +\hat{\mathcal{H}}_{\bm{k}}^\mathrm{I}\right)$. They can be expressed as
\begin{eqnarray}
\hat{\mathcal{H}}_{\bm{k}}^\mathrm{R,I} &=&
-\frac{1}{2}\frac{\partial^2}{\partial \left(v_{\bm k}^{\rm R,I}\right)^2}
+\frac12\omega^2(\eta,{\bm k})
\left(\hat{v}_{\bm{k}}^\mathrm{R,I}\right)^2,
\label{eq:hamiltoniandensity}
\end{eqnarray}
where we have made use of the representations~(\ref{eq:ElemActions1}).

\par

We are now in a position where we can solve the Schr\"odinger
equation. Let us consider the following Gaussian state
\begin{equation}
\label{eq:gaussianpsi}
\Psi_{\bm k}^{\rm R,I}\left(\eta,v_{\bm k}^{\rm R,I}\right)
=N_{\bm k}(\eta){\rm e}^{-\Omega_{\bm k}(\eta)\left(v_{\bm k}^{\rm R,I}\right)^2}.
\end{equation}
The functions $N_{\bm k}(\eta)$ and $\Omega _{\bm k}(\eta)$ are time
dependent and do not carry the subscripts ``${\rm R}$'' and/or ``${\rm
  I}$'' because they are the same for the wavefunctions of the real
and imaginary parts of the Mukhanov-Sasaki variable (see below). Then,
inserting $\Psi_{\bm k}$ given by Eq.~(\ref{eq:gaussianpsi}) into the
Schr\"odinger equation~(\ref{eq:schrodinger}) implies that $N_{\bm k}$
and $\Omega _{\bm k}$ obey the differential equations
\begin{equation}
i\frac{N_{\bm k}'}{N_{\bm k}}=\Omega _{\bm k}, \quad 
\Omega_{\bm k}'=-2i\Omega_{\bm k}^2+\frac{i}{2}\omega^2(\eta,{\bm k}).
\label{eq:equapsi}
\end{equation}
The solutions can be easily found and read
\begin{equation}
\label{eq:solpsi}
\left \vert N_{\bm k}\right \vert 
=\left(\frac{2\Rea  \Omega _{\bm k}}{\pi}\right)^{1/4}, \quad 
\Omega_{\bm k}=-\frac{i}{2}\frac{f_{\bm k}'}{f_{\bm k}},
\end{equation}
where $f_{\bm k}$ is a function obeying the equation $f_{\bm
  k}''+\omega^2f_{\bm k}=0$, that is to say exactly
Eq.~(\ref{eq:eomv}). The first equation~(\ref{eq:solpsi}) guarantees
that the wavefunction is properly normalized, \ie
\begin{equation}
\int \Psi_{\bm k}^{\rm R,I}\Psi_{\bm k}^{\rm R,I}{}^*{\rm d}v_{\bm k}^{\rm R,I}=1.
\end{equation}

\par

Let us now discuss the initial conditions. The fundamental assumption
of inflation is that the perturbations are initially in their ground
state. At the beginning of inflation, all the modes of astrophysical
interest today have a physical wavelength smaller than the Hubble
radius, \ie $k/(aH)\rightarrow \infty$. In this regime, one has
$\omega^2(\eta,{\bm k})\rightarrow k^2$ and each mode now behaves as
an harmonic oscillator (as opposed to a parametric oscillator in the
generic case) with frequency $\omega=k$. As a consequence, the
differential equation for $f_{\bm k}(\eta)$ can easily be solved and
the solution reads $f_{\bm k}=A_{\bm k}{\rm e}^{ik\eta}+B_{\bm k}{\rm
  e}^{-ik\eta}$, $A_{\bm k}$ and $B_{\bm k}$ being integration
constants. Upon using the second equation~(\ref{eq:solpsi}), one has
\begin{equation}
\Omega_{\bm k}\rightarrow \frac{k}{2}\frac{A_{\bm
  k}{\rm e}^{ik\eta}-B_{\bm k}{\rm e}^{-ik\eta}}{A_{\bm k}{\rm e}^{ik\eta}
+B_{\bm k}{\rm e}^{-ik\eta}}.
\end{equation}
The wavefunction~(\ref{eq:gaussianpsi}) represents the ground state
wavefunction of an harmonic oscillator if $\Omega_{\bm
  k}=k/2$. Therefore, one must choose the initial conditions such that
$B_{\bm k}=0$. Moreover, it is easy to check that the Wronskian
$W\equiv f_{\bm k}'f_{\bm k}^*-f_{\bm k}'^*f_{\bm k}$ is a conserved
quantity, ${\rm d}W/{\rm d}\eta=0$, thanks to the equation of motion
of $f_{\bm k}$. Straightforward calculation leads to $W=2ik\left\vert
  A_{\bm k}\right \vert ^2$.  In the Heisenberg picture the canonical
commutation relations require that $W=i$. Even if in the Schr\"odinger
picture presently used, the specific value of $W$ is irrelevant since
it cancels out on all calculable physical quantities, this value is
conventionally adopted, which amounts to setting $A_{\bm
  k}=1/\sqrt{2k}$. As announced, requiring the initial state to be the
ground state has completely fixed the initial conditions. We see that
Eq.~(\ref{eq:eomv}) (or, equivalently, the equation for $f_{\bm k}$)
should thus be solved with the boundary condition
\begin{equation}
\label{eq:asymp}
\lim _{k/(aH)\rightarrow +\infty}f_{\bm k}=\frac{1}{\sqrt{2k}}{\rm e}^{ik\eta}.
\end{equation}
This choice of initial conditions is referred to as the Bunch-Davies vacuum.

\subsection{The Power Spectrum}
\label{subsec:powerspectrum}

Let us now turn to the calculation of the power spectrum and
first introduce the two-point correlation function,
defined by
\begin{widetext}
\begin{equation}
\left \langle \Psi\left \vert \hat{v}(\eta ,{\bm x})
\hat{v}(\eta,{\bm x}+{\bm r})\right \vert \Psi\right \rangle
=\int \prod _{\bm k}{\rm d}v_{\bm k}^{\rm R}{\rm d}v_{\bm k}^{\rm I}
\Psi_{\bm k}^*(v_{\bm k}^{\rm R},v_{\bm k}^{\rm I})
v(\eta ,{\bm x})
v(\eta,{\bm x}+{\bm r})
\Psi_{\bm k}(v_{\bm k}^{\rm R},v_{\bm k}^{\rm I}).
\end{equation}
The next step consists in using the Fourier transform of the
Mukhanov-Sasaki variable, see Eq.~(\ref{eq:tfv}) and the explicit form
of the wavefunction of Eq.~(\ref{eq:gaussianpsi}). One arrives at
\begin{equation}
\left \langle \Psi\left \vert \hat{v}(\eta ,{\bm x})
\hat{v}(\eta,{\bm x}+{\bm r})\right \vert \Psi\right \rangle
=\frac{1}{(2\pi)^3}
\int {\rm d}{\bm p}\, {\rm d}{\bm q}\,
{\rm e}^{i{\bm p}\cdot {\bm x}}\,
{\rm e}^{i{\bm q}\cdot ({\bm x}+{\bm r})}
\prod _{\bm k}\left(\frac{2\Rea  \Omega _{\bm k}}{\pi}\right)
\int \prod _{\bm k}{\rm d}v_{\bm k}^{\rm R}{\rm d}v_{\bm k}^{\rm I}
{\rm e}^{-2\sum_{\bm k} \Rea \Omega_{\bm k}
\left[(v_{\bm k}^{\rm R})^2+(v_{\bm k}^{\rm I})^2\right]}
v_{\bm p}v_{\bm q}.
\end{equation}
If $\bm{p}\neq \pm \bm{q}$, the result of the integration is zero since the integrand
(up to the Gaussian weight) becomes linear in $v_{\bm
  p}^{\rm R,I}$ or $v_{\bm q}^{\rm R,I}$. If ${\bm p}={\bm q}$, then
the only non-linear term in the integrand is given by $\left[\left(v_{\bm
      p}^{\rm R}\right)^2-\left(v_{\bm p}^{\rm I}\right)^2\right]/2$.
Each term contributes the same amount, so the difference
vanishes. The only possibility left is therefore ${\bm p}=-{\bm
  q}$, such that $v_{\bm p}v_{\bm q}=\left[\left(v_{\bm p}^{\rm
      R}\right)^2+\left(v_{\bm p}^{\rm I}\right)^2\right]/2$, the
factor $1/2$ coming from the definition of $v_{\bm k}^{\rm R,I}$, see
Eqs.~(\ref{eq:defvRI}). This leads to
\begin{equation}
\left \langle \Psi\left \vert \hat{v}(\eta ,{\bm x})
\hat{v}(\eta,{\bm x}+{\bm r})\right \vert \Psi\right \rangle
=\frac{2}{(2\pi)^3}\frac12
\int {\rm d}{\bm p}\, 
{\rm e}^{-i{\bm p}\cdot {\bm r}}
\prod _{\bm k}^N\left(\frac{2\Rea  \Omega _{\bm k}}{\pi}\right)
\int \prod _{\bm k}^N{\rm d}v_{\bm k}^{\rm R}{\rm d}v_{\bm k}^{\rm I}
{\rm e}^{-2\sum_{\bm k} \Rea \Omega_{\bm k}
\left[(v_{\bm k}^{\rm R})^2+(v_{\bm k}^{\rm I})^2\right]}
\left(v_{\bm p}^{\rm R}\right)^2,
\end{equation}
the factor of $2$ originating from the fact that we have two
contributions, one given by the term $\left(v_{\bm p}^{\rm
    R}\right)^2$ and the other by $\left(v_{\bm p}^{\rm I}\right)^2$.
The Gaussian integrals can easily be carried out. They are all of the
form ``$\int {\rm d}x\, {\rm e}^{-\alpha x^2}$'', except of course the
one over $v_{\bm p}^{\rm R}$ which is of the form ``$\int {\rm d}x\,
x^2\, {\rm e}^{-\alpha x^2}$''. As a consequence, one obtains
\begin{equation}
\left \langle \Psi\left \vert \hat{v}(\eta ,{\bm x})
\hat{v}(\eta,{\bm x}+{\bm r})\right \vert \Psi\right \rangle
=\frac{1}{(2\pi)^3}
\int {\rm d}{\bm p}\, 
{\rm e}^{-i{\bm p}\cdot {\bm r}}
\prod _{\bm k}^N\left(\frac{2\Rea  \Omega _{\bm k}}{\pi}\right)
\frac12\left[\frac{\sqrt{\pi}}{(\sqrt{2\Rea  \Omega _{\bm p}})^3}\right]
\prod _{\bm k}^N\left(\frac{\sqrt{\pi}}{\sqrt{2\Rea  \Omega _{\bm k}}}\right)
\prod _{\bm k}^{N-1}\left(\frac{\sqrt{\pi}}{\sqrt{2\Rea  \Omega _{\bm k}}}\right)\, .
\end{equation}
\end{widetext}
The infinite product ``$\prod_{\bm k}^{N-1}$'' means a product over
all the wavevectors but ${\bm p}$. One can always write this product
as ``$\sqrt{2\Rea  \Omega _{\bm p}/\pi}\prod_{\bm k}^N$'', then the two
last infinite products in the above expression exactly cancel the
first one. Therefore, we are left with
\begin{equation}
\left \langle \Psi\left \vert \hat{v}(\eta ,{\bm x})
\hat{v}(\eta,{\bm x}+{\bm r})\right \vert \Psi\right \rangle
=\frac{1}{(2\pi)^3}
\int {\rm d}{\bm p}\, 
{\rm e}^{-i{\bm p}\cdot {\bm r}}\frac{1}{4\Rea  \Omega_{\bm p}}\, .
\label{eq:calculmeanv2}
\end{equation}
We now need to express $\Rea  \Omega_{\bm p}$ in terms of the function
$f_{\bm p}$. From the second Eq.~(\ref{eq:solpsi}), one easily shows
that
\begin{equation}
\Rea  \Omega_{\bm p}=-\frac{i}{4}\frac{W}{\vert f_{\bm p}\vert^2},
\end{equation}
and we obtain our final expression for the two-point correlation
function
\begin{eqnarray}
\left \langle \Psi\left \vert \hat{v}(\eta ,{\bm x})
\hat{v}(\eta,{\bm x}+{\bm r})\right \vert \Psi\right \rangle
&\hskip-2mm=&\hskip-2mm\frac{1}{(2\pi)^3}
\int {\rm d}{\bm p}\, 
{\rm e}^{-i{\bm p}\cdot {\bm r}}\frac{i}{W}\vert f_{\bm p}\vert^2
\cr
&\hskip-2mm=&\hskip-2mm \frac{1}{2\pi^2}\int_0^{+\infty}
\frac{{\rm d}p}{p}\frac{\sin pr}{pr}p^3\vert f_{\bm p}\vert^2\, ,
\cr
&&
\end{eqnarray}
where, in the last expression, we have used our choice $W=i$. The
power spectrum is just defined as the square of the Fourier amplitude
per logarithmic interval at a given scale, \ie
\begin{equation}
{\cal P}_v(k)=\frac{k^3}{2\pi^2}\vert f_{\bm k}\vert^2.
\end{equation}

The same manipulations allow us to express the two-point correlation
of two Fourier amplitudes. It can be written as 
\begin{equation}
\left \langle \Psi\left \vert \hat{v}_{\bm k}
\hat{v}_{\bm p}^*\right \vert \Psi\right \rangle 
=\int \prod _{\bm q}{\rm d}v_{\bm q}^{\rm R}{\rm d}v_{\bm q}^{\rm I}
\Psi_{\bm q}^*\, \hat{v}_{\bm k}\, 
\hat{v}_{\bm p}^*\, \Psi _{\bm q}.
\end{equation}
This integral is non-vanishing only if ${\bm k}={\bm p}$ (otherwise
one has to integrate an odd function) and receives two contributions,
one from $\left(v_{\bm k}^{\rm R}\right)^2$ and the other from
$\left(v_{\bm k}^{\rm I}\right)^2$. Repeating calculations already
performed before, one finally arrives at
\begin{equation}
\label{eq:twopointfourier}
\left \langle \Psi\left \vert \hat{v}_{\bm k}
\hat{v}_{\bm p}^*\right \vert \Psi\right \rangle 
=\frac{2\pi^2}{k^3}{\cal P}_v(k)\delta\left({\bm k}-{\bm p}\right).
\end{equation}

We now need to explain how the cosmological perturbations of
quantum-mechanical origin studied above are related to observables in
cosmology. This is the goal of the next section.

\subsection{From Quantum Fluctuations to CMB Anisotropies}
\label{subsec:quantumcmb}

The presence of quantum fluctuations in the inflaton and gravitational
fields has many observational implications. Here, we focus on one of
them, namely the existence of CMB temperature anisotropies. The
importance of this observable is that we now have at our disposal very
high accuracy measurements of those
anisotropies~\cite{Larson:2010gs,Komatsu:2010fb}. Moreover, even more
accurate data will be released soon~\cite{Bouchet:2009tr}. The
relation between the temperature fluctuations along a given direction
${\bm e}$ and the cosmological perturbations is expressed by the
so-called Sachs-Wolfe effect~\cite{Sachs:1967er,Panek:1986wr}. A
simplified version of this result, valid on large angular scales only,
can be written as~\cite{Panek:1986wr}
\begin{equation}
\label{eq:sw}
\frac{\delta T}{T}\left({\bm e}\right)
=\frac{1}{5}\zeta\left[\eta_{\rm \ell ss},-{\bm e}
\left(\eta_{\rm \ell ss}-\eta_0\right)+{\bm x}_0\right],
\end{equation}
where $T$ represents the averaged background temperature, \ie $T\simeq
2.7\, \mbox{K}$, $\eta _{\ell \mathrm{ss}}$ is the conformal time at emission
(that is to say at the surface of last scattering) and $\eta _0$ is
the present conformal time. The vector ${\bm x}_0$ landmarks the place
of reception, in the present case Earth (or a satellite orbiting the
Earth). The quantity $\zeta $ denotes the curvature perturbation. It
is related to the Bardeen potential defined in
Eq.~(\ref{eq:defbardeen}) through the following expression
\cite{Mukhanov:1990me,Lyth:1984gv,Martin:1997zd}
\begin{equation}
  \zeta=\frac{2}{3}\frac{{\cal H}^{-1}\Phi_{_{\rm B}}'
+\Phi_{_{\rm B}}}{1+w}+\Phi_{_{\rm B}},
\end{equation}
where $w\equiv p/\rho$ is the equation of state parameter, that is to
say the energy density to pressure ratio of the dominant fluid. For
instance, for the matter dominated era ($w=0$), during which
recombination takes place (at a redshift of $z_{\rm \ell ss}\simeq
1100$), on large scales, one simply has $\zeta\simeq 5\Phi_{_{\rm
    B}}/3$ since the Bardeen potential is constant. The importance of
$\zeta $ lies in the fact that it is a conserved quantity on large
scales~\cite{Lyth:1984gv,Martin:1997zd}. Therefore, its spectrum,
calculated at the end of inflation, can directly be propagated to the
recombination time as it is not sensitive to the details of the
cosmological evolution, in particular to those of the
complicated reheating
era~\cite{Traschen:1990sw,Shtanov:1994ce,Kofman:1997yn,Finelli:1998bu,Jedamzik:2010dq}. The
curvature perturbation can also be expressed in terms of the
Mukhanov-Sasaki variable as
\begin{equation}
\label{eq:linkzetav}
\zeta =\frac{1}{a\sqrt{2\epsilon_1}}\frac{v}{\Mp}.
\end{equation}
Finally, in the framework of the theory of inflationary cosmological
perturbations of quantum-mechanical origin, we have seen that $v$ is
in fact an operator. This implies that $\zeta $ and $\delta T/T$ are
also quantum operators and, for this reason, from now on, we will
denote them with a hat.

\par

Since the operator $\widehat{\delta T}/T$ lives on the celestial
sphere, it can be expanded over the spherical harmonic basis
according to
\begin{equation}
\label{eq:decompsh}
\frac{\widehat{\delta T}}{T}({\bm e})=\sum _{\ell=2}^{\infty }
\sum_{m=-\ell}^{m=\ell}\hat{a}_{\ell m}Y_{\ell m}(\theta, \phi),
\end{equation}
where $\theta $ and $\phi$ are the angles defining the direction along
which the vector ${\bm e}$ is pointing. Then, the angular two-point
correlation function can be expressed in terms of the multipole
moments $C_{\ell}$ as
\begin{equation}
\label{eq:corralm}
\left\langle \Psi\left \vert \hat{a}_{\ell m}\hat{a}_{\ell 'm'}^{*}
\right \vert \Psi\right \rangle=C_{\ell }\delta _{\ell \ell'}\delta _{mm'},
\end{equation}
and, as a consequence, the two-point correlation function of the
temperature fluctuations operator can be written as
\begin{equation}
\left\langle \Psi\left \vert \frac{\widehat{\delta T}}{T}({\bm e}_1)
\frac{\widehat{\delta T}}{T}({\bm e}_2)
\right \vert \Psi\right \rangle=\frac{1}{4\pi}\sum_{\ell=2}^{\infty}
(2\ell +1)C_{\ell }P_{\ell }({\bm e}_1\cdot {\bm e}_2),
\end{equation}
the quantity $P_{\ell }$ denoting Legendre polynomials.

\par

In order to pursue our demonstration that the CMB anisotropies are
entirely determined by the quantum fluctuations, let us now express
the multipole moments in term of the cosmological perturbation power
spectrum. Upon using Eqs.~(\ref{eq:sw}) and~(\ref{eq:decompsh}), one
obtains
\begin{equation}
\label{eq:alm}
\hat{a}_{\ell m}=\frac{1}{(2\pi)^{3/2}}\!\!\int\!\!{\rm d}\Omega_{\bm e}
\dd {\bm k}\, \frac{\hat{\zeta}_{\bm k}(\eta _{\rm \ell ss})}{5} 
{\rm e}^{-i{\bm k}\cdot \left[{\bm e}(\eta _{\rm \ell ss}-\eta_0)-{\bm x}_0\right]}
Y^*_{\ell m}({\bm e})
\end{equation}
and, from this expression, it is easy to show that
\begin{equation}
\label{eq:cl}
C_{\ell }=\frac{1}{2a^2\Mp^2\epsilon_1}
\frac{4\pi}{25}\int \frac{{\rm d}k}{k}j_{\ell}^2
\left[k\left(\eta _0-\eta_{\rm \ell ss}\right)\right]
{\cal P}_v(k),
\end{equation}
where $j_{\ell }$ is a spherical Bessel function and where we used
Eq.~(\ref{eq:twopointfourier}) to show that
\begin{equation}
\left \langle \Psi\left \vert \hat{\zeta}_{\bm k}\, 
\hat{\zeta}_{\bm p}^*\right \vert \Psi\right \rangle 
=\frac{1}{2a^2\Mp^2\epsilon_1}
\frac{2\pi^2}{k^3}{\cal P}_v(k)\delta\left({\bm k}-{\bm p}\right).
\end{equation}
We see that $C_{\ell}$ is given by an integral over wavenumbers of the
Mukhanov-Sasaki power spectrum times a quantity that can be viewed as
a ``transfer matrix $j_{l{\bm k}}\equiv j_{\ell}^2\left[k\left(\eta
    _0-\eta_{\rm \ell ss}\right)\right]$'' which allows us to
``translate'' a three dimensional spatial frequency ${\bm k}$ into a
two-dimensional spatial frequency $\ell $ on the celestial sphere. We
emphasize again that the above result is valid on large scales only;
otherwise the integral in Eq.~(\ref{eq:cl}) contains another transfer
function $T_{\zeta}(k)$ which takes into account the subsequent
evolution of the modes when they re-enter the Hubble radius after
inflation. Since $\zeta $ is a conserved quantity, we have
$T_{\zeta}(k\rightarrow 0)=1$.

\par

Finally, let us also notice that Eq.~(\ref{eq:alm}) implies that
$\langle \Psi \vert \hat{a}_{\ell m}\vert \Psi \rangle =0$ since
$\langle \Psi \vert \hat{\zeta}_{\bm k}\vert \Psi \rangle =0$. Of
course, this also means that $\langle \Psi \vert \widehat{\delta
  T}/T\vert \Psi \rangle =0$.

\subsection{Inflationary Predictions}
\label{subsec:prediction}

We have just seen that, in order to calculate the CMB multipole
moments, we need to evaluate the curvature perturbation power
spectrum. In this section, we calculate this quantity for power
law inflation.

\par

The first step consists in solving the equation of
motion~(\ref{eq:eomv}). Upon using Eq.~(\ref{eq:defomega}), one
obtains the time dependence of the frequency of the parametric
oscillator, which reads
\begin{equation}
\omega^2(\eta,{\bm k})=k^2-\frac{\beta (\beta +1)}{\eta^2}.
\end{equation}
From this expression, one sees that there are two regimes depending on
whether the first term is dominant or subdominant. The Hubble radius
is given by $\ell_{_{\rm H}}\equiv 1/H=a\eta/(1+\beta)$ and the
Fourier mode wavelength can be expressed in terms of the co-moving
wavenumber as $\lambda =2\pi a/k$. The first terms dominates if $\vert
k\eta \vert\gg 1$ or, equivalently, $\lambda \ll \ell_{_{\rm H}}$. In
this case $\omega \simeq k$ and we expect the mode function to
oscillate as it would in Minkowski spacetime since, at those scales,
spacetime curvature is negligible for the mode evolution.  On the
contrary, if $\vert k\eta \vert \ll 1$, or $\lambda\gg \ell_{_{\rm
    H}}$, one has $\omega \sim 1/\eta$, so curvature dominates and one
obtains one growing mode and one decaying mode.  These arguments are
confirmed when one studies the exact solution for the mode function
$f_{\bm k}$. It can be expressed in terms of Bessel functions
$J_{\nu}(z)$ as~\cite{Abramovitz:1970aa,Gradshteyn:1965aa}
\begin{equation}
\label{eq:solbessel}
f_{\bm k}=\left(-k\eta\right)^{1/2}
\left[C_{\bm k}J_{\beta+1/2}(-k\eta)+D_{\bm k}J_{-(\beta +1/2)}(-k\eta)\right],
\end{equation}
where $C_{\bm k}$ and $D_{\bm k}$ are two integration constants. In
order to match the initial vacuum behavior~(\ref{eq:asymp}), one must
choose
\begin{equation}
\label{eq:bunchdaviesstandard}
C_{\bm k}=-D_{\bm k}{\rm e}^{i\pi (\beta +1/2)}, \quad 
D_{\bm k}=\frac{i}{2}\sqrt{\frac{\pi}{k}}\frac{{\rm e}^{-i\pi/4-i\pi(\beta +1/2)/2}}
{\sin[\pi (\beta +1/2)]}.
\end{equation}
In particular, one notices that both coefficients $C_{\bm k}$ and
$D_{\bm k}$ scale as $1/\sqrt{k}$.

\par

Since we want to evaluate the power spectrum on large scales, it is
sufficient to take the limit $k\eta \rightarrow 0$ in
Eq.~(\ref{eq:solbessel}). Then, one is led to
\begin{eqnarray}
{\cal P}_{\zeta}\bigl \vert _{\rm stand}&=&
\frac{1}{2a^2\Mp^2\epsilon_1}{\cal P}_{v}(k)\nonumber\\
&=&\frac{1}{\pi\epsilon_1m_{_{\rm Pl}}^2\ell_0^2}f(\beta)k^{2\beta +4}
\equiv A_{_{\rm S}}k^{n_{_{\rm S}}-1},
\label{eq:powerstandard}
\end{eqnarray}
where $\Mp=m_{_{\rm Pl}}/\sqrt{8\pi}$ and the function $f(\beta)$ is
defined by~\cite{Martin:1999wa}
\begin{equation}
\label{eq:deffuncf}
f(\beta)\equiv \frac{1}{\pi}\left[\frac{\Gamma\left(-\beta -1/2\right)}
{2^{1+\beta}}\right]^2,
\end{equation}
where $\Gamma (z)$ is the Euler integral of the first
kind~\cite{Abramovitz:1970aa,Gradshteyn:1965aa}. This function is such
that, for the de Sitter case $\beta=-2$, one has $f(\beta=-2)=1$. The
scalar spectral index $n_{_{\rm S}}=2\beta +5$ and, for solutions
close to the de Sitter solutions, one has $n_{_{\rm S}}\simeq 1$, \ie
we have an almost scale invariant power spectrum. As discussed before,
the deviations from scale invariance, are related to the deviation
from the de Sitter case $\beta=-2$. This conclusion is in fact valid
for any slow-roll models. The amplitude $A_{_{\rm S}}$ determines the
level of the temperature fluctuations observed in the sky, namely
$\delta T/T\sim 10^{-5}$.

\par

Finally, let us evaluate the multipole moments explicitly. Upon using
Eq.~(\ref{eq:cl}) and the expression of the power spectrum established
above, one arrives at
\begin{equation}
C_{\ell}=\frac{\pi^{3/2}\Gamma\left[(3-n_{_{\rm S}})/2\right]
\Gamma\left[\ell +\left(n_{_{\rm S}}-1\right)/2\right]}
{\Gamma\left[(4-n_{_{\rm S}})/2\right]
\Gamma\left[\ell +2-\left(n_{_{\rm S}}-1\right)/2\right]}
\left(r_{\rm \ell ss}\right)^{1-n_{_{\rm S}}}\frac{A_{_{\rm S}}}{25},
\end{equation}
where we have defined $r_{\rm \ell ss}\equiv \eta _0-\eta _{\rm \ell
  ss}$. Since this equation has been derived for large scales, roughly
speaking one can estimate it to be valid in the regime $\ell \ll
20$. For $n_{_{\rm S}}\simeq 1$, the above expression implies that
$C_{\ell}\propto 1/[\ell (\ell+1)]$.

\par

Of course, in the real world, the argument goes the other way
round. From measurements of the CMB anisotropies, we observe that, on
large scales, $C_{\ell}\propto 1/[\ell (\ell+1)]$ and, therefore, we
deduce that the corresponding power spectrum is close to scale
invariance, \ie $n_{_{\rm S}}\simeq 1$. Obviously, this also means
that a spectrum that is not very close to scale invariance is now
ruled out (more precisely, the WMAP data indicate that $1-n_{_{\rm
    S}}=0.018^{+0.019}_{-0.02}$~\cite{Larson:2010gs,Komatsu:2010fb,Martin:2006rs}). As
already emphasized, the great success of inflation is that it
precisely leads to such a power spectrum.

\par

It should also be clear that the above discussion, although perfectly
correct at the level of principles, is oversimplified at the technical
level. The multipole moments are in fact computed at any scale (\ie
for any value of $\ell$) by means of numerical calculations (since, in
the most general case, they are solution of more involved differential
equations)~\cite{Ringeval:2007am}. Moreover, their shape is not only
determined by the spectral index but is also affected by the other
cosmological parameters. The constraints on the different inflationary
models are then obtained by a Markov Chain exploration of the
parameter space~\cite{Lewis:2002ah}. But these technical
considerations do not affect the considerations presented in this
paper. Once again, as far as physical principles are concerned, the
discussion presented is this section is accurate.

\section{The Cosmological Measurement Problem}
\label{sec:measureproblem}

\subsection{Squeezed State}
\label{subsec:squeezed}

In this section, we study in more detail the properties of the quantum
state in which the cosmological perturbations are placed
\cite{Grishchuk:1990bj,Polarski:1995jg,Kiefer:1998pb,Kiefer:2008ku}. As
already mentioned around Eq.~(\ref{eq:gaussianpsi}), it is described by
the wavefunction
\begin{eqnarray}
\label{eq:wavefunction}
\hskip-4mm
\Psi_{\bm k}\left(\eta ,v_{\bm k}^{\rm R},v_{\bm k}^{\rm I}\right)
&=& \left(\frac{2\Rea  \Omega _{\bm k}}{\pi}\right)^{1/2}
{\rm e}^{-\Omega _{\bm k}\left[\left(v_{\bm k}^{\rm R}\right)^2+
\left(v_{\bm k}^{\rm I}\right)^2\right]} \\ 
&=& \left(\frac{2\Rea  \Omega _{\bm k}}{\pi}\right)^{1/2}
{\rm e}^{-2\Omega _{\bm k}(\eta)v_{\bm k}v_{\bm k}^*}.
\end{eqnarray}
We see that this quantum state is completely known once the time
dependence of $\Omega_{\bm k}(\eta)$ has been determined. The
differential equation controlling the evolution of $\Omega_{\bm
  k}(\eta)$ is given by the second of Eqs.~(\ref{eq:equapsi}). This
equation is a Ricatti equation (\ie a first order, non-linear,
differential equation). As is well known, it can always be reduced to
a second order but linear differential equation. As already mentioned,
this is achieved through the change of variable $\Omega_{\bm
  k}=-if_{\bm k}'/(2f_{\bm k})$. The function $f_{\bm k}(\eta)$ obeys
$f_{\bm k}''+\omega ^2f_{\bm k}=0$ and has been solved in
Eq.~(\ref{eq:solbessel}). In the small-scale limit, one has $\Omega
_{\bm k}\rightarrow k/2$ and the wavefunction~(\ref{eq:wavefunction})
is the ground state of an harmonic oscillator. In the large-scale
limit, a lengthy but straightforward calculation leads to
\begin{widetext}
\begin{equation}
\label{eq:omegasuperhubble}
\frac{\Omega_{\bm k}(\eta)}{k}=-\frac{i}{2k\eta}(1+\beta)
-\frac{i}{4(\beta +3/2)}(-k\eta)-\frac{i}{\pi}2^{2\beta}
\sin\left(2\pi \beta\right)
\Gamma^2\left(\beta +\frac32\right)
(-k\eta)^{-2\beta -2}
+\frac{\pi \, 2^{2\beta +1}}{\Gamma^2(-\beta -1/2)}
(-k\eta)^{-2\beta -2}
+\cdots .
\end{equation}
\end{widetext}
{}From this expression, one deduces that
\begin{equation}
\label{eq:reomegaeta}
\Rea  \Omega _{\bm k}(\eta)=\frac{k\, \pi \, 2^{2\beta +1}}{\Gamma^2(-\beta -1/2)}
(-k\eta)^{-2\beta -2}
+\cdots \rightarrow 0, 
\end{equation}
and
\begin{equation}
\label{eq:imomegaeta}
\Ima  \Omega _{\bm k}(\eta) = -\frac{1}{2\eta}(1+\beta)+\cdots
=-\frac{a'}{2a}\rightarrow \infty,
\end{equation}
where the limits are taken in the super-Hubble regime in which $k\eta
\rightarrow 0$.

\par

We have mentioned above that the Ricatti equation~(\ref{eq:equapsi})
can always be reduced to a linear second order differential
equation. Of course, it can also be expressed as two linear, first
order, differential equations. Therefore, one can introduce the
functions $u_{\bm k}(\eta)$ and $v_{\bm k}(\eta)$ such that $f_{\bm
  k}\equiv \left(u_{\bm k}+v_{\bm k}^*\right)/\sqrt{2k}$, the
normalization $1/\sqrt{2k}$ being introduced for convenience. Then it
is easy to show that these two functions obey
\begin{eqnarray}
u_{\bm k}' &=& iku_{\bm k}+\frac{(a\sqrt{\epsilon_1})'}{a\sqrt{\epsilon_1}}
v_{\bm k}^*, \\
v_{\bm k}' &=& ikv_{\bm k}+\frac{(a\sqrt{\epsilon_1})'}{a\sqrt{\epsilon_1}}
u_{\bm k}^*.
\end{eqnarray}
The Wronskian $W=f_{\bm k}'f_{\bm k}^*-f_{\bm k}'^*f_{\bm k}$ can be
straightforwardly evaluated as $W=i\left(\vert u_{\bm k}\vert^2-\vert
  v_{\bm k}\vert^2\right)$. This means that, if we want to work with
the choice $W=i$, one must have $\vert u_{\bm k}\vert^2-\vert v_{\bm
  k}\vert^2=1$. This suggests to introduce the following
parametrization
\begin{eqnarray}
\label{eq:eom:u}
u_{\bm k}(\eta) &=& {\rm e}^{i\theta_{\bm k}}\cosh r_{\bm k},\\
\label{eq:eom:v}
v_{\bm k}(\eta ) &=& {\rm e}^{-i\theta_{\bm k}+2i\phi_{\bm k}}\sinh r_{\bm k}.
\end{eqnarray}
The three functions $r_{\bm k}(\eta)$, $\theta_{\bm k}(\eta)$ and
$\phi_{\bm k}(\eta)$ are called the squeezing parameter, rotation
angle and squeezing angle respectively. It is clear that the knowledge of
these three functions is equivalent to that of the function
$\Omega_{\bm k}(\eta)$ and, therefore, of the wavefunction. Upon using
Eqs.~(\ref{eq:eom:u}) and~(\ref{eq:eom:v}), it is easy to show that
\begin{eqnarray}
\label{eq:r}
r_{\bm k}' &=& \frac{\left(a\sqrt{\epsilon_1}\right)'}{a\sqrt{\epsilon_1}}
\cos\left(2\phi_{\bm k}\right), \\
\label{eq:phi}
\phi_{\bm k}' &=& k-
\frac{\left(a\sqrt{\epsilon_1}\right)'}{a\sqrt{\epsilon_1}}
\coth \left(2r_{\bm k}\right)\sin \left(2\phi_{\bm k}\right),\\
\label{eq:theta}
\theta_{\bm k}' &=& k-
\frac{\left(a\sqrt{\epsilon_1}\right)'}{a\sqrt{\epsilon_1}}
\tanh r_{\bm k}\sin \left(2\phi_{\bm k}\right).
\end{eqnarray}
The explicit relation between $\Omega _{\bm k}$ and the three squeezing 
parameters is given by
\begin{equation}
\label{eq:grandomega}
\Omega _{\bm k}=\frac{k}{2}\frac{\cosh r_{\bm k}-{\rm e}^{-2i\phi_{\bm k}}
\sinh r_{\bm k}}{\cosh r_{\bm k}+{\rm e}^{-2i\phi_{\bm k}}
\sinh r_{\bm k}}-i\frac{a'}{2a},
\end{equation}
from which one deduces that
\begin{eqnarray}
\label{eq:reomega}
\Rea  \Omega _{\bm k} &=& \frac{k}{2}\frac{1}{
\cosh \left(2r_{\bm k}\right)+\cos \left(2 \phi_{\bm k}\right)
\sinh \left(2r_{\bm k}\right)}, \\
\label{eq:imomega}
\Ima  \Omega _{\bm k} &=& 
\frac{k}{2}\frac{\sin \left(2\phi_{\bm k}\right)
\sinh\left(2r_{\bm k}\right)}{
\cosh \left(2r_{\bm k}\right)+\cos \left(2 \phi_{\bm k}\right)
\sinh \left(2r_{\bm k}\right)}-\frac{a'}{2a}.\nonumber\\
\end{eqnarray}

Eqs.~(\ref{eq:r}), (\ref{eq:phi}) and~(\ref{eq:theta}) are highly non
linear differential equations and cannot be solved in general. We
notice that Eqs.~(\ref{eq:r}) and~(\ref{eq:phi}) are in fact decoupled
from Eq.~(\ref{eq:theta}). Therefore, they can be solved in a first
step and then the solutions can be inserted in Eq.~(\ref{eq:theta}) to
find the behavior of $\theta_{\bm k}$. In the case of power-law
inflation, one can find explicit solutions for the de Sitter case,
$\beta =-2$. Although this is not a solution for an arbitrary value of
$\beta$, it is sufficient to understand the main features of the
phenomenon of squeezing. One obtains
\begin{eqnarray}
\label{eq:solr}
r_{\bm k}(\eta) &=& -\arg\sinh\left(\frac{1}{2k\eta}\right), \\
\label{eq:solphi}
\phi_{\bm k}(\eta) &=& \frac{\pi}{4}+\frac12 
\arctan\left(\frac{1}{2k\eta}\right).
\end{eqnarray}
Therefore, we see that, initially in the sub-Hubble limit, $r_{\bm
  k}=0$ (and $\phi_{\bm k}=\pi/4$) while the super-Hubble limit
corresponds to the limit of strong squeezing $r_{\bm k}\rightarrow
+\infty$ (and $\phi_{\bm k}\rightarrow 0$). 

\par

Based on the previous considerations, it is clear that the super
Hubble limit is always associated with strong squeezing, even if we do
not deal with the exact de Sitter solution. Indeed, now for an
arbitrary $\beta$, Eq.~(\ref{eq:phi}) can be written as $\phi_{\bm
  k}'\simeq -(\beta +1) \sin (2\phi_{\bm k})/\eta$ which can be
integrated and leads to $\phi_{\bm k}\simeq \arctan\left[C\vert
  \eta\vert^{-2(\beta +1)}\right]$. For $\beta \lesssim -2$, this
confirms the fact that $\phi_{\bm k}\rightarrow 0$. In the same limit,
one has $r_{\bm k}'\simeq 1/\eta$ from which one obtains $r_{\bm
  k}\propto (1+\beta)\ln a$. This confirms that the super-Hubble limit
is the strong squeezing limit and, given the fact that modes of
astrophysical interest today leave the Hubble scale $50$-$60$ e-folds
before the end of inflation, one can deduce that $r_{\bm k}\simeq 120$
for those modes~\cite{Grishchuk:1990bj,Grishchuk:1992tw}. Compared to
what can be achieved in the laboratory in quantum optics, this is a
very large value~\cite{2008PhRvL.100c3602V}.

\par

In order to understand better the features of the quantum
state~(\ref{eq:wavefunction}), it is also interesting to calculate the
mean values and dispersion of various quantities. First of all, it is
clear that
\begin{equation}
\label{eq:vandp}
\left \langle \Psi \left \vert \hat{v}_{\bm k}^{\rm R,I}\right \vert \Psi 
\right \rangle 
=\left \langle \Psi \left \vert \hat{p}_{\bm k}^{\rm R,I}\right \vert \Psi 
\right \rangle
=0.
\end{equation}
Secondly, we also have 
\begin{eqnarray}
\label{eq:v2}
\left \langle \Psi \left \vert 
\left(\hat{v}_{\bm k}^{\rm R,I}\right)^2\right \vert \Psi \right \rangle 
&=&\frac{1}{4\Rea  \Omega _{\bm k}}, \\
\label{eq:p2} 
\left \langle \Psi \left \vert \left(\hat{p}_{\bm k}^{\rm R,I}\right)^2
\right \vert \Psi \right \rangle 
&=& \Rea \Omega _{\bm k}+\frac{\left(\Ima  \Omega _{\bm k}\right)^2}
{\Rea  \Omega _{\bm k}}.
\end{eqnarray}
Finally, the cross-products can be expressed as
\begin{eqnarray}
\label{eq:vp}
\left \langle \Psi \left \vert \hat{v}_{\bm k}^{\rm R}\hat{p}_{\bm k}^{\rm R}
\right \vert \Psi \right \rangle &=&
\frac{i\Omega _{\bm k}}{2\Rea  \Omega _{\bm k}}, \\
\label{eq:pv}
\left \langle \Psi \left \vert \hat{p}_{\bm k}^{\rm R}\hat{v}_{\bm k}^{\rm R}
\right \vert \Psi \right \rangle &=& 
-i+\frac{i\Omega _{\bm k}}{2\Rea  \Omega _{\bm k}},
\end{eqnarray}
and, of course, similar expressions for the operators $\hat{v}_{\bm
  k}^{\rm I}$ and $\hat{p}_{\bm k}^{\rm I}$. It is also interesting to
notice that $\left \langle \Psi \left \vert \hat{v}_{\bm k}^{\rm
      R}\hat{p}_{\bm k}^{\rm I} \right \vert \Psi \right \rangle=\left
  \langle \Psi \left \vert \hat{v}_{\bm k}^{\rm I}\hat{p}_{\bm k}^{\rm
      R} \right \vert \Psi \right \rangle =0$.

\par

At this point, it is worth digressing about the definition of the
conjugate momentum. The action~(\ref{eq:action}) is of course defined
up to a total derivative. In Ref.~\cite{Martin:2007bw}, it was shown
that adding the term ${\rm d}\left[(a'/a)(v_{\bm k}^{\rm
    R})^2+(a'/a)(v_{\bm k}^{\rm I})^2\right]/(2{\rm d}\eta)$ can also
be viewed as a canonical transformation. This generates an additional
term $(a'/a)(p_{\rm k}^{\rm R,I}v_{\rm k}^{\rm R,I}{}^*+p_{\rm k}^{\rm
  R,I}{}^*v_{\bm k}^{\rm R,I})$ in the Hamiltonian. A complete study
was presented in Ref.~\cite{Martin:2007bw} and, here, we only quote
the main results. It was shown that, at the quantum level, this
canonical transformation leaves the amplitude $\hat{v}_{\bm k}^{\rm
  R,I}$ invariant but induces the following transformations for the
momentum: $\hat{p}_{\bm k}^{\rm R,I}\rightarrow \hat{\pi}_{\bm k}^{\rm
  R,I}$ with
\begin{eqnarray}
\label{eq:transp}
  \hat{\pi}_{\bm k}^{\rm R,I} &=& \hat{p}_{\bm k}^{\rm R,I}
  -\frac{a'}{a}\hat{v}_{\bm k}^{\rm R,I}.
\end{eqnarray}
On the other hand, the wavefunction is also modified, $\Psi_{\bm k}
\rightarrow \bar{\Psi}_{\bm k}$, and the function $\Omega_{\bm k}$
changes according to $\Omega_{\bm k}\rightarrow \bar{\Omega}_{\bm k}$,
where
\begin{eqnarray}
\label{eq:transomega}
\bar{\Omega}_{\bm k} &=& \Omega_{\bm k}+i\frac{a'}{2a}.
\end{eqnarray}
In particular, we see that the canonical transformation is such that
the term $ia'/(2a)$ in the expression~(\ref{eq:grandomega}) of the
function $\Omega _{\bm k}(\eta)$ is exactly canceled. The factor
$N_{\bm k}$ of the wavefunction is not modified and is still given by
the first of Eqs.~(\ref{eq:solpsi}) (but of course should be used
either with $\Omega _{\bm k}$ or $\bar{\Omega}_{\bm k}$ according to
which set of variables is used). This also means that when the
averages~(\ref{eq:vandp}), (\ref{eq:v2}), (\ref{eq:p2}), (\ref{eq:vp})
and~(\ref{eq:pv}) are computed in the state $\vert \bar{\Psi}\rangle$,
one obtains exactly the same expression, $\Omega_{\bm k}$ being just
replaced with $\bar{\Omega}_{\bm k}$ (of course, $\vert \Psi \rangle$
and $\vert \bar{\Psi}\rangle$, being related by a canonical
transformation, represent the same physical state).

\par

We now come back to our calculation of the dispersion of amplitude
operator and its conjugate momentum. Upon using Eqs.~(\ref{eq:v2})
and~(\ref{eq:reomega}), one obtains
\begin{equation}
\left \langle \bar{\Psi}\left \vert 
\left(\hat{v}_{\bm k}^{\rm R,I}\right)^2\right \vert \bar{\Psi}\right \rangle 
=\frac{1}{2k}\left[\cosh \left(2r_{\bm k}\right)+\cos \left(2 \phi_{\bm k}\right)
\sinh \left(2r_{\bm k}\right)\right].
\end{equation}
In the same manner, the dispersion of the operator $\hat{\pi}_{\bm
  k}^{\rm R,I}$ is given by
\begin{equation}
\left \langle \bar{\Psi}\left \vert 
\left(\hat{\pi}_{\bm k}^{\rm R,I}\right)^2\right \vert \bar{\Psi}
\right \rangle 
=\frac{k}{2}\frac{1+\sin^2(2\phi_{\bm k})\sinh^2(2r_{\bm k})}
{\cosh \left(2r_{\bm k}\right)+\cos \left(2 \phi_{\bm k}\right)
\sinh \left(2r_{\bm k}\right)}.
\end{equation}
Let us now consider two new operators $\hat{\cal A}_{\bm k}^{\rm R,I}$
and $\hat{\cal B}_{\bm k}^{\rm R,I}$, defined from $\hat{\pi}_{\bm
  k}^{\rm R,I}/\sqrt{k}$ and $\sqrt{k}\hat{v}_{\bm k}^{\rm R,I}$
through a rotation by the squeezing angle $\phi_{\bm k}$:
\begin{eqnarray}
\hat{\cal A}_{\bm k}^{\rm R,I} &=&\frac{\hat{\pi}_{\bm k}^{\rm R,I}}{\sqrt{k}}
\cos \phi_{\bm k}+\sqrt{k}\hat{v}_{\bm k}^{\rm R,I}\sin \phi_{\bm k}, \\
\hat{\cal B}_{\bm k}^{\rm R,I} &=&\frac{\hat{\pi}_{\bm k}^{\rm R,I}}{\sqrt{k}}
\sin \phi_{\bm k}-\sqrt{k}\hat{v}_{\bm k}^{\rm R,I}\cos \phi_{\bm k}.
\end{eqnarray}
It is easy to check that $\left[\hat{\cal A}_{\bm k},\hat{\cal B}_{\bm
    k}\right]=i$. Then, a lengthy but straightforward calculation
leads to
\begin{eqnarray}
\left \langle \bar{\Psi}\left \vert 
\hat{\cal A}_{\bm k}^{\rm R,I}\right \vert \bar{\Psi}\right \rangle 
&=& \frac{{\rm e}^{-2r_{\bm k}}}{2}, \\
\left \langle \bar{\Psi}\left \vert 
\hat{\cal B}_{\bm k}^{\rm R,I}\right \vert \bar{\Psi}\right \rangle 
&=& \frac{{\rm e}^{2r_{\bm k}}}{2}.
\end{eqnarray}
Therefore, we see that there exists a direction in the plane
$(\pi_{\bm k},v_{\bm k})$ where the dispersion is extremely
small. This is why the corresponding state is called a squeezed
state. In order to satisfy the Heisenberg inequality, the dispersion
along the direction perpendicular to the previous one becomes very
large. As already mentioned, the phenomenon of squeezing is widely
studied in many different branches of physics, in particular in
quantum optics. Squeezing occurs each time the quantization of a
parametric oscillator is carried out. It is remarkable that the
quantization of small fluctuations on top of an expanding universe
also leads to that concept (squeezing here, \ie $r_{\bm{k}}\not= 0$,
does not require an accelerated expansion, only a dynamical background
is necessary).

\subsection{The Classical Limit}
\label{subsec:classicallimit}

We have seen in the last section that the super-Hubble limit
corresponds to a limit where the squeezing parameter $r_{\bm k}$ is
large. In the literature, this regime is very often described as a
regime where the cosmological perturbations have
classicalized~\cite{Lesgourgues:1996jc,Polarski:1995jg,Kiefer:1998qe,Kiefer:1998qe,Kiefer:1999gt}. Since
this concept is subtle in quantum mechanics (and particularly when
quantum mechanics is applied to cosmology), we need to come back to
this issue and to describe accurately what is meant by a ``classical
limit'' in this context. In particular, it may seem strange at first
sight that a quantum system placed in a strongly squeezed state can be
described as a classical state since, in the context of, say, quantum
optics, a similar situation would precisely be described as a
non-classical situation~\cite{Caves:1985zz,Schumaker:1985zz}.

\par

A convenient tool to study this question is the Wigner function, defined by
\begin{widetext}
\begin{equation}
W\left(v_{\bm k}^{\rm R},v_{\bm k}^{\rm I},p_{\bm k}^{\rm R},p_{\bm k}^{\rm I}\right)
= \frac{1}{(2\pi)^2}
\int {\rm d}x{\rm d}y \, \Psi^*\left(v_{\bm k}^{\rm R}-\frac{x}{2}, 
v_{\bm k}^{\rm I}-\frac{y}{2}\right)
{\rm e}^{-ip_{\bm k}^{\rm R}x-ip_{\bm k}^{\rm I}y}\, 
\Psi\left(v_{\bm k}^{\rm R}+\frac{x}{2}, 
v_{\bm k}^{\rm I}+\frac{y}{2}\right).
\end{equation}
Indeed, it is well known that the Wigner function can be understood as
a classical probability distribution function whenever it is positive
definite. Then, upon using the quantum state~(\ref{eq:wavefunction}),
the following explicit form is obtained
\begin{equation}
\label{eq:wignersqueeze}
W\left(v_{\bm k}^{\rm R},v_{\bm k}^{\rm I},p_{\bm k}^{\rm R},p_{\bm k}^{\rm I}\right)
=\Psi\Psi^*\frac{1}{2\pi \Rea  \Omega _{\bm k}}
\exp\left[-\frac{1}{2\Rea  \Omega _{\bm k}}\left(p_{\bm k}^{\rm R}+2\Ima  \Omega _{\bm k}
v_{\bm k}^{\rm R}\right)^2\right]
\exp\left[-\frac{1}{2\Rea  \Omega _{\bm k}}\left(p_{\bm k}^{\rm I}+2\Ima  \Omega _{\bm k}
v_{\bm k}^{\rm I}\right)^2\right].
\end{equation}
\end{widetext}
The following remark is in order at this stage. One could have
calculated the Wigner function with the state $\bar{\Psi}_{\bm
  k}$. Obviously, one would have obtained exactly the same expression
except that all the $\Omega_{\bm k}$ terms would have been replaced
with $\bar{\Omega}_{\bm k}$ and $p_{\bm k}^{\rm R,I}$ with $\pi_{\bm
  k}^{\rm R,I}$. In particular, this means that the term in
parenthesis in the argument of the exponentials would have read
$\pi_{\bm k}^{\rm R,I}+2\Ima \bar{\Omega}_{\bm k}v_{\bm k}^{\rm
  R,I}$. But, thanks to Eqs.~(\ref{eq:transp})
and~(\ref{eq:transomega}), this is precisely $p_{\bm k}^{\rm R,I}+2\Ima 
\Omega _{\bm k}v_{\bm k}^{\rm R,I}$ since the two terms proportional
to $a'/a$ exactly cancel out. This is of course related to the fact
that the Wigner function is invariant under a canonical
transformation.

\begin{figure*}[t]
\begin{center}
\includegraphics[width=7.cm]{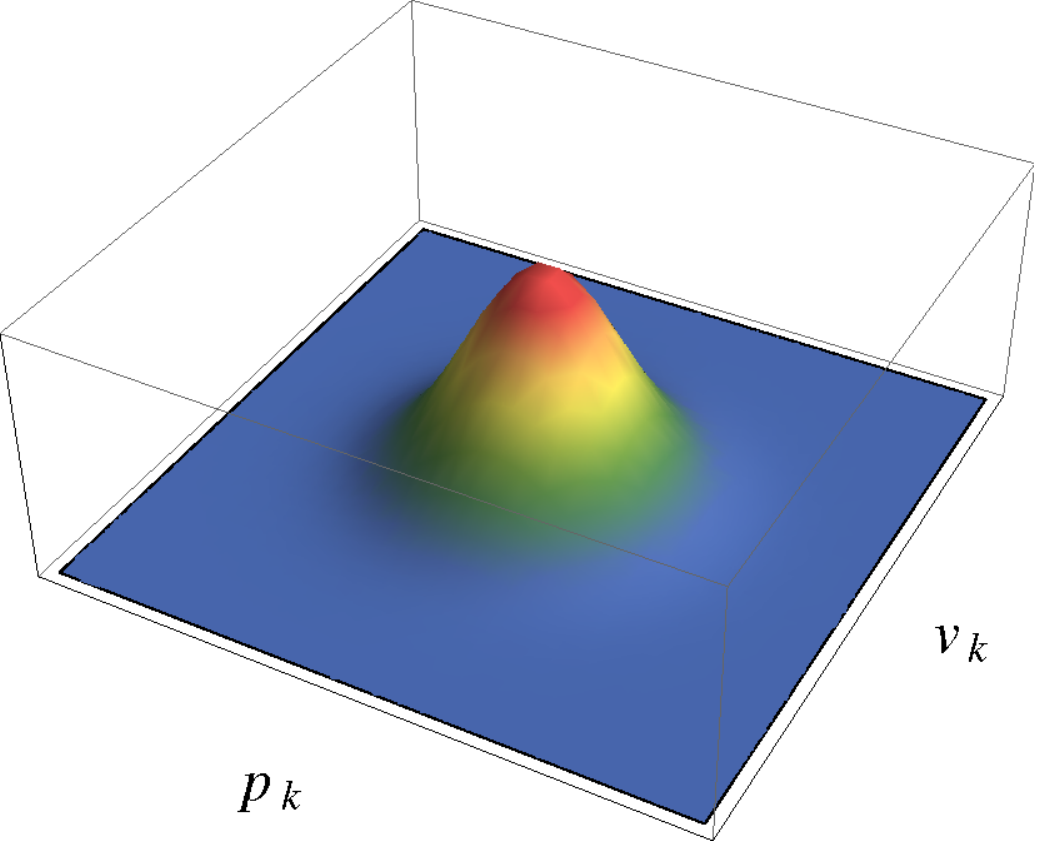}
\includegraphics[width=7.cm]{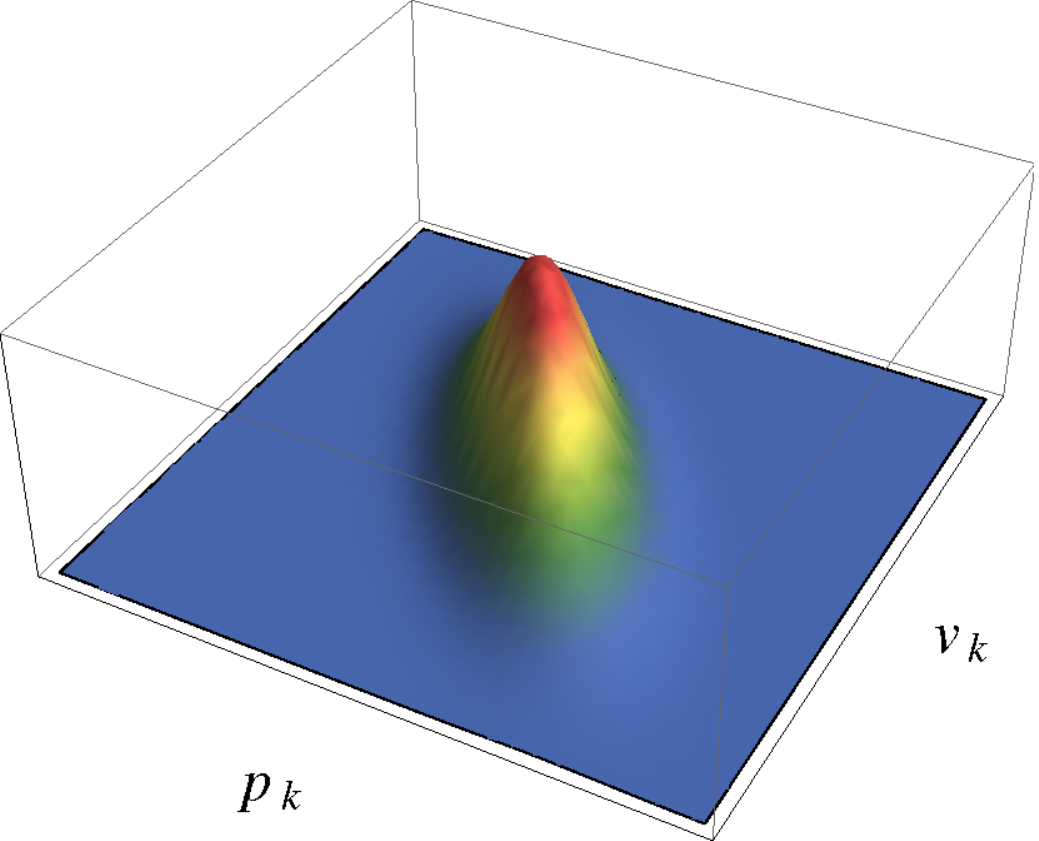}
\includegraphics[width=7.cm]{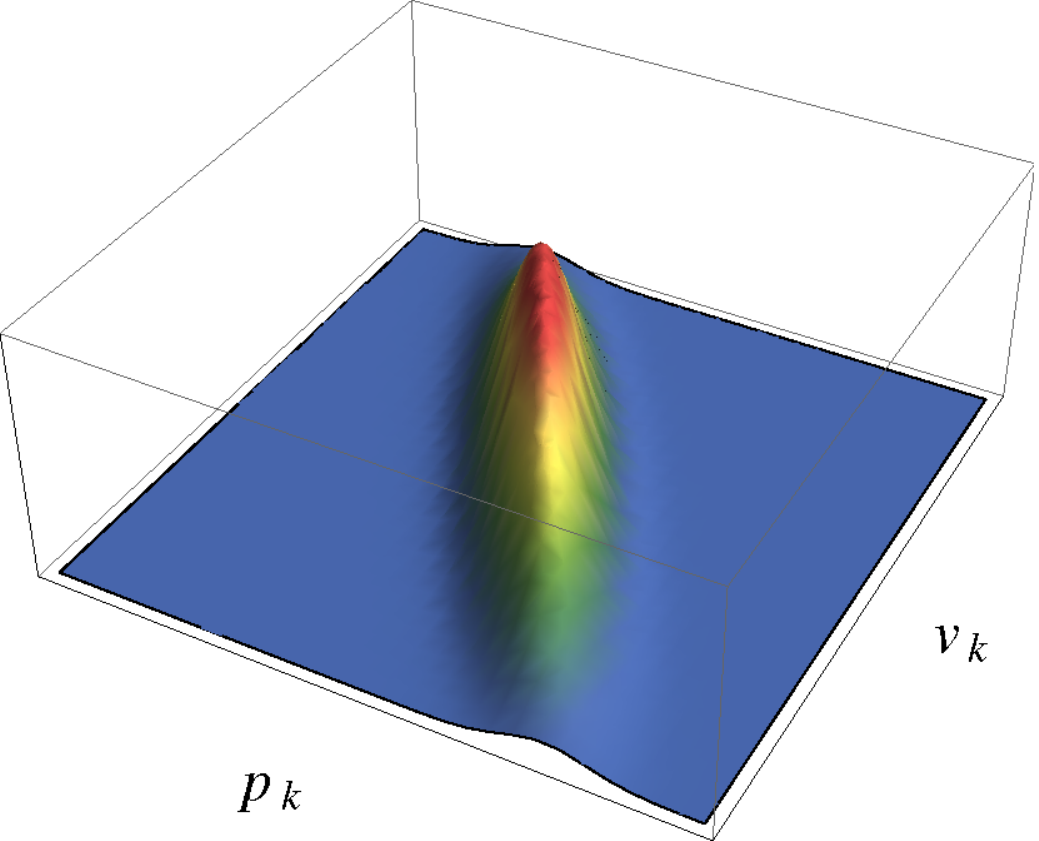}
\includegraphics[width=7.cm]{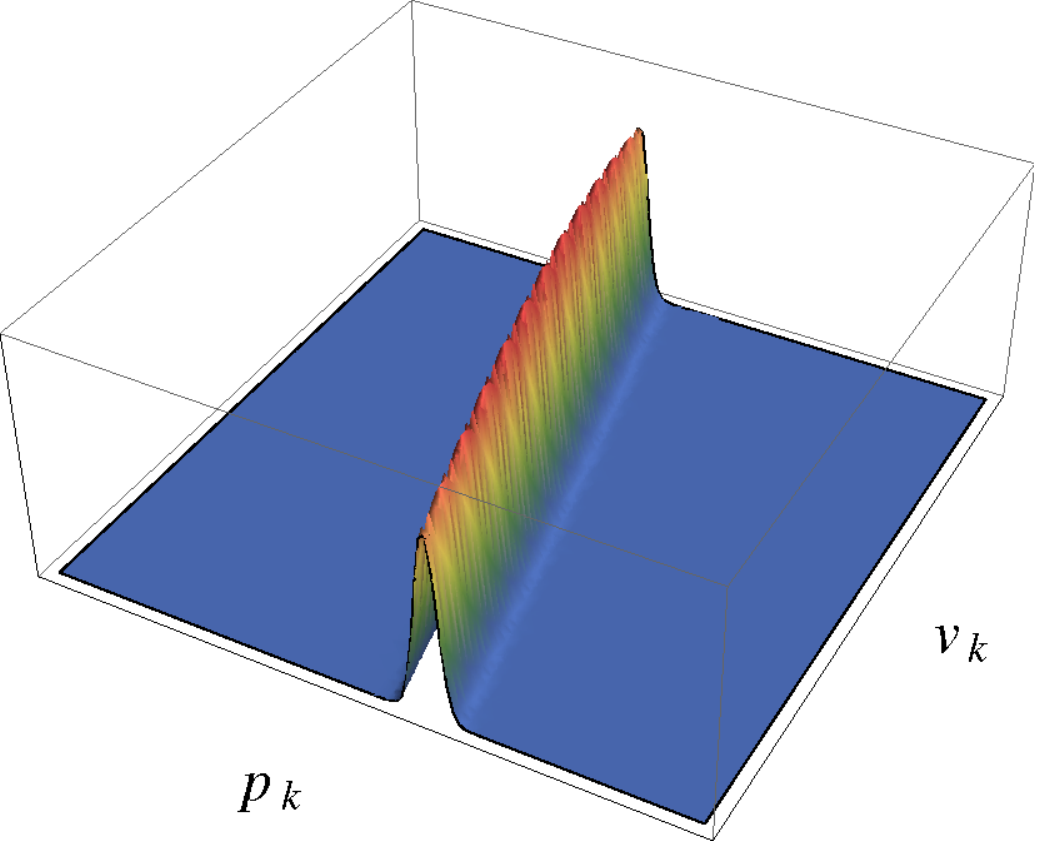}
\end{center}
\caption{Wigner function of a squeezed quantum state at different
  times during inflation. Only the two-dimensional function
  corresponding to the set of variables $\left(v_{\bm k}^{\rm
      R},p_{\bm k}^{\rm R}\right)$ has been represented, see
  Eq.~(\ref{eq:wignersqueeze}). The time evolution of $\Rea  \Omega
  _{\bm k}$ and $\Ima  \Omega _{\bm k}$ has been expressed in terms of
  the two squeezing parameters $r_{\bm k}$ and $\phi_{\bm k}$. These
  ones are given by the solutions~(\ref{eq:solr})
  and~(\ref{eq:solphi}). The left upper panel corresponds to $r_{\rm
    k}=0.0005$ and the corresponding state is almost a coherent
  one. The right upper panel corresponds to $r_{\bm k}=0.48$, the left
  bottom one to $r_{\bm k}=0.88$ and, finally, the right bottom one to
  $r_{\bm k}=2.31$. The effect of the squeezing and the cigar shape of
  Eq.~(\ref{eq:wignerstrongr}) are clearly visible.}
\label{fig:wigner}
\end{figure*}

The Wigner function~(\ref{eq:wignersqueeze}) is represented in
Fig.~\ref{fig:wigner} at different times or, equivalently, at
different values of $r_{\bm k}$ ($r_{\bm k}=0.0005$, $0.48$, $0.88$
and $2.31$). The effect of the strong squeezing is clearly
visible. Initially, in the sub-Hubble regime, $r_{\bm k}$ is small and
the Wigner function is peaked over of small region in phase space. As
inflation proceeds, the modes become super Hubble and $r_{\bm k}$
increases. As a consequence, the Wigner function spreads and acquires
a cigar shape typical of squeezed states. In fact, in the strong
squeezing limit, one has $\Rea \bar{\Omega}_{\bm k}\rightarrow 0$ and
$\Ima \bar{\Omega} _{\bm k}\rightarrow k \sin \phi_{\bm k}/(2
\cos\phi_{\bm k})\rightarrow 0$, see Eq.~(\ref{eq:reomega})
and~(\ref{eq:imomega}). Let us notice in passing that this last
equation is consistent with Eq.~(\ref{eq:imomegaeta}). On the other
hand, if one considers $\Ima \bar{\Omega}_{\bm k}$, then the leading
term $a'/(2a)$ is absent and one has to go to the next order in
Eq.~(\ref{eq:imomega}). This one is given by $k/[4(\beta +3/2)](-k\eta
)$ and represents the leading term of $\Ima \bar{\Omega}_{\bm k}$. It
goes to zero in agreement with the fact that $\phi_{\bm k}\rightarrow
0$ in the strong squeezing limit. In this regime, the Wigner function
can be written as
\begin{eqnarray}
\label{eq:wignerstrongr}
W\left(v_{\bm k}^{\rm R},v_{\bm k}^{\rm I},p_{\bm k}^{\rm R},p_{\bm k}^{\rm I}\right)
& \rightarrow &
\Psi\Psi^*\delta\left(p_{\bm k}^{\rm R}+k
\frac{\sin \phi_{\bm k}}{\cos \phi_{\bm k}}
v_{\bm k}^{\rm R}\right)
\nonumber \\ & & \times
\delta\left(p_{\bm k}^{\rm I}+k
\frac{\sin \phi_{\bm k}}{\cos \phi_{\bm k}}
v_{\bm k}^{\rm I}\right).
\end{eqnarray}
This last equation represents the mathematical formulation of the
cigar shape mentioned above. 

\par

\begin{figure*}[t]
\begin{center}
\includegraphics[width=7.cm]{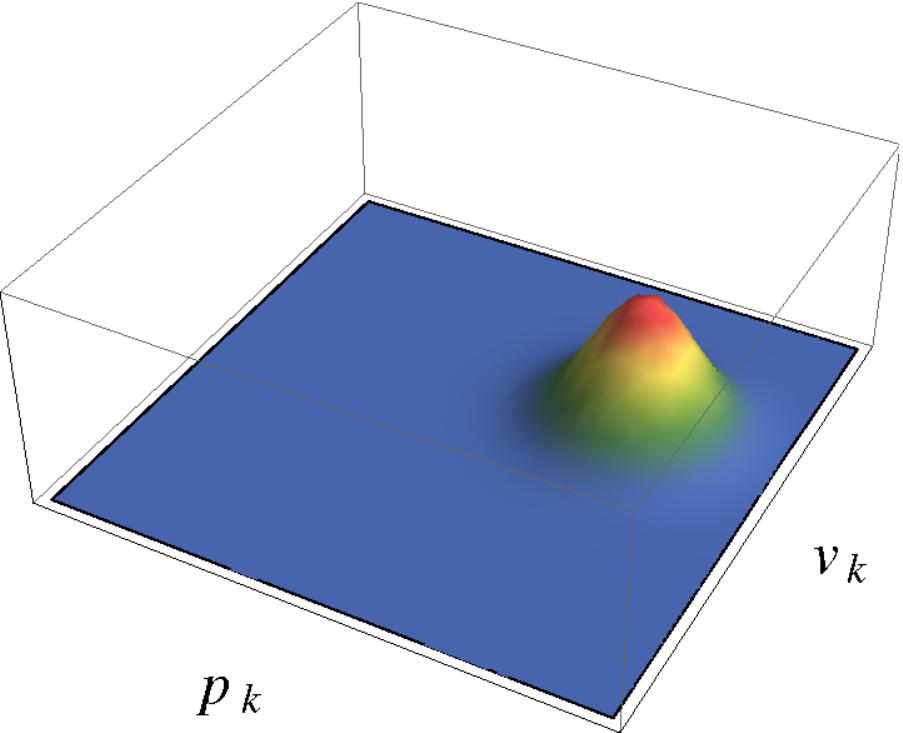}
\includegraphics[width=7.cm]{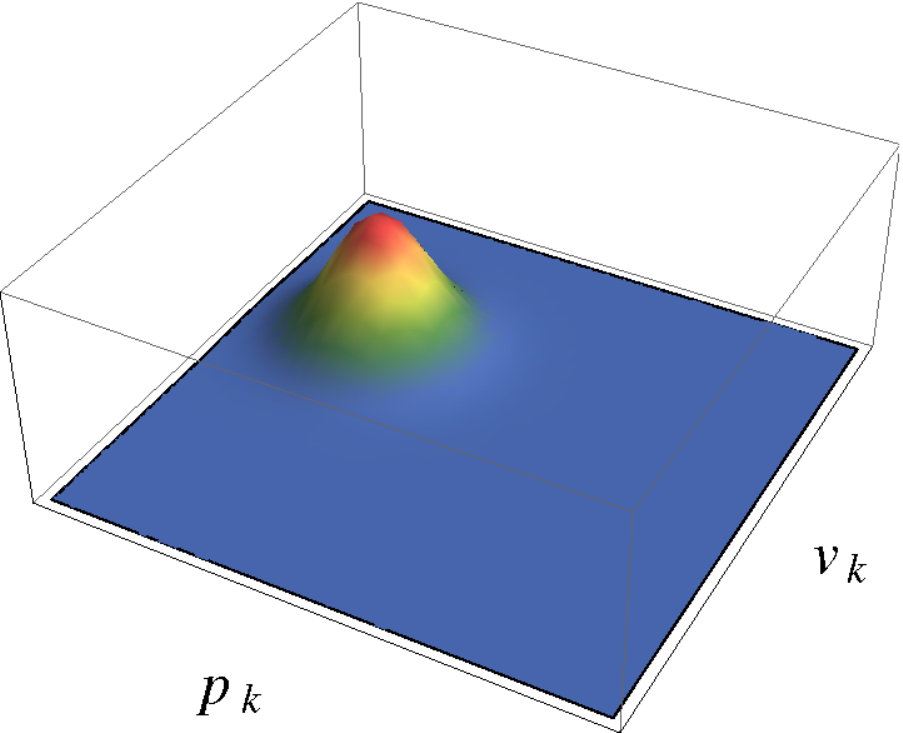}
\includegraphics[width=7.cm]{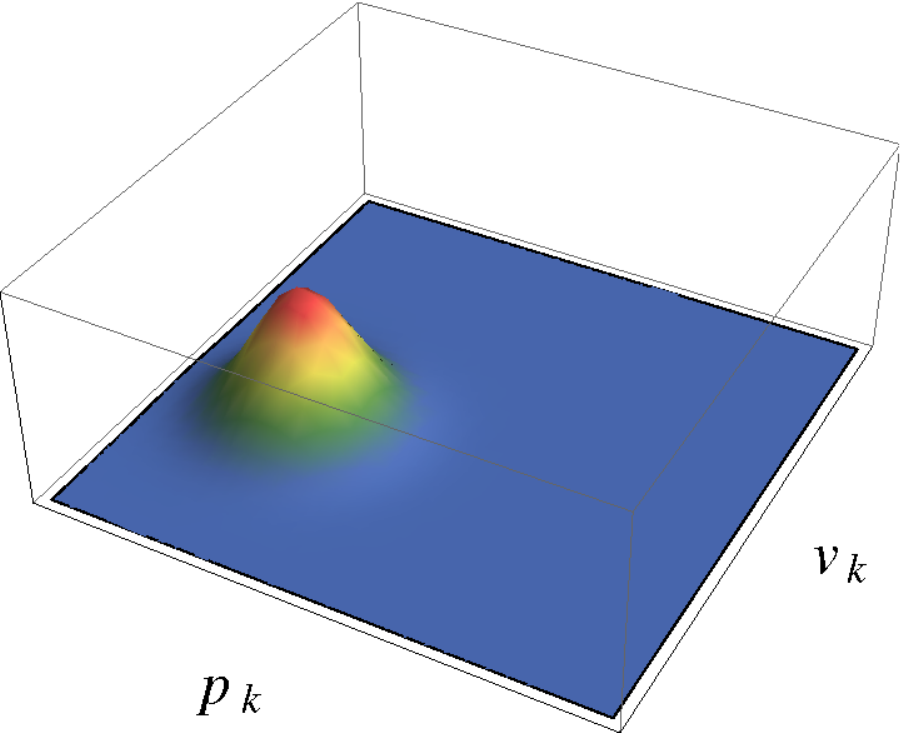}
\includegraphics[width=7.cm]{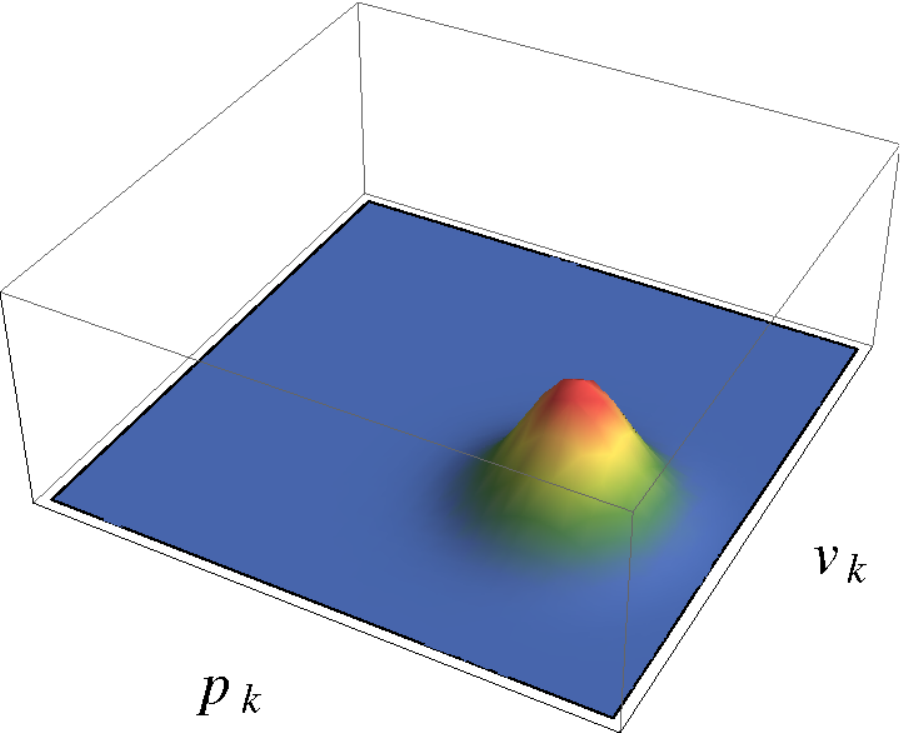}
\end{center}
\caption{Wigner function of a coherent 
state~(\ref{eq:wignercoh}), represented at different times during
  inflation. Contrary to the Wigner function of a squeezed state of
  Fig.~\ref{fig:wigner}, the shape remains unchanged during the
  cosmological evolution. The Wigner function just follows the
  classical trajectory, an ellipse here since we deal with an harmonic
  oscillator. This justifies the fact that a coherent state can be
  viewed as the ``most classical quantum state''.}
\label{fig:wignercoh}
\end{figure*}

It is important to notice that the behavior described above is very
different from the behavior of the Wigner function of a coherent
state. The coherent states are usually considered as the ``most
classical'' states and their Wigner function is given by
\begin{equation}
\label{eq:wignercoh}
W\left(v_{\bm k}^{\rm R},p_{\bm k}^{\rm R}\right)
=\frac{1}{\pi}{\rm e}^{-k\left[v_{\bm k}^{\rm R}
-v_{\bm k}^{\rm R,cl}(\eta)\right]^2}
{\rm e}^{-\left[p_{\bm k}^{\rm R}
-p_{\bm k}^{\rm R,cl}(\eta)\right]^2/k},
\end{equation}
where $v_{\bm k}^{\rm R,cl}$ and $p_{\bm k}^{\rm R,cl}$ represent the
classical solutions. The typical shape is plotted in
Fig.~\ref{fig:wignercoh}. One sees that the Wigner functions remain
peaked over a small region in phase space and that this packet follows
the classical trajectory (an ellipse in this context). Comparing
Figs.~\ref{fig:wigner} and~\ref{fig:wignercoh}, we understand why a
coherent state is usually considered as classical while a squeezed
state is considered as highly non classical. In the case of the
coherent state, if one is given, say, the value of $v_{\bm k}^{\rm
  R}$, then one obtains a value for the momentum, $p_{\bm k}^{\rm R}$,
which is very close to the one one would have inferred in the
classical case. This is of course due to the fact that the Wigner
function follows the classical trajectory and has minimal spread
around it in all phase space directions. On the contrary, in the case
of the squeezed state, if one is given $p_{\bm k}^{\rm R}$ then the
value of $v_{\bm k}^{\rm R}$ is very uncertain since the Wigner
function is spread over a large region in phase space. Therefore, we
conclude that the cosmological perturbations do not behave classically
in the usual sense.

Given the previous discussion, it may seem relatively easy to observe
genuine quantum effects in the CMB. Unfortunately this is not so,
essentially because, in the strong squeezed limit, all quantum
predictions can be in fact obtained from averages performed by mean of
a classical stochastic process. 

\par

Let us first study how this question is usually treated. For this
purpose, let us consider again the expectation of the operator
$\left(\hat{\pi}_{\bm k}^{\rm R}\right)^2 $ [of course, one could also
treat the case of $\left(\hat{\pi}_{\bm k}^{\rm R}\right)^n$]. The
quantum average is given by Eq.~(\ref{eq:p2}), namely
\begin{equation}
\label{eq:pi2}
\left \langle \bar{\Psi}\left \vert \left(\hat{\pi}_{\bm k}^{\rm
    R}\right)^2 \right \vert \bar{\Psi}\right \rangle =\Rea  \bar{\Omega}_{\bm k}
+\frac{(\Ima  \bar{\Omega}_{\bm k})^2}{\Rea  \bar{\Omega}_{\bm k}}.
\end{equation}
On the other hand, if one computes the quantity 
\begin{equation}
\label{eq:pi2wignerstrongr}
\int {\rm d}v_{\bm k}^{\rm R}{\rm d}\pi_{\bm k}^{\rm R}
W_{r_{\bm k}\rightarrow \infty}\left(v_{\bm k}^{\rm R},
\pi_{\bm k}^{\rm R}\right)
\left(\pi_{\bm k}^{\rm R}\right)^2,
\end{equation}
where $W_{r_{\bm k}\rightarrow \infty}\left(v_{\bm k}^{\rm R},\pi_{\bm
    k}^{\rm R}\right)$ refers to the Wigner function in the strong
squeezing limit~(\ref{eq:wignerstrongr}), then one obtains $\left(\Ima
  \bar{\Omega}_{\bm k}\right)^2/\Rea \bar{\Omega }_{\bm k}$, which
coincides with Eq.~(\ref{eq:pi2}) in the limit $r_{\bm k}\rightarrow
\infty$. This result is often taken as a proof that a strongly
squeezed state can be described as classical stochastic
process. However, this argument is not very convincing since it is a
theorem~\cite{Polarski:1995jg} that the exact Wigner function [we
stress again that, in Eq.~(\ref{eq:pi2wignerstrongr}), we have not
used the general Wigner function but its limit when $r_{\bm k}$ is
large] satisfies the following property
\begin{equation}
\left \langle \hat{A}\left(\hat{v}_{\bm k}^{\rm R},
\hat{\pi}_{\bm k}^{\rm R}\right) \right \rangle =\int {\rm d}v_{\bm k}^{\rm R}
{\rm d}\pi_{\bm k}^{\rm R}
W\left(v_{\bm k}^{\rm R},\pi_{\bm k}^{\rm R}\right)
A\left(v_{\bm k}^{\rm R},
\pi_{\bm k}^{\rm R}\right),
\end{equation} 
where $\hat{A}$ is an arbitrary operator. Therefore, it does not come
as a surprise that an expression like Eq.~(\ref{eq:pi2wignerstrongr})
reproduces the corresponding quantum average in the limit $r_{\bm
  k}\rightarrow \infty$.

\par

In fact, as was discussed in
Refs.~\cite{Kiefer:1998jk,Kiefer:2008ku,Kiefer:1998qe}, what makes the
situation so peculiar is something different. The point is
that, in the limit $r_{\bm k}\rightarrow \infty$, all the quantum
predictions can be reproduced if one assumes that the system always
followed classical laws but had random initial conditions with a given
probably density function. This can be easily understood on the
example of a free
particle~\cite{Kiefer:1998jk,Kiefer:2008ku,Kiefer:1998qe}. Let us
assume that, initially (at $t=0$), the probability to find the
particle at $x$ is given by
\begin{equation}
\label{eq:freepart:distribx}
\left \vert \Psi\left(x,0\right)\right\vert ^2
=\sqrt{\frac{2}{\pi b^2}}{\rm e}^{-2x^2/b^2},
\end{equation}
where $b$ is a parameter that characterizes the width of the
distribution. At time $t$, this probability is given by
\begin{eqnarray}
\label{eq:wfunctionfree}
\left \vert \Psi\left(x,t\right)\right\vert ^2
&=&
\sqrt{\frac{2}{\pi b^2}}
\frac{1}{\sqrt{1+4t^2/(m^2b^4)}}
\nonumber \\ & & \times 
\exp\left[
-\frac{2b^2\left(x-k_0t/m\right)^2}{b^4+4t^2/m^2}\right],
\end{eqnarray}
where $m$ is the mass of the particle and $k_0$ the center of the
Gaussian wave packet in Fourier space.

\par

Now let us consider a situation where we repeat many times an
experiment consisting in sending a classical particle from the origin
with a velocity $v$ (equivalently, instead of repeating the
experiments many times, one could also consider an ensemble of
classical particles) and detecting it at a position $x\neq 0$. By
definition, the particle follows the laws of classical physics which
means that its motion can be described by the equation: $x=vt$ (they
all start from $x=0$ at $t=0$). Then, let us assume that the
velocities are classical random variables with a probability
distribution function given by
\begin{equation}
\label{eq:freepart:distribv}
P(v)=\frac{1}{\sqrt{\pi}\Delta v}{\rm e}^{-v^2/(\Delta v)^2}.
\end{equation}
This means, that according to the particle considered, the velocity is
in fact not always the same. But because different particles have
different velocities, they will not reach the position $x$ at the same
time. It is important to stress that, here, only the initial
conditions are random and that the trajectory is purely
classical. From the above distribution, we can easily infer that the
probability of finding a particle at $x$, at time $t$, is
\begin{equation}
P(x,t)=\frac{1}{\sqrt{\pi}t\Delta v }{\rm e}^{-(x-vt)^2/(t\Delta v )^2}.
\end{equation}
This distribution is in fact exactly $\vert \Psi(x,t)\vert^2$ in the
limit $t\rightarrow \infty$ provided we identify $v=k_0/m$ and $\Delta
v=\sqrt{2}/(mb)$. Let us notice that this last
  relation is exactly what is obtained at the quantum level since $x$
  and $v$ are conjugated variables. As a matter of fact,
  Eqs.~(\ref{eq:freepart:distribx}) and~(\ref{eq:freepart:distribv})
  are Fourier transforms of each other. We conclude that, provided we
detect the particles far from the origin, the quantum predictions for
the particles can be completely mimicked by means of a classical
stochastic process.

\par

As discussed in Ref.~\cite{Kiefer:1998jk}, the situation is exactly similar
for the inflationary perturbations. The limit $r_{\bm k}\rightarrow
\infty$ is in fact equivalent to the limit of large times in the
example above. One can even calculate the Wigner function of the free
particle described by the wavefunction~(\ref{eq:wfunctionfree}) and
show that it takes the same form as the one of
Eq.~(\ref{eq:wignersqueeze}). Therefore, the inflationary
perturbations are said to be classical in the sense explained
before: they can be described by a classical stochastic process. In
practice, for instance, one can consider the $\hat{a}_{\ell m}$ in
Eqs.~(\ref{eq:decompsh}) and~(\ref{eq:corralm}) as classical random
variables with probability density functions given by
\begin{eqnarray}
\label{eq:palo}
P\left(a_{\ell 0}^{\rm R}\right)&=&\frac{1}{\sqrt{2\pi}C_{\ell}}
{\rm e}^{-\left(a_{\ell 0}^{\rm R}\right)^2/(2C_{\ell})}, \\
\label{eq:palmr}
P\left(a_{\ell m}^{\rm R}\right)&=&\frac{1}{\sqrt{\pi}C_{\ell}}
{\rm e}^{-\left(a_{\ell m}^{\rm R}\right)^2/C_{\ell}}, \quad m\neq 0, \\
\label{eq:palmi}
P\left(a_{\ell m}^{\rm I}\right)&=&\frac{1}{\sqrt{\pi}C_{\ell}}
{\rm e}^{-\left(a_{\ell m}^{\rm I}\right)^2/C_{\ell}}, \quad m\neq 0.
\end{eqnarray}
Of course one can check that $\left \langle a_{\ell m}a_{\ell
    'm'}\right \rangle =C_{\ell }\delta _{\ell \ell'}\delta _{mm'}$
where, now, the bracket means a classical average calculated by means
of the above distributions.

\par

Finally, we conclude this section by a few words on the density matrix
$\hat{\rho}_{\bm k}^{\rm R}$. In fact, the density matrix is nothing
but the Fourier transform of the Wigner function. Let us denote by
$\vert v_{\bm k}^{\rm R}\rangle$ the eigenstates of the operator
$\hat{v}_{\bm k}^{\rm R}$. Then, we have
\begin{equation}
\left\langle v_{\bm k}^{\rm R}{}'\left \vert \hat{\rho}_{\bm k}^{\rm R}
\right \vert  v_{\bm k}^{\rm R}\right\rangle =\int _{-\infty}^{\infty}
  {\rm d}y\, {\rm e}^{iy(v_{\bm
      k}^{\rm R}{}'-v_{\bm
      k}^{\rm R})}W\left(\frac{v_{\bm
        k}^{\rm R}{}'+v_{\bm
        k}^{\rm R}}{2},y\right).
\end{equation}
Upon using Eq.~(\ref{eq:wignersqueeze}) in the above equation, one
arrives at
\begin{eqnarray}
\left\langle v_{\bm k}^{\rm R}{}'\left \vert \hat{\rho}_{\bm k}^{\rm R}
\right \vert  v_{\bm k}^{\rm R}\right\rangle &=&
\left(\frac{2\Rea  \Omega _{\bm k}}{\pi}
\right)^{1/2}
{\rm e}^{-\Rea  \Omega _{\bm k}\left[\left(v_{\bm
        k}^{\rm R}{}'\right)^2+\left(v_{\bm
        k}^{\rm R}\right)^2\right]}\nonumber\\
\nonumber \\ & & \times
{\rm e}^{-i\Ima  \Omega _{\bm k}\left[\left(v_{\bm
        k}^{\rm R}{}'\right)^2-\left(v_{\bm
        k}^{\rm R}\right)^2\right]}\, .
\end{eqnarray}
We notice that the off-diagonal terms, $v_{\bm k}^{\rm R}{}' \neq
v_{\bm k}^{\rm R}$, oscillate very rapidly in the strong squeezing
limit. This means that decoherence (defined as the disappearance of
those off-diagonal terms) does not occur without taking into account
an environment for the perturbations. Various discussions on what this
environment may be can be found in
Refs.~\cite{Anderson:2005hi,Burgess:2006jn,Martineau:2006ki}.

\subsection{Ergodicity}
\label{subsec:ergodicity}

Let us now discuss how, in practice, we can check the predictions of
the theory previously reviewed. Initially, the system is placed in an
eigenstate of the Hamiltonian (the vacuum state) of
Eq.~(\ref{eq:gaussianpsi}), which can also be expressed as a
superposition in the basis of the states $\vert v_{\bm k}^{\rm
  R}\rangle$, namely
\begin{equation}
  \vert \Psi \rangle =\int {\rm d}v_{\bm k}^{\rm R}\, N_{\bm k}(\eta)\, 
{\rm e}^{-\Omega_{\bm k}(\eta) 
\left(v_{\bm k}^{\rm R}\right)^2}\vert v_{\bm k}^{\rm R}\rangle.
\end{equation}
The corresponding mean value of the Hamiltonian operator can be
expressed as
\begin{equation}
\langle \Psi\vert{\cal H}_{\bm k}^{\rm R}\vert \Psi\rangle 
=\frac12\Rea  \Omega _{\bm k}
+\frac12\frac{\left(\Ima  \Omega_{\bm k}\right)^2}{2\Rea  \Omega _{\bm k}}
+\frac{\omega ^2}{2}\frac{1}{4\Rea  \Omega _{\bm k}}.
\end{equation}
Of course, initially $\Omega_{\bm k}=k/2$ and the energy is nothing
but $\omega/2$ as expected for the vacuum state.

\par

In the real world, we measure the temperature anisotropies. As we have
seen (and as is appropriate for an observable in the
quantum-mechanical framework), this quantity is represented by an
operator. According to Eq.~(\ref{eq:decompsh}), measuring the
temperature anisotropies is equivalent to measuring the observables
$\hat{a}_{\ell m}$ which, in turn, according to Eq.~(\ref{eq:alm}), is
equivalent to measuring the observables $\hat{\zeta}_{\bm k}$ or
$\hat{v}_{\bm k}$ (that is to say $\hat{v}_{\bm k}^{\rm R}$ and
$\hat{v}_{\bm k}^{\rm I}$).

\par

According to the postulates of quantum mechanics, measuring the observable
$\hat{v}_{\bm k}^{\rm R}$ gives an eigenvalue $v_{\bm k}^{\rm R}$ (no
hat, it is a number) with probability $\vert \langle v_{\bm k}^{\rm
  R}\vert \Psi \rangle \vert^2$ and, immediately after this
measurement, the system is placed in the eigenstate $\vert v_{\bm
  k}^{\rm R}\rangle $. More concretely, after the measurement, we
``see'' a specific CMB map and we say that the measurement has
produced a specific ``realization''. The result is given in terms of
coefficients $a_{\ell m}$ (again, no hat) expressed in terms of the
numbers $v_{\bm k}^{\rm R}$ through Eq.~(\ref{eq:alm}) (except, of
course, that this equation should now be used with no hat on both
sides). Equivalently, we see a specific temperature pattern $\delta
T({\bm e})/T$ (no hat) corresponding to the set of numbers $a_{\ell
  m}$, see also Eq.~(\ref{eq:decompsh}). In conclusion, the CMB map
observed, say, by the WMAP satellite corresponds to one measurement
(or one ``realization'') of the operator $\widehat{\delta T}({\bm
  e})/T$.

\par

Then comes the question of how one can operationally verify these
theoretical predictions. In quantum mechanics, in an ordinary
laboratory situation, one would check that the theory is correct by
repeating the experiment many times. In this way, one would generate
many realizations of $\hat{v}_{\bm k}^{\rm R}$ (or, equivalently, of
$\hat{a}_{\ell m}$ or $\widehat{\delta T}/T$) \ie one would obtain
$N_{\rm real}$ numbers $v_{\bm k}^{\rm R}{}^i, i=1,\cdots ,N_{\rm
  real}$ [or $a_{\ell m}^i$ or $(\delta T/T)^i$] where $N_{\rm real}$
is the number of realizations (that is to say the number of times the
experiments have been performed). With these $N_{\rm real}$ CMB maps,
one could then check that the $v_{\bm k}^{\rm R}{}^i$ are indeed
distributed with a Gaussian probability density function in agreement
with Eq.~(\ref{eq:wavefunction}) or, with the $N_{\rm real}$ sets of
numbers $a_{\ell m}^i$, one could infer whether they follow
Eqs.~(\ref{eq:palo}), (\ref{eq:palmr}) and~(\ref{eq:palmi}), determine
the corresponding variance and check that it is given by the
$C_{\ell}$ predicted by the theory. Let us notice that the above
discussion is independent from the fact that the perturbations can be
described classically or not. If we are in the classical limit (in the
restricted sense defined in the previous section), then we showed
that measuring the observable $\hat{a}_{\ell m}$ can be viewed as
measuring a classical system with random initial conditions but this
does not change the fact that we need many realizations to check that
the probability density function predicted by the theory is the
correct one.

\par

Clearly, in cosmology, the program described above cannot be carried
out because one cannot repeat the experiment many times since we are
given only one CMB map~\cite{Grishchuk:1997pk}. How, then, can we
check the predictions of the theory of cosmological perturbations? To
discuss this question, let us be more accurate about the operator
$\widehat{\delta T}/T({\bm e})$. In the large-squeezing limit, we have
seen that it can be viewed as a classical stochastic process and,
therefore, it is convenient to write it as
\begin{equation}
\frac{\delta T}{T}(\xi, {\bm e}),
\end{equation}
where the symbol $\xi$ labels the realizations. A given realization of
a stochastic process is a function of ${\bm e}$. By contrast, a given
realization of a random variable is not a function but a number. This
is for instance the case of $a_{\ell m}(\xi)$. The idea is then to
replace ensemble averages by spatial averages (\ie averages over
different directions ${\bm e}$)~\cite{Grishchuk:1997pk}.  If the
process is ergodic, these two types of averages are
equal~\cite{Grishchuk:1997pk}. In that case, one can check the
predictions of the theory even if one has only one realization at our
disposal. Unfortunately, one can also show that a stochastic process
living on a sphere (here, of course, the celestial sphere) cannot be
ergodic~\cite{Grishchuk:1997pk}. Therefore, we are left with the task
of constructing unbiased estimators with minimal variances. For
instance, let us assume that we have calculated the number $C_{\ell}$
in some inflationary scenario and that we would like to compare its
value to an actual measurement. How would we proceed?  We would
consider the random variable ${\cal C}_{\ell}(\xi)$ defined by the
following expression~\cite{Grishchuk:1997pk}
\begin{equation}
\label{eq:defestimator}
{\cal C}_{\ell}(\xi)=\frac{1}{4\pi}%\int _{S^2}
\int _{S^2}{\rm d}\Omega _1
{\rm d}\Omega_2P_{\ell }(\cos \delta _{12})
\frac{\delta T}{T}(\xi, {\bm e}_1)
\frac{\delta T}{T}(\xi, {\bm e}_2),
\end{equation}
where $\delta _{12}$ is the angle between the direction ${\bm e}_1$
and ${\bm e}_2$. As announced, the estimator ${\cal C}_{\ell}(\xi)$ is
expressed as a spatial average of the stochastic process $\delta
T/T$. It is easy to show that it is unbiased,
$\left\langle\left\langle {\cal C}_{\ell}\right\rangle\right\rangle
=C_{\ell}$ and has the minimum variance \cite{Grishchuk:1997pk}
(called the ``cosmic variance'') given by
$\sqrt{2/(2\ell+1)}C_{\ell}$. The double brackets
$\left\langle\left\langle\, \right\rangle\right\rangle$ means an
ensemble average, which amounts to a quantum average in the high
squeezing limit as mentioned before. One should be careful that this
ensemble average has nothing to do with the one introduced below
(denoted $\mathbb E$) for the CSL modifications of the Schr\"odinger
equation, since these two stochasticities have completely different
natures, the former being effective and the later intrinsic. 

\par

In practice, we would proceed as follows. From our CMB map $\delta
T(\xi, {\bm e})/T$, we compute the integral in
Eq.~(\ref{eq:defestimator}) and this gives a number representing one
realization of the estimator ${\cal C}_{\ell}$, the only one we can
have access to. It is unlikely that this number be $C_{\ell}$ because
it is unlikely that one realization of a random variable be exactly
equal to the mean value of that variable. However, if the variance is
small (\ie if the estimator is good), the corresponding probability
density function will be sharply peaked around the mean value and any
realization will therefore be close to the mean (and, in our case, it
is not possible to decrease the value of the variance since we work
with the best estimator). Therefore, we can study where the number we
have obtained by following the above described procedure falls,
compared to the interval $C_{\ell}\pm \sqrt{2/(2\ell+1)}C_{\ell}$,
where $C_{\ell}$ is the theoretically predicted multipole
moment. Then, for instance, one can start a calculation of the
$\chi^2$ to assess to which confidence we have verified the theory. In
fact, the cosmic variance can simply be seen as another source of
error, besides those coming from the instruments.

\par

Given the previous discussion, there is one issue that one can raise
and which is the subject of the present paper. The question is how a
specific outcome (a realization) is produced. Above, we have just
assumed that this happens without discussing this point. According to
the postulates of quantum mechanics in the Copenhagen interpretation,
this ``macro-objectification'' takes place when a measurement is
performed. Since the CMB anisotropies were produced at a redshift of
$z_{\ell\mathrm{ss}}\simeq 1100$, this means that it should have
happened prior to that epoch (possibly during inflation itself). But,
clearly, there was no observer at those early times. We face here the
conventional measurement problem of quantum mechanics which is, in the
context of cosmology, exacerbated.

\section{The Pearle-Ghirardi-Rimini-Weber Theory}
\label{sec:grw}

\subsection{A Dynamical Collapse Model}
\label{subsec:reviewcsl}

Although one can manage to obtain, based on primordial vacuum quantum
fluctuations, a set of correlation functions that are formally
indistinguishable from a classical stochastic distribution, one still
has to face the problem of reaching a specific realization before
cosmological perturbations can start to grow in a classical way. This
amounts to the question of the measurement problem in quantum
mechanics, namely that there are two distinct evolution processes: the
unitary and linear Schr\"odinger time evolution on the one hand, and
the stochastic and non linear wavepacket reduction on the other

\par

In what follows, we briefly present the collapse theories and explain
how the Schr\"odinger equation can be modified in order to allow a
dynamical description of the wave-packet reduction. In fact, to be
more precise, we shall restrict attention to the case of
CSL~\cite{Ghirardi:1985mt,Pearle:1988uh,Ghirardi:1989cn,Bassi:2003gd}.

\par

The CSL model relies on the idea that to the Schr\"odinger linear
evolution should be added an extra stochastic behavior, encoded
through a Wiener process $W_t$, whose differential acts as a random
square root of that of time, namely
\begin{equation}
\mathbb{E}\left(\dd W_t\right) = 0, \ \ \  \hbox{and} \ \ \ \ \ 
\mathbb{E}\left(\dd W_t\, \dd W_{t^\prime}\right) 
= \delta\left(t-t^\prime\right)\dd t^2,
\label{dW2}
\end{equation}
where $\mathbb{E}$ stands for an ensemble average. One then expands
the state vector variation $\dd |\chi\rangle$ up to first order in
time through
\begin{equation}
  \dd |\chi\rangle = \left(\hat{A}\dd t + \hat{B}\dd W_t \right) |\chi\rangle,
\label{modSch}
\end{equation}
where $\hat{A}$ and $\hat{B}$ are operators acting on the Hilbert
space of available states. One then demands that, on average, the
wavefunction be normalized, \ie
\begin{equation}
\mathbb{E}\left(\langle \chi | \chi\rangle\right) = 1 \ \ \ \ \Longrightarrow
\ \ \ \ \mathbb{E}\left[\dd\left(\langle \chi | \chi\rangle\right)\right] = 0,
\end{equation}
which, upon using It\^o calculus\footnote{This means that for two
  functions $f$ and $g$ of the stochastic variable $W$, one has $\dd
  (fg) = f\dd g + (\dd f) g + \mathbb{E}\left[ \left(\dd f\right)
    \left( \dd g\right) \right]$ and $\dd f(W) = f'(W) \dd W + \frac12
  f''(W)\mathbb{E}\left[ \left( \dd W\right)^2 \right]$, where a prime
  stands for ordinary derivative with respect to the argument $W$. It
  is necessary to expand up to second order in the noise because
  Eq.~(\ref{dW2}) means $\mathbb{E}(\dd W_t^2)=\dd t$.}.
%$X_t=X_0+\int_0^t
%\sigma_s\, \dd W_s + \int_0^t\mu_s\, \dd s$ being an It\^o process,
%  for any regular function $f$, $f\left(X_t\right)$ is also an It\^o
%  process satisfying $\dd
%  f\left(X_t\right)=f^\prime\left(X_t\right)\dd
%  X_t+\frac{1}{2}f^{\prime\prime} \left(X_t\right)\sigma_t^2\dd t$.}
for the differentials and Eq. (\ref{dW2}), yields
\begin{equation}
\hat{A}^\dagger + \hat{A} = - \hat{B}^\dagger \hat{B},
\label{AABB}
\end{equation}
since the state $|\chi\rangle$ is arbitrary. The general solution of
Eq. (\ref{AABB}) is $\hat{A} = -i \hat{H} -\frac12 \hat{B}^\dagger
\hat{B}$, where $\hat{H}$ is hermitian and to be identified with the
Hamiltonian leading to the usual Schr\"odinger dynamics.

\par

In order to assign a probabilistic meaning to the norm of the
wavefunction, it should be normalized. However, according to
Eq. (\ref{modSch}), although this is true on average, it varies
stochastically according to
\begin{equation}
  \dd || \chi ||^2 =\langle\chi|\left(\hat{B}+\hat{B}^\dagger\right)
|\chi\rangle \dd W_t= 2 
  \langle\chi|\hat{B}|\chi\rangle \dd W_t,
\label{dChi2}
\end{equation}
where from now on we assume that $\hat{B}$ is hermitian.

\par

Eq.~(\ref{dChi2}) implies that the state $|\chi\rangle$ is not
normalized, and one can define a normalized one whose probability
distribution will thus be interpretable in terms of measurements. We
then set
\begin{equation}
|\psi\rangle \equiv \frac{|\chi\rangle}{||\chi||},
\end{equation}
whose dynamics can be computed using the previously derived rules. One
finds
\begin{eqnarray}
\dd |\psi\rangle &=& \biggl\{ \left[ -i\hat{H}
-\frac12 \left(\hat{B}-\langle \hat{B} \rangle\right)^2 \right] \dd t
\nonumber \\ & &
+ \left(\hat{B}-\langle \hat{B} \rangle\right) 
\dd W_t \biggr\} |\psi\rangle,
\label{SchMod}
\end{eqnarray}
where the quantum expectation value is taken on the normalized state
vector and thus defined as
\begin{equation}
\langle \hat{B} \rangle\equiv \langle\psi|\hat{B}|\psi\rangle.
\end{equation}
The operator $\hat{B}$ can be decomposed as $\hat{B}=\sqrt{\gamma}
\hat{Q}$.  The coupling constant $\gamma$ is the product of the
localization rate with the width of the Gaussian wavefunction inducing
the localizations \cite{Ghirardi:1985mt}, and sets the strength of the
nonlinear effects and therefore the characteristic time scale over
which these are measurable. The observable $\hat{Q}$, for instance the
position operator, is the basis on which the states are to
spontaneously collapse to (in the following, we also call the operator
$\hat{Q}$, the ``collapse operator'').

\par

As it turns out, and this is exemplified later in the case where the
operator $\hat{Q}$ is identified with a cosmological perturbation
Fourier mode (see Sec.~\ref{subsec:modifiedforv}), the natural
evolution of Eq.~(\ref{SchMod}) is to project an initial state
$|\psi_0\rangle$ on an eigenstate $|\alpha\rangle$ of the operator
$\hat{Q}$: setting
$$
\hat{Q}=\sum_\alpha q_\alpha |\alpha\rangle\langle\alpha|,
$$
(the sum being replaced by an integral in the case of a continuous
spectrum for $\hat{Q}$) such that
$\hat{Q}|\alpha\rangle=q_\alpha|\alpha\rangle$, one finds that
$\lim_{t\to\infty} \vert \Psi(t)\rangle = |\alpha\rangle$ for a given value
of $\alpha$, and this with a probability $P(\alpha) =
|\langle\Psi|\alpha\rangle|^2$. In other words, the Born rule is
naturally implemented as a dynamical consequence instead of being
imposed as an extra hypothesis.

\par

Finally, defining the density operator as
\begin{equation}
\hat{\rho}\equiv \mathbb{E}\left(|\Psi\rangle\langle\Psi |\right),
\end{equation}
one obtains, using Eq.~(\ref{SchMod}) the so-called Lindblad equation,
namely
\begin{equation}
\frac{\dd\hat{\rho}}{\dd t} = -i\left[ \hat{H},\hat{\rho}\right] 
-\frac{\gamma}{2}
\left[\hat{Q},\left[\hat{Q},\hat{\rho}\right]\right]\, .
\end{equation}
providing its time development. 

\par

Let us now come to another very important aspect of the CSL theory and
describe the so-called ``amplification mechanism'' which enables to
understand why the dynamics of microscopic systems is not much altered
by the extra stochastic and non linear terms in
Eq.~(\ref{SchMod}). This is phenomenologically very important since
this means that the laboratory experiments performed on ``small''
quantum systems are still accurately predicted by the standard
Schr\"odinger equation while the macroscopic objects are quickly and
efficiently localized. Let us consider an ensemble of $N$ identical
particles, assuming that, for each of them, the collapse operator is
the physical position in space. Therefore, we can identify the
operator and Wiener processes according to
\begin{equation}
  \hat{B} \to \sqrt{\gamma}\,\sum_{i=1}^N \hat{x}_i \ \ \ \ \hbox{and}
  \ \ \ \ \ \dd W_t \to \dd W_t^{(i)}
\label{B7}
\end{equation}
in Eq.~(\ref{SchMod}), with $\hat{x}_i$ the position operator for the
$i^\mathrm{th}$ particle. Note that in this case, one has as many
independent Wiener processes as there are particles; they satisfy
\begin{equation}
\mathbb{E}\left[ \dd W_t^{(i)} \dd W_{t'}^{(j)} \right]
= \delta^{ij}\delta\left(t-t^\prime\right)\dd t^2.
\end{equation}
This naturally generalizes Eq.~(\ref{SchMod}) to a set of operators
and particles on which to project the relevant states.

\par

We now assume that one can decompose the total wave vector
$|\Psi\rangle$ in the form
\begin{equation}
|\Psi\left(\left\{ x_i\right\} \right)\rangle = |\Psi_{_\mathrm{CM}} 
\left(R\right)\rangle \otimes
|\Psi_\mathrm{rel} \left(\left\{ r_i\right\} \right)\rangle\, ,
\label{psi3}
\end{equation}
where the total wavefunction depends on the set of all the position
operators $\left\{ x_i\right\}$, while the "macroscopic" part of it,
$\vert \Psi_{_\mathrm{CM}}\rangle $, depends only on the position
$R\equiv N^{-1} \sum_i x_i$ of the center of mass, and the rest is a
function only of the relative coordinates $r_i$ defined through $x_i =
R+r_i$.

\par

Using It\^o calculus to evaluate the differential of the tensor
product in Eq.~({\ref{psi3}), it is easily checked that
  $|\Psi\left(\left\{ x_i\right\} \right)\rangle$ satisfies
  Eq.~(\ref{SchMod}) with $\hat{B}$ and $\dd W_t$ given by
  Eq.~(\ref{B7}) if the components of the product respectively satisfy
\begin{widetext}
\begin{equation}
\dd |\Psi_{_\mathrm{CM}}\left(R\right)\rangle =
\left\{ \left[ -i\hat{H}_{_\mathrm{CM}} - \frac{\gamma_{_\mathrm{CM}}}{2}
\left(\hat{R}-\langle \hat{R} \rangle\right)^2 \right] \dd t
+\sqrt{\gamma_{_\mathrm{CM}}} \left(\hat{R}-\langle 
\hat{R} \rangle\right) \dd W_t \right\}
|\Psi_{_\mathrm{CM}}\left(R\right) \rangle\, ,
\label{PsiCM}
\end{equation}
and
\begin{equation}
\dd |\Psi_\mathrm{rel} \left(\left\{ r_i\right\} \right)\rangle =
\left\{ \left[ -i\hat{H}_\mathrm{rel} - \frac{\gamma}{2}
\sum_{i=1}^{N-1} \left(\hat{r}_i-\langle \hat{r}_i \rangle\right)^2 
\right] \dd t
+\sqrt{\gamma}\sum_{i=1}^{N-1}  \left(\hat{r}_i-\langle \hat{r}_i
\rangle\right) \dd W_t^{(i)} \right\}
|\Psi_\mathrm{rel} \left(\left\{ r_i\right\} \right)\rangle\, ,
\label{psirel}
\end{equation}
\end{widetext}
where we have assumed the total Hamiltonian could be split into
$\hat{H}= \hat{H}_{_\mathrm{CM}}(\hat{R})+\hat{H}_\mathrm{rel}
\left(\left\{ \hat{r}_i\right\} \right)$ and the new constant
$\gamma_{_\mathrm{CM}}$ appearing in Eq.~(\ref{PsiCM}) is given by
$\gamma_{_\mathrm{CM}} = N\gamma$. This illustrates the mechanism
thanks to which localization is amplified for a macroscopic object
containing a large number (in practice $N\sim 10^{23}\gg 1$ for usual
classical systems) of particles, while the usual quantum spread is
mostly conserved for the internal degrees of freedom. A recent
inventory of all the constraints derived so far in various physical
situations on the CSL parameter $\gamma$ can be found in
Ref.~\cite{Feldmann:2011rt}.

\subsection{An Illustrative Example: the Harmonic Oscillator}
\label{sec:HarmOsc}

In this section, we illustrate how the CSL theory works on the example
of the harmonic oscillator resetting the Planck constant $\hbar$ for
easier comparison with previous works. This is an interesting case
because it represents the prototypical example of a quantum system
and, to our knowledge, this case has not been solved explicitly in the
case of the CSL theory. Moreover, in cosmology, as explained before,
we deal with a parametric oscillator, a case which shares some
similarities with an harmonic oscillator, at least in some regimes. It
is therefore important to understand first this simplest case in the
CSL framework.  In the following, we assume that the operator
$\hat{B}$ introduced in the previous section is the position operator
$\hat{x}$. As a consequence, the modified Schr\"odinger equation can
be written as
\begin{eqnarray}
\label{eq:Schrooscillo}
\dd\Psi &=&\left[-\frac{i}{\hbar}\hat{H}\dd t
+\sqrt{\gamma}\left(\hat{x}-\mean{\hat{x}}\right)
\dd W_t\right.\nonumber\\ 
& &-\left.
\frac{\gamma}{2}\left(\hat{x}-\mean{\hat{x}}\right)^ 2\dd t\right]
\Psi\, ,
\end{eqnarray} 
where $\hat{H}=\hat{p}^2/(2m)+m\omega^2\hat{x}^2/2$ is the
Hamiltonian. The parameter $\gamma $ sets the strength of the collapse
mechanism and, since we have chosen the position as the preferred
basis, it has dimension $L^{-2}\times T^{-1}$. Following
Ref.~\cite{Bassi:2005fp}, the wavefunction can be taken as a Gaussian
state and the most general form can be expressed as
\begin{eqnarray}
\label{eq:gaussianoh}
\Psi \left(t,x\right)&=&
\vert N\left(t\right)\vert \exp\Bigl\lbrace
-\Rea  \Omega \left(t \right)
\left[x-\bar{x}\left(t\right)\right]^2
+i\sigma (t)
\nonumber \\ & &
+i\chi (t)x
-i\Ima  \Omega (t) x ^2\Bigr\rbrace\, 
\end{eqnarray}
where, a priori, $\vert N\vert $, $\Rea  \Omega$, $\bar{x}$, $\sigma$,
$\chi$ and $\Ima  \Omega $ are real stochastic variables. Introducing
this wavefunction in Eq.~(\ref{eq:Schrooscillo}), one obtains the
following set of equations
\begin{eqnarray}
\label{eq:noh}
\frac{\vert N\vert ^\prime}{\vert N\vert}&=&
\frac14\frac{\left(\Rea  \Omega\right)^\prime}{\Rea  \Omega}
=\frac{\hbar}{m}\Ima  \Omega+\frac{\gamma}{4\Rea  \Omega}\, ,\\
\label{eq:reooh}
\left(\Rea  \Omega\right)^\prime &=&\gamma +4\frac{\hbar}{m}
\left(\Rea  \Omega \right)\left(\Ima  \Omega\right)\, ,\\
\label{eq:imooh}
\left(\Ima  \Omega\right)^\prime &=&-\frac{\hbar}{m}
\left[2\left(\Rea  \Omega\right)^2
-2\left(\Ima  \Omega\right)^2\right]+\frac{m}{\hbar}
\frac{\omega^2}{2}\, ,\cr
&&\\
\label{eq:xbaroh}
\bar{x}^\prime&=&\frac{\hbar}{m}\left[
\chi -2\left(\Ima  \Omega\right)\bar{x}\right]
+\frac{\sqrt{\gamma}}{2\Rea  \Omega}\frac{\dd W_t}{\dd t}\, ,\\
\label{eq:sigmaoh}
\sigma ^\prime&=&\frac{\hbar}{m}\left[
-\Rea  \Omega+2\left(\Rea  \Omega\right)^2
\bar{x}^2
-\frac12 \chi^2\right]\, ,\\
\label{eq:chioh}
\chi ^\prime&=&-\frac{\hbar}{m}\left[
4\left(\Rea  \Omega\right)^2\bar{x}
-2\chi \Ima  \Omega\right]\, ,
\end{eqnarray}
where a prime means a derivative with respect to time. We see that the
first equation can be integrated to give $\vert N\vert =(2\Rea
\Omega/\pi)^{1/4}$, which ensures that the wavefunction is properly
normalized. Then, the two following equations, Eqs.~(\ref{eq:reooh})
and~(\ref{eq:imooh}) ``decouple'' from the other equations and can be
integrated separately. In particular, if we add them up, we arrive at
\begin{equation}
\label{eq:ricattioh}
\Omega '=-2i\frac{\hbar}{m}\Omega ^2+\gamma +\frac{im}{2\hbar}\omega ^2.
\end{equation}
This equation should be compared to Eq.~(\ref{eq:solpsi}). As
expected, there are identical provided we take $\hbar=m=1$ and $\gamma
=0$. Of course, in the present case, the frequency $\omega$ is
constant since we deal with an harmonic oscillator rather than a
parametric oscillator as it is the case for cosmological
perturbations. Eq.~(\ref{eq:ricattioh}) is a Ricatti equation and we
have already seen that the appropriate change of variable to transform
it into a linear second order differential equation is
$\Omega=-imf'/(2\hbar f)$, where the function $f(t)$ obeys the
equation
\begin{equation}
\label{eq:modeoh}
f^{\prime\prime}
+\left(\omega^2-2i\frac{\hbar}{m}\gamma\right)
f=0\, .
\end{equation}
This equation admits simple solutions that can be expressed in terms
of exponentials, namely $f(t)\propto\exp\left(\pm \alpha \,t\right)$ where
$\alpha$ is defined by
\begin{equation}
\label{eq:defalphaoh}
\alpha\equiv \sqrt{\frac{2i\gamma\hbar}{m}-\omega^2}.
\end{equation}
As a consequence, the solution for $\Omega (t)$ can be written as
\begin{equation}
\label{eq:harmSol}
\Omega(t)=-\frac{im}{2\hbar}\alpha \tanh\left(\alpha\, t+\phi\right)\, ,
\end{equation}
where $\phi$ is an integration constant that can be expressed in terms 
of the initial value of the function $\Omega(t)$
\begin{equation}
\phi=\mathrm{argtanh}\left[
-\frac{2\hbar}{im}\frac{\Omega\left(t=0\right)}
{\alpha}\right]\, .
\end{equation}
This solution resembles the formula obtained in the case of the free
particle, see Ref.~\cite{Bassi:2005fp}.

\par

At this stage, we need to discuss the initial conditions. Our
assumption is that, at $t=0$, the quantum state is simply given by the
ground state of the harmonic oscillator in conventional quantum
mechanics. Technically, this means that we require the wavefunction
to be
\begin{equation}
\label{eq:ground}
\Psi\left(t=0,x\right)=\left(\frac{m\omega}{\pi\hbar}\right)^{1/4}
{\rm e}^{-m\omega x^2/(2\hbar)}\, ,
\end{equation}
which implies that $\Rea \Omega=m\omega /(2\hbar)$ and $\Ima \Omega=0$
or, equivalently, $\phi=\arg\tanh(i\omega/\alpha)$. Notice that this
choice is fully compatible with the normalization established above,
$\vert N\vert =(2\Rea \Omega/\pi)^{1/4}$. Of course, our choice also
amounts to imposing $\bar{x}(t=0)=\sigma(t=0)=\chi(t=0)=0$.

\par

Since the evolution of the stochastic wavefunction is controlled by
the function $\Omega (t)$, it is interesting to study how it evolves
with time. Writing the number $\alpha$ as $\alpha\equiv \alpha ^{\rm
  R}+i\alpha^{\rm I}$, where it is easy to show that
\begin{eqnarray}
\label{eq:defalphar}
\alpha^{\rm R} &=& \frac{\omega}{\sqrt{2}}
\left(\sqrt{1+4\frac{\hbar^2\gamma ^2}{m^2\omega^4}}-1\right)^{1/2},
\\
\label{eq:defalphai}
\alpha^{\rm I} &=& \frac{\sqrt{2}}{\omega}\frac{\hbar\gamma}{m}
\left(\sqrt{1+4\frac{\hbar^2\gamma ^2}{m^2\omega^4}}-1\right)^{-1/2},
\end{eqnarray}
and $\phi\equiv \phi ^{\rm R}+i\phi^{\rm I}$, straightforward
algebraic manipulations lead to the following expressions for $\Rea 
\Omega$ and $\Ima  \Omega $:
\begin{widetext}
\begin{eqnarray}
\label{eq:sol:Omega}
\Rea  \Omega(t)&=&\frac{m}{2\hbar}\frac{\alpha^\mathrm{I}
\sinh\left[2\left(\alpha^\mathrm{R} t
+\phi^\mathrm{R}\right)\right]+\alpha^\mathrm{R}
\sin\left[2\left(\alpha^\mathrm{I} t
+\phi^\mathrm{I}\right)\right]}
{\cos\left[2\left(\alpha^\mathrm{I} t
+\phi^\mathrm{I}\right)\right]+
\cosh\left[2\left(\alpha^\mathrm{R} t
+\phi^\mathrm{R}\right)\right]}
\, ,\\
\label{eq:sol:h}
\Ima  \Omega (t) &=&\frac{m}{2\hbar}\frac{\alpha^\mathrm{I}
\sin\left[2\left(\alpha^\mathrm{I} t
+\phi^\mathrm{I}\right)\right]-\alpha^\mathrm{R}
\sinh\left[2\left(\alpha^\mathrm{R} t
+\phi^\mathrm{R}\right)\right]}
{\cos\left[2\left(\alpha^\mathrm{I} t
+\phi^\mathrm{I}\right)\right]+
\cosh\left[2\left(\alpha^\mathrm{R} t
+\phi^\mathrm{R}\right)\right]}
\, .
\end{eqnarray}
\end{widetext}
In particular, the function $\Rea  \Omega (t)$, with the initial
condition specified above, is always positive. Notice also that there
is a sign ambiguity in the definitions of the quantities $\alpha^{\rm
  R}$ and $\alpha ^{\rm I}$ in Eqs.~(\ref{eq:defalphar})
and~(\ref{eq:defalphai}) but one can show that this does not affect
the physical predictions of the model. It is also interesting to
calculate the limit for large times of the two functions in
Eqs~(\ref{eq:sol:Omega}) and~(\ref{eq:sol:h}). One obtains
\begin{eqnarray}
\label{eq:Omega:limit+}
\lim_{t\rightarrow \infty} \Rea  \Omega&=&\frac{m\alpha^\mathrm{I}}{2\hbar}
\simeq \frac{m\omega}{2\hbar}
\left(1+\frac12\frac{\hbar^2\gamma^2}{m^2\omega^4}+\cdots \right)\, , 
\nonumber\\ \\
\lim_{t\rightarrow +\infty}\Ima  \Omega &=&-\frac{m\alpha^\mathrm{R}}{2\hbar}
\simeq -\frac{\gamma}{2\omega}
\left(1-\frac12\frac{\hbar^2\gamma^2}{m^2\omega^4}+\cdots 
\right)\, , \nonumber\\
\label{eq:Omega:limit-}
\end{eqnarray}
where the dots indicate an expansion in the small dimensionless
parameter $\hbar \gamma /(m\omega ^2)$. We see that, if $\gamma=0$, we
obtain the ground state given by Eq.~(\ref{eq:ground}). Deviations
from that solution are controlled by the parameter $\hbar
\gamma/(m\omega ^2)$.

\par

We are now in a position where one can investigate the physical
properties of the quantum state~(\ref{eq:gaussianoh}). In
particular, it is easy to show that $\mean{\hat{x}}=\bar{x}$ and
$\mean{\hat{p}}=\chi-2\left(\Ima  \Omega\right)\bar{x}$. Initially,
$\bar{x}=0$ and the position operator has a vanishing mean value as
expected for the ground state of the harmonic oscillator but, at later
times, due to the stochastic evolution of the wavefunction, it
acquires a non zero value. It is also possible to calculate the spread
in position and momentum. One obtains
\begin{eqnarray}
\label{eq:sigmanu}
\sigma_{x}\equiv
\sqrt{\mean{\hat{x}^2}
-\mean{\hat{x}}^2}&=&
\frac{1}{2}\frac{1}{\sqrt{\Rea  \Omega}}\, ,\\
\sigma_{p}\equiv
\sqrt{\mean{\hat{p}^2}
-\mean{\hat{p}}^2}&=&
\hbar \sqrt{\frac{(\Rea  \Omega)^2
+(\Ima  \Omega)^2}{\Rea  \Omega}}\, .\cr &&
\end{eqnarray}
We see that these quantities only depend on $\Rea \Omega$ and $\Ima
\Omega$. As a consequence, inserting Eq.~(\ref{eq:sol:Omega}) and
(\ref{eq:sol:h}) in the above expressions of $\sigma_x$ and $\sigma
_p$, one arrives at
\begin{widetext}
\begin{eqnarray}
\label{eq:spreads:HO}
\sigma_{x}&=&\sqrt{\frac{\hbar}{2m}}
\sqrt{
\frac{\cos\left[2\left(\alpha^\mathrm{I} t
+\phi^\mathrm{I}\right)\right]+
\cosh\left[2\left(\alpha^\mathrm{R} t
+\phi^\mathrm{R}\right)\right]}{\alpha^\mathrm{I}
\sinh\left[2\left(\alpha^\mathrm{R} t
+\phi^\mathrm{R}\right)\right]+\alpha^\mathrm{R}
\sin\left[2\left(\alpha^\mathrm{I} t
+\phi^\mathrm{I}\right)\right]}}
\, ,\nonumber\\
\sigma_{p}&=&\sqrt{\frac{m\hbar}{2}}
\sqrt{\left(\alpha^\mathrm{R}\right)^2+\left(\alpha^\mathrm{I}\right)^2}
\sqrt{\frac{\cosh\left[2\left(\alpha^\mathrm{R} t
+\phi^\mathrm{R}\right)\right]
-\cos\left[2\left(\alpha^\mathrm{I} t
+\phi^\mathrm{I}\right)\right]}
{\alpha^\mathrm{I}\sinh\left[2\left(\alpha^\mathrm{R} t
+\phi^\mathrm{R}\right)\right]+
\alpha^\mathrm{R}\sin\left[2\left(\alpha^\mathrm{I} t
+\phi^\mathrm{I}\right)\right]}}
\, . 
\end{eqnarray}
\end{widetext}
\begin{figure*}
\begin{center}
\includegraphics[width=0.5\textwidth,clip=true]{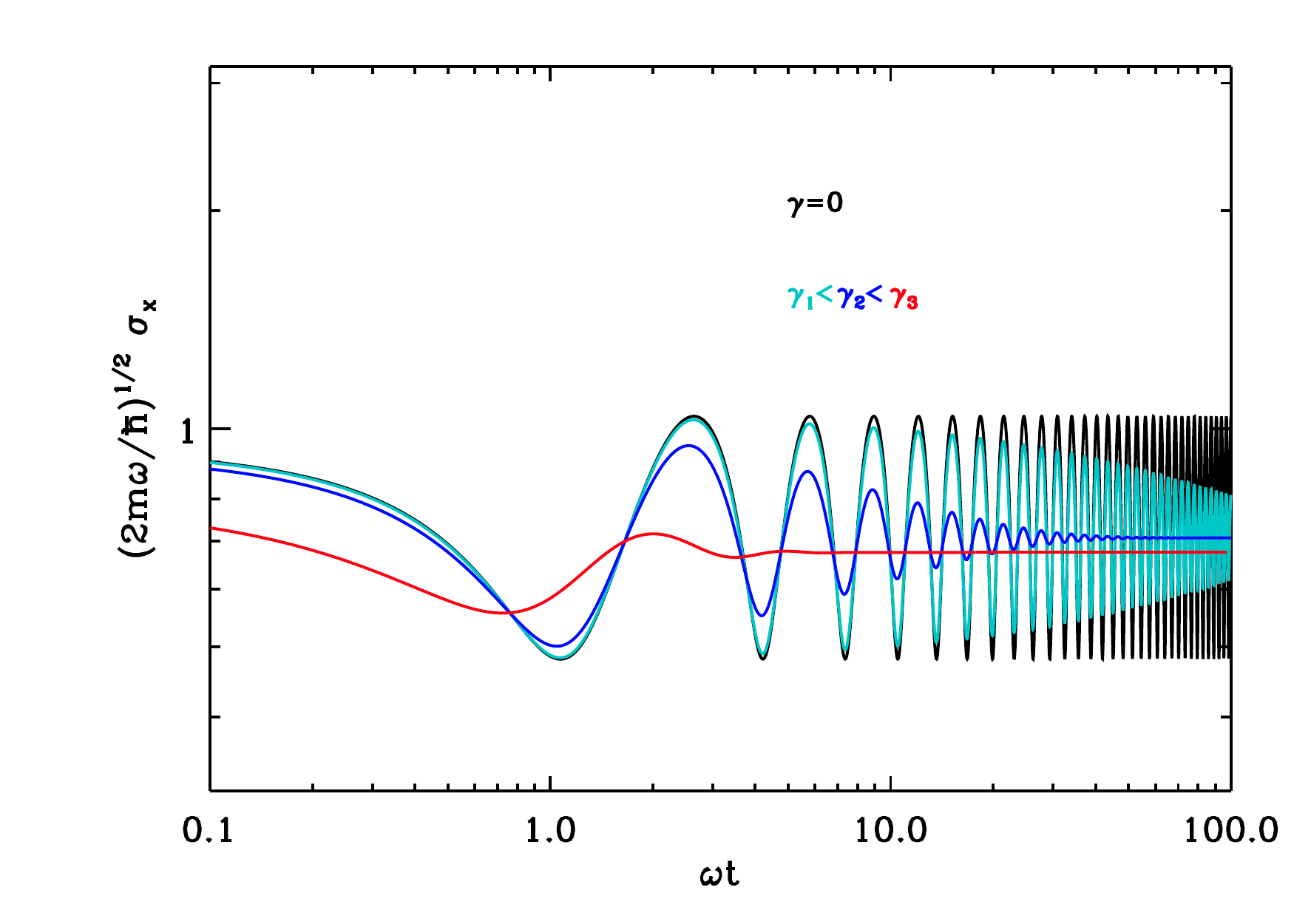}\hskip-5mm
\includegraphics[width=0.5\textwidth,clip=true]{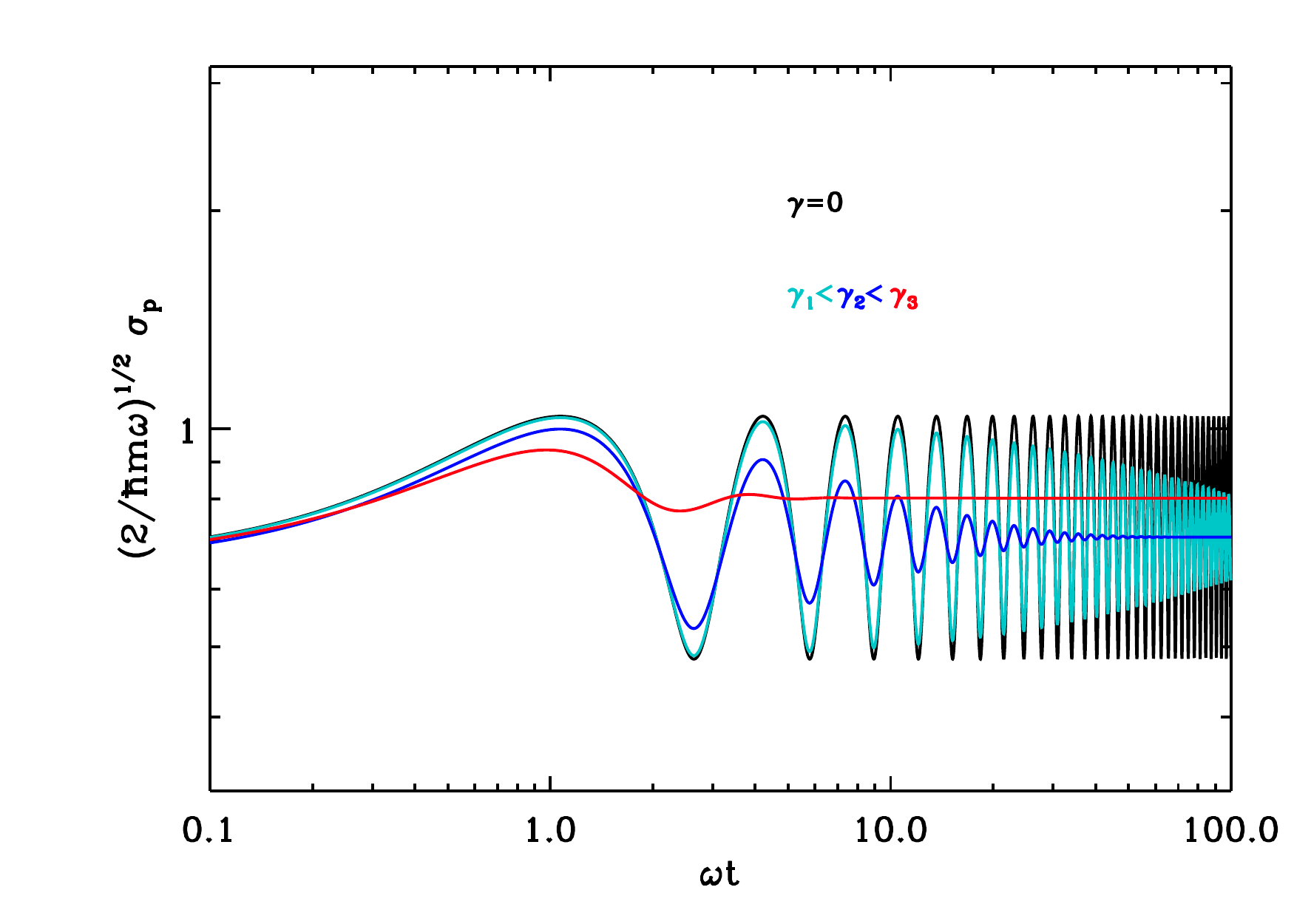}
\caption{Spread in position and momentum for different values of
  $\gamma$ in the case of the harmonic oscillator, see
  Eq.~(\ref{eq:spreads:HO}). The conventional Schr\"odinger
  evolution corresponds to $\gamma=0$ and is represented by the black
  curve which oscillates. On the contrary, when the collapse mechanism
  is turned on, the oscillations are damped (blue and red curves), the
  spreads tend toward a constant value and localization occurs.}
\label{fig:spread}
\end{center}
\end{figure*}
The time evolution of these quantities is displayed in
Fig.~\ref{fig:spread}. The black curves correspond to the conventional
Schr\"odinger evolution, \ie the case $\gamma=0$.  They show the usual
oscillatory behavior. On the contrary, when $\gamma \neq 0$, we see
that the oscillations are damped (see the red and blue curves).  Then,
the spreads converge towards a constant value, which only depends on
$\gamma$, $\omega$ and $m$. This value is easy to evaluate and one
finds
\begin{eqnarray}
\hskip-2mm
\lim_{t\rightarrow\infty}\sigma_{x}&=&
\frac{1}{2^{3/4}}\sqrt{\frac{\omega}{\gamma}}
\left(\sqrt{1+4\frac{\hbar^2\gamma^2}{m^2\omega^4}}-1\right)^{1/4}
, \\
\hskip-2mm
\lim_{t\rightarrow\infty}\sigma_{p}&=&
m\omega \left(1+4\frac{\hbar^2\gamma^2}{m^2\omega^4}\right)^{1/4}\times
\lim_{t\rightarrow\infty}\sigma_{x}. 
\end{eqnarray}
{}From these formulas one can see that the spread in position at
infinity decreases with $\gamma $, from $\sqrt{\hbar/(2m\omega)}$ for
$\gamma =0$ to $0$ for $\gamma \rightarrow \infty$. We see that the
modified Schr\"odinger equation, as expected, implies a localization
in position. We also notice that the microscopic behavior of the
system is altered by the non linear and stochastic terms added to the
theory. By contrast, in order to satisfy the Heisenberg uncertainty
relation, the spread in momentum increases with $\gamma $, from
$\sqrt{m\omega \hbar/2}$ for $\gamma =0$ to infinity for $\gamma
\to\infty$. For $\gamma =0$ and at large times, one finds that the
Heisenberg relation is saturated, $\sigma _x\sigma_p=\hbar/2$, as
appropriate for a coherent state. In the limit $\gamma \rightarrow
\infty$ , one finds a larger value $\sigma
_x\sigma_p=\hbar/\sqrt{2}$. Let us also remark that an exact
eigenstate of the operator $\hat{x}$ is given by a Dirac function
$\delta\left(x-\bar{x}\right)$ centered at some value $\bar{x}$. On
the other hand, we see that adding non linear and stochastic terms
results in a spreading of the Dirac function into a Gaussian
wavefunction with a finite width decreasing for increading
$\gamma$. Therefore, the modified Schr\"odinger equation does not
exactly lead to an eigenstate of the position operator. In fact, the
asymptotic value of $\sigma _x$ obtained above defines the
``precision" of the collapse and characterizes how close to an
eigenstate of the collapse operator the final state is. In that sense,
since $\sigma_x$ decreases with $\gamma$, the bigger $\gamma$, the
more ``precise" the collapse.

\par

To conclude this section, it is also interesting to calculate the time
derivative of the quantum mean value of the Hamiltonian operator. One obtains
\begin{eqnarray}
\frac{{\rm d}\mean{\hat{H}}}{{\rm d}t}
&=&\frac{\hbar^2}{2m}\gamma -\frac{\hbar}{m}\sqrt{\gamma}
\frac{\Ima  \Omega}{\Rea  \Omega}
\mean{\hat{p}}\frac{{\rm d}W_t}{{\rm d}t}
\nonumber \\ & &
+\frac12m\omega^2\mean{\hat{x}}\frac{\sqrt{\gamma}}{\Rea  \Omega}
\frac{{\rm d}W_t}{{\rm d}t}.
\end{eqnarray}
This equation implies that
\begin{eqnarray}
\label{eq:nonconservationH}
\frac{{\rm d}\, \mathbb{E}\left[\mean{\hat{H}}\right]}{{\rm d}t}
&=&\frac{\hbar^2}{2m}\gamma \, .
\end{eqnarray}
As is well know, this formula expresses the non conservation of energy
in the CSL theory. From a phenomenological point of view, this
increase of energy is usually so small (given the values of $\gamma $
usually considered) that it cannot be detected. Put it differently,
the non conservation of energy in the CSL theory cannot be used to
rule out this theory~\cite{Bassi:2012bg}.

\section{The Inflationary CSL theory}
\label{sec:infcsl}

The dynamical collapse model of the previous sections should apply to
any quantum system, and hence in particular to cosmological
perturbations as they arise from vacuum fluctuations. Spontaneously
collapsing these happens to be a tremendously complicated task for
many reasons discussed below, so in what follows, we suggest a much
simplified modeling method which we then apply to the inflationary
situation.

\subsection{The Modified Schr\"odinger Equation for the
  Mukhanov-Sasaki Variable}
\label{subsec:modifiedforv}

The first obvious problem one encounters when dealing with quantum
cosmological perturbations is that the underlying theory ought to be
relativistic. The straightforward relativistic generalization of the
CSL model for quantum field theory, starting with the action
(\ref{eq:action}) in the Tomonaga picture for instance, leads to
unremovable divergences \cite{Ghirardi:1989cn} (see however
\cite{Bedingham:2010hz}), even more so when non linearities inherent
to general relativity are taken into account.

\par

The second next option which happens to lead to a model in which
calculations are actually possible consists in noting, as mentioned
earlier in Sec.~\ref{subsec:quantumcmb}, that the spectrum of
primordial perturbations depends on the wavenumber $\bm{k}$. In other
words, once the Fourier spectrum is known, all the observable
quantities related with the CMB can be computed and compared with
actual data.  This means that mere knowledge of the modes
$\hat{v}_{\bm{k}}$ ought to be enough in order for a complete
description of the possible observations to be realized.

\par

We shall therefore accordingly assume in what follows that the
modified Schr\"odinger equation of motion for the wave function will
be done at the level of the Fourier mode $\Psi_{\bm{k}}$ , with
spontaneous localization on the $\hat{v}_{\bm{k}}$
eigenmanifolds. This is consistent with previous approaches aimed at
studying decoherence of cosmological perturbations where the pointer
basis is often assumed to be precisely the Mukhanov-Sasaki operators,
see Ref.~\cite{Kiefer:2006je}. Separating as before into real and
imaginary parts, we shall thus assume the following basic equation
\begin{eqnarray}
\dd\Psi_{\bm{k}}^\mathrm{R}&=&\left[-i
\hat{\mathcal{H}}_{\bm{k}}^\mathrm{R}\dd\eta
+\sqrt{\gamma}\left(\hat{v}_{\bm{k}}^\mathrm{R}-
\mean{\hat{v}_{\bm{k}}^\mathrm{R}}\right)
\dd W_\eta\right.\nonumber\\ 
& &-\left.
\frac{\gamma}{2}\left(\hat{v}_{\bm{k}}^\mathrm{R}-
\mean{\hat{v}_{\bm{k}}^\mathrm{R}}\right)^ 2\dd\eta\right]
\Psi_{\bm{k}}^\mathrm{R}\, ,
\label{eq:SchroModes}
\end{eqnarray} 
and a similar equation for $\Psi_{\bm{k}}^\mathrm{I}$. Here, the
quantity $\gamma $ is a positive constant with mass dimension $2$ if
the scale factor is chosen to be dimensionless but is dimensionless if
the scale factor is chosen to have mass dimension $-1$, which is the
convention adopted here. As in Sec.~\ref{sec:grw}, the parameter
$\gamma $ sets the strength of the collapse mechanism.

\par

\begin{figure*}[t]
\begin{center}
\includegraphics[width=0.85\textwidth,clip=true]{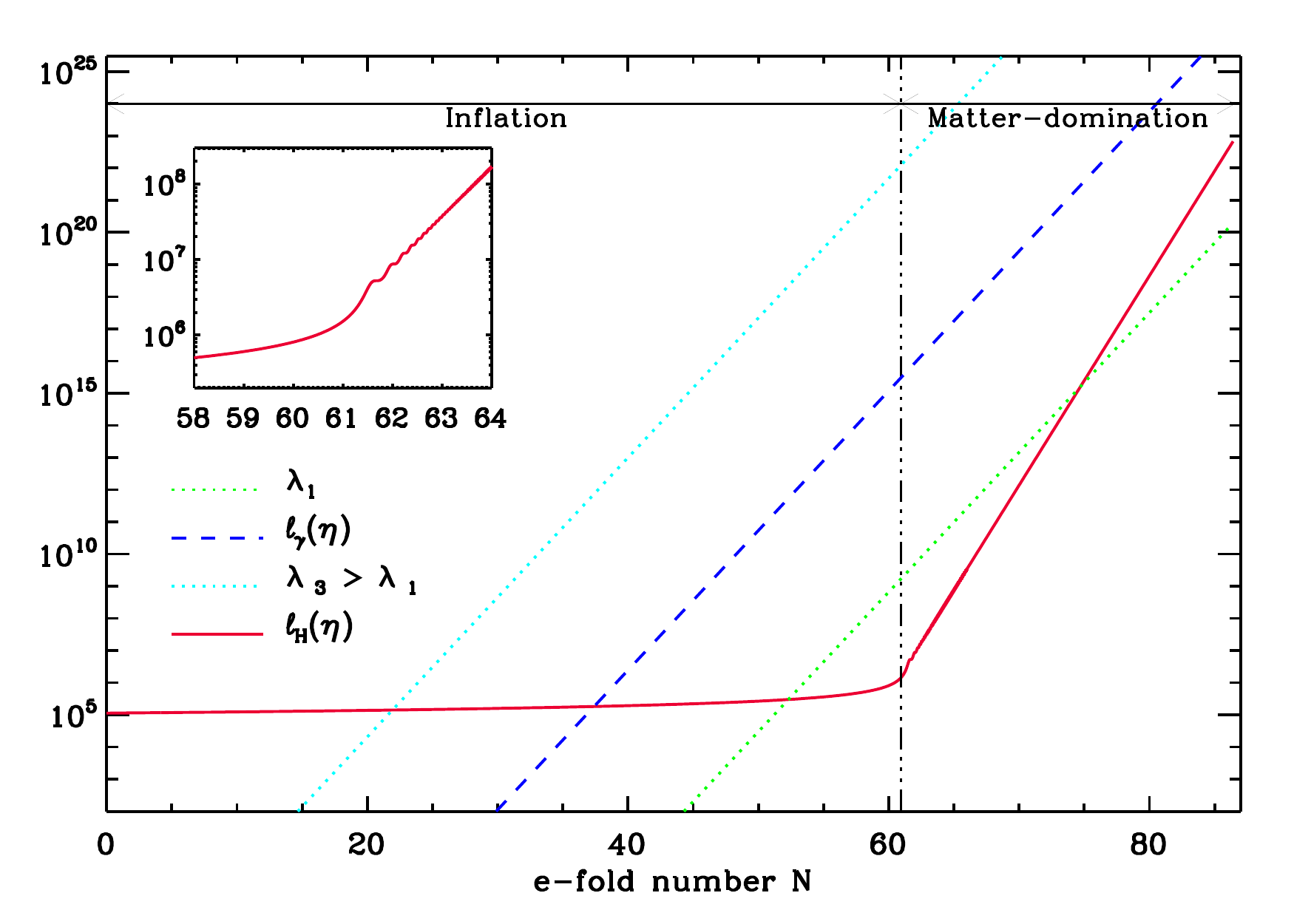}
\caption{Evolution of various physical length scales with time during
  the cosmic history in the CSL model described by
  Eq.~(\ref{eq:SchroModes}) with a zoom on the transition from
  inflation to reheating inserted (see the concluding section).  The
  solid line represents the Hubble radius $\ell_{_{\rm H}}$ and the
  dashed-dotted green and red lines, the physical wavelengths of two
  Fourier modes of cosmological relevance today. The solid blue line
  represents the built-in CSL scale $\ell_{\gamma }$, see the
  discussion above Eq.~(\ref{eq:argbesselcsl}). It is a preferred
  comoving scale and can also be viewed as a time dependent preferred
  physical scale. Therefore, when a mode is below (above)
  $\ell_{\gamma}$ it remains so during the whole history of the
  universe as is clear from the plot. This means that, contrary to the
  Hubble scale, there is no ``$\ell_{\gamma }$ crossing'' during the
  cosmic evolution. As a consequence, one expects the power spectrum
  to acquire a broken power-law shape, with two different branches, an
  expectation confirmed by the calculations in
  Sec.~\ref{subsec:cslps}.}
\label{fig:scalecsl}
\end{center}
\end{figure*}

\par

Let us now review all the limitations of postulating an ansatz
equation such as Eq.~(\ref{eq:SchroModes}). First, one should note
that the constant $\gamma$ in Eq.~(\ref{eq:SchroModes}) cannot be the
same as the one associated with the choice of the position operator as
the collapse operator appearing in Eq.~(\ref{eq:Schrooscillo}),
despite our choice of the same notation. It is clear that each time
one considers different ``collapse operators'', this leads to
different CSL parameters with different mass dimension. The same
phenomenon is observed in Ref.~\cite{Bassi:2003vf} where the
``collapse operator'' is chosen to be a spin operator. In this case,
it is clear that the corresponding CSL parameter cannot be the same as
the one corresponding to the case where the ``collapse operator'' is
the position (as it is for the case of the free
particle~\cite{Bassi:2005fp}). This is unfortunate when it comes to a
comparison of the constraints obtained in the laboratory with the
constraints obtained in cosmology. In fact, what could be done is to
consider the strict CSL theory where the ``collapse operator'' is
usually taken to be the averaged density operator. In the language of
cosmological perturbations, this amounts to assuming that there is
spontaneous localization on the $\widehat{\delta \rho}(\eta,{\bm x})$
eigenmanifold, where $\delta \rho (\eta ,{\bm x})$ is the perturbed
energy density. This would have the advantage to introduce a universal
$\gamma $ with always the same dimension. Unfortunately, $\delta \rho
(\eta,{\bm x})$ is a complicated functional of $v_{\bm k}$ and this
would probably render the whole approach untractable. Let us also
notice that $\gamma $ could be taken as a function of the wavenumber
${\bm k}$, \ie different CSL parameters for different modes. In this
article, for simplicity, we do not follow this route.

\par

Another issue is that we moved from real to reciprocal space while
keeping the structure of the equation unchanged. In so doing, we also
avoid from the outset any mode mixing that would be naturally arising
from a real space modified Schr\"odinger equation: its stochastic
version being non linear, one would expect a coupling of the Fourier
modes, which is here automatically set to zero. Note this
approximation is justified by data observations of the CMB.

\par

Another important limitation of our treatment is the fact that the
collapse concerns the modes independently. As a result, the
amplification mechanism, so crucial to explain why the quantum
behavior becomes increasingly less important for increasingly large
systems (the effective collapse time being inversely proportional to
the number of particles involved and, hence, to the size of the
system), is simply not operating here! Therefore, even though one
might consider cosmological size effects, the collapse will occur just
as it would for an independent quantum particle. As we will see, that
implies a severe constraint on the constant $\gamma$ when comparison
of the modified spectrum is made to actual observations on Hubble-size
scales.

\par

Finally, Eq.~(\ref{eq:SchroModes}) is written in terms of the
conformal Fourier mode of the original action. Because its
normalization implies the equation be non linear, this means the
constant $\gamma$ can be translated, as we will show later, into a
privileged {\sl conformal} scale, and hence a time-dependent
privileged length $\ell_{\gamma }$, as shown in
Fig.~\ref{fig:scalecsl} and the discussion above
Eq.~(\ref{eq:argbesselcsl}). This is somehow similar to the fact,
except at the perturbative and conformal levels, that considering non
flat spatial sections permits to define a curvature length and thus
forbids to renormalize the scale factor arbitrarily. However, as shown
in Appendix \ref{sec:appendix}, this last limitation does not affect
the general conclusions drawn here.

\subsection{Gaussian State}
\label{subsec:gaussianstate}

Our goal is now to solve Eq.~(\ref{eq:SchroModes}). As was done for
the standard case~(\ref{eq:wavefunction}), one considers that the
wavefunction assumes a Gaussian shape. Concretely, we take the most
general form, namely
\begin{widetext}
\begin{equation}
\label{eq:SingleGaussianR}
\Psi_{\bm{k}}^\mathrm{R,I}\left(\eta,v_{\bm{k}}^\mathrm{R,I}\right)=
\vert N_{\bm{k}}\left(\eta\right)\vert \exp\Bigl\lbrace
-\Rea  \Omega_{\bm{k}}\left(\eta\right)
\left[v_{\bm{k}}^\mathrm{R,I}-\bar{v}_{\bm{k}}^\mathrm{R,I}\left(\eta\right)\right]^2
+i\sigma_{\bm k}^\mathrm{R,I}(\eta)+i\chi_{\bm k}^\mathrm{R,I}(\eta)
v_{\bm{k}}^\mathrm{R,I}
-i\Ima  \Omega_{\bm k}(\eta )
\left(v_{\bm{k}}^\mathrm{R,I}\right)^2\Bigr\rbrace\, ,
\end{equation}
where $\bar{v}_{\bm{k}}^\mathrm{R,I}$, $\sigma_{\bm{k}}^\mathrm{R,I}$
and $\chi_{\bm k}^\mathrm{R,I}$ are real numbers.  The fact that one
can assume $\vert N_{\bm{k}}\vert $ and $\Omega_{\bm{k}}$ to be
independent of ``$\mathrm{R}$'' or ``$\mathrm{I}$'' will be justified
in the following. Compared to Eq.~(\ref{eq:wavefunction}), we see that
Eq.~(\ref{eq:SingleGaussianR}) is more general and, therefore,
contains more parameters. The case of Eq.~(\ref{eq:wavefunction})
corresponds to $\bar{v}_{\bm{k}}^\mathrm{R,I}=0$,
$\chi_{\bm{k}}^\mathrm{R,I}=0$ and $\arg N_{\bm k}=\sigma _{\bm k}$.
Of course, the above Gaussian is similar to the wavefunction
considered in the case of the harmonic oscillator of
Eq.~(\ref{eq:gaussianoh}). The only difference is that the stochastic
functions characterizing the wavefunction now depend on the wavenumber
${\bm k}$ and the role of the position is played by the Fourier
amplitude of the Mukhanov-Sasaki variable.

\par

The next step is to insert Eq.~(\ref{eq:SingleGaussianR}) into
Eq.~(\ref{eq:SchroModes}) in order to derive the differential
equations obeyed by the functions parameterizing the Gaussian state.
Straightforward manipulations making use of the It\^o calculus lead to
the following expressions
\begin{eqnarray}
\label{eq:evol:N}
\frac{\vert N_{\bm{k}}\vert ^\prime}{\vert N_{\bm{k}}\vert}&=&
\frac14\frac{\left(\Rea  \Omega_{\bm{k}}\right)^\prime}{\Rea  \Omega_{\bm{k}}}
=\Ima  \Omega_{\bm{k}}+\frac{\gamma}{4\Rea  \Omega_{\bm{k}}},\\
\label{eq:evol:Omega}
\left(\Rea  \Omega_{\bm{k}}\right)^\prime &=&\gamma+4\left(\Rea \Omega_{\bm{k}}\right)
\left(\Ima  \Omega_{\bm k}\right),\\
\label{eq:evol:h}
\left(\Ima  \Omega_{\bm{k}}\right)^\prime &=&-2\left(\Rea  \Omega_{\bm{k}}\right)^2
+2\left(\Ima  \Omega_{\bm{k}}\right)^2+\frac12\omega^2
\left(\eta, \bm{k}\right),\\
\label{eq:evol:barnu}
\left(\bar{v}_{\bm{k}}^\mathrm{R,I}\right)^\prime&=&
\chi_{\bm{k}}^\mathrm{R,I}+\frac{\sqrt{\gamma}}
{2\Rea  \Omega_{\bm{k}}}\frac{\dd W_\eta}{\dd \eta}
-2\left(\Ima  \Omega_{\bm{k}}\right)\bar{v}_{\bm{k}}^\mathrm{R,I},\\
\label{eq:evol:f}
\left(\sigma_{\bm{k}}^\mathrm{R,I}\right)^\prime&=&
-\Rea  \Omega_{\bm{k}}+2\left(\Rea  \Omega_{\bm{k}}\right)^2
\left(\bar{v}_{\bm{k}}^\mathrm{R,I}\right)^2
-\frac12\left(\chi_{\bm{k}}^\mathrm{R,I}\right)^2, \\
\label{eq:evol:g}
\left(\chi_{\bm{k}}^\mathrm{R,I}\right)^\prime&=&
-4\left(\Rea  \Omega_{\bm{k}}\right)^2\bar{v}_{\bm{k}}^\mathrm{R,I}
+2\chi_{\bm{k}}^\mathrm{R,I}\left(\Ima  \Omega_{\bm{k}}\right).
\end{eqnarray}
\end{widetext}
Several remarks are in order at this point. Firstly, we see that the
evolution equations for $\vert N_{\bm k}\vert$, $\Rea  \Omega_{\bm{k}}$
and $\Ima  \Omega_{\bm{k}}$ are deterministic and independent of that of
$\bar{v}_{\bm{k}}^\mathrm{R,I}$, $\sigma_{\bm{k}}^\mathrm{R,I}$ or
$\chi_{\bm{k}}^\mathrm{R,I}$. This justifies the fact that one can
assume these quantities to be independent on $\mathrm{R,I}$ provided
similar initial conditions are chosen for $\mathrm{R,I}$. This also
means that these three quantities are not random (but their evolution
is still explicitly modified by the stochastic dynamics when $\gamma
\neq0$). Secondly, Eq.~(\ref{eq:evol:N}) explicitly implies the
conservation of the wavefunction norm: if one initially has a
normalized state, \ie
\begin{equation}
\vert N_{\bm{k}}\vert =\left(\frac{2\Rea  \Omega_{\bm{k}}}{\pi}\right)^{1/4}\, ,
\end{equation}
it will remains so at any time. In fact, this equation is similar to
Eq.~(\ref{eq:solpsi}) which is therefore not modified by the
introduction of the non linear stochastic terms. Moreover, in the
present case where the wavefunction is given by a single Gaussian,
$\sigma _{\bm{k}}^\mathrm{R,I}$ is just an irrelevant global phase and
can be ignored (this will no longer be the case when the quantum state
is the sum of two Gaussians, see below). Thirdly, it is easy to check
that Eqs.~(\ref{eq:evol:N}), (\ref{eq:evol:Omega}), (\ref{eq:evol:h}),
(\ref{eq:evol:barnu}), (\ref{eq:evol:f}) and~(\ref{eq:evol:g}) are the
exact counter parts of Eqs.~(\ref{eq:noh}), (\ref{eq:reooh}),
(\ref{eq:imooh}), (\ref{eq:xbaroh}), (\ref{eq:sigmaoh})
and~(\ref{eq:chioh}). The only difference is that $\omega$ is now a
time-dependent quantity as expected since we deal with a parametric
oscillator. We conclude that, instead of six coupled stochastic
differential equations, we have in fact to solve two sets of two
coupled differential equations, the first one being deterministic and
the second one being stochastic. In particular
Eqs.~(\ref{eq:evol:Omega}) and~(\ref{eq:evol:h}) can be combined and
lead to the following Ricatti equation for the quantity $\Rea 
\Omega_{\bm{k}}+i\Ima  \Omega_{\bm{k}}=\Omega_{\bm k}$:
\begin{equation}
\Omega_{\bm{k}}^\prime=-2i\Omega_{\bm{k}}^2 
+\gamma+\frac{i}{2}\omega^2\left(\eta,\bm{k}\right)\, .
\end{equation}
This equation is similar to Eq.~(\ref{eq:ricattioh}) obtained for the
harmonic oscillator. Of course, if $\gamma =0$, then one exactly
recovers the Ricatti equation~(\ref{eq:equapsi}). As discussed before,
a Ricatti equation can always be reduced to a linear but second order
differential equation: this is achieved through the transformation
$\Omega_{\bm{k}}=-if_{\bm{k}}^\prime/(2f_{\bm{k}})$, where
$f_{\bm{k}}$ is a solution of the following linear differential
equation:
\begin{equation}
\label{eq:u}
f_{\bm{k}}^{\prime\prime}
+\left[\omega^2\left(\eta,\bm{k}\right)-2i\gamma\right]
f_{\bm{k}}=0\, .
\end{equation}
This equation is very similar to the equation for the mode function
considered before. The only difference is the appearance of the term
$-2i\gamma $ in the effective frequency. Obviously, if $\gamma =0$,
then one recovers the conventional case. Moreover, the fact that this
is the counterpart of Eq.~(\ref{eq:modeoh}) is obvious.

\subsection{Evolution of the Stochastic Wave-function during Inflation}
\label{subsec:evolpsi}

We now study the time evolution of the quantum
state~(\ref{eq:SingleGaussianR}) in more detail. We start with the
evolution of $\Rea  \Omega_{\bm k}$ and $\Ima  \Omega _{\bm k}$ since we
have shown in the last section that it decouples from the other
equations of motion. To derive the corresponding solutions, it is
sufficient to solve Eq.~(\ref{eq:u}). If the background is driven by a
phase of power-law inflation, $\omega\left(\eta, \bm{k}\right)$ is
given by
$\omega\left(\eta,\bm{k}\right)=k^2-\beta\left(\beta+1\right)/\eta^2$
and the differential equation~(\ref{eq:u}) reads
\begin{equation}
\label{eq:modecsl}
f_{\bm{k}}^{\prime\prime}
+\left[k^2-\frac{\beta\left(\beta+1\right)}
{\eta^2}-2i\gamma\right]
f_{\bm{k}}=0\, .
\end{equation}
We see that the only effect of the CSL term $-2i\gamma $ is to modify
the comoving wave number $k^2\rightarrow k^2-2i\gamma $. The solution
of Eq.~(\ref{eq:modecsl}) can be written in terms of Bessel functions
\begin{eqnarray}
\label{eq:solmodecsl}
f_{\bm{k}}(\eta) &=&\left(-z_{\bm{k}}\, k\eta\right)^{1/2}\bigl[ 
C_{\bm{k}}J_{\beta+\frac{1}{2}}\left(-z_{\bm{k}}\, k\eta\right)
\nonumber \\ & & + 
D_{\bm{k}}J_{-(\beta+1/2)}\left(-z_{\bm{k}}\, k\eta\right)\bigr]\, ,
\end{eqnarray}
where $C_{\bm{k}}$ and $D_{\bm{k}}$ are integration constants and
where the complex number $z_{\bm k}$ is defined by
\begin{equation}
\label{eq:defz}
z_{\bm{k}}\equiv\sqrt{1-i\frac{2\gamma}{k^2}}=
\left(1+4\frac{\gamma^2}{k^4}\right)^{1/4}
\mathrm{e}^{-\frac{i}{2}\mathrm{arctan}\left(2\gamma/k^2\right)}.
\end{equation}
Eq.~(\ref{eq:solmodecsl}) should be compared to its non-CSL
counterpart, Eq.~(\ref{eq:solbessel}). The only difference is the
appearance of the $z_{\bm{k}}$ factor. This is consistent with the
remark made above since this factor always multiplies the expression
$k\eta$ and can, therefore, be viewed as a ``renormalization'' of the
wavenumber $k$. In the non-CSL case where $\gamma =0$, one obviously
has $z_{\bm{k}}=1$ and Eq.~(\ref{eq:solmodecsl}) reduces to
Eq.~(\ref{eq:solbessel}). It is interesting to remark that $z_{\bm k}$
for the parametric oscillator plays a role similar to that of $\alpha$ for
the harmonic oscillator, see the definition~(\ref{eq:defalphaoh}). In
fact, strictly following this last definition, one can introduce a
mode-dependent $\alpha _{\bm k}$ parameter, namely $\alpha_{\bm
  k}\equiv \sqrt{2i\gamma -k^2}$ (using $\omega=k$ for massless
perturbations) and, then, $z_{\bm k}$ appears to be just a rescaled
$\alpha_{\bm k}$ parameter: $\alpha_{\bm k}=ikz_{\bm k}$. Finally,
notice also that the sign ambiguity in the definition of $z_{\bm{k}}$
due to the presence of a square root has absolutely no impact on the
results presented below.

\par

Let us now discuss the solution $f_{\bm k}(\eta)$ and what this
implies for the behavior of the wavefunction. In presence of the CSL
term, the problem is characterized by three scales: the wavelength of
the Fourier mode given by $\lambda_{\bm{k}}(\eta)= a(\eta)/k $, the
Hubble radius $\ell_{\mathrm{H}}\left(\eta\right)= a^2/a^{\prime}$ and
a new scale associated with the parameter $\gamma$ defined by
$\ell_{\gamma}\equiv a\left(\eta\right)/\sqrt{\gamma
  }$ or, in terms of mass scale, $M_{\gamma}\equiv
  \sqrt{\gamma}/a(\eta)$. Notice that $\ell _{\gamma }$ is a new,
  time-dependent, physical scale that is built in the inflationary CSL
  theory, see Fig.~\ref{fig:scalecsl}. In terms of these three
  physical scales, the quantity $z_{\bm k}k\eta$ which appears in
  Eq.~(\ref{eq:solmodecsl}) can be written as
\begin{equation}
\label{eq:argbesselcsl}
z_{\bm k}k\eta=(1+\beta)\frac{\ell_{_{\rm H}}}{\lambda_{\bm k}}
\sqrt{1-2i\frac{M_{\gamma}^2}{k^2_{\rm phys}}}
\end{equation}
where $k_{\rm phys}=k/a$ is the physical wavenumber. At the beginning
of inflation, the modes of cosmological interest today laid far inside
the Hubble radius, which means $\lambda_{\bm{k}}\ll
\ell_{\mathrm{H}}$, \ie $k\eta\rightarrow -\infty$. Notice that these
considerations are independent of the value of $M_{\gamma}$. Indeed,
if $k_{\rm phys}\gg M_{\gamma}$, then $z_{\bm k}\simeq 1$ and the
previous limit is not changed. On the contrary, if $k_{\rm phys}\ll
M_{\gamma }$, then the condition $\vert z_{\bm k}\vert \gg 1$ is even
better satisfied. It is also interesting to remark that, in this last
case, $z_{\bm k}k\eta$ does not go to $-\infty$ along the real axis
but along a direction that is inclined in the complex plane. However,
this does not change the asymptotic behavior of the Bessel functions
in this regime. Upon using Eq.~(\ref{eq:solmodecsl}), one obtains
\begin{eqnarray}
\label{eq:superHcsl}
\lim_{\lambda_{\bm k}/\ell_{\rm H}\rightarrow 0}f_{\bm{k}}(\eta)&=&
\sqrt{\frac{2}{\pi}}
\biggl[C_{\bm{k}}\sin\left(-z_{\bm{k}}\, k\eta-\frac{\pi}{2}\beta\right)
\nonumber \\ & &
+D_{\bm{k}}\cos\left(-z_{\bm{k}}\, k\eta+\frac{\pi}{2}\beta\right)\biggr]\, .
\end{eqnarray}
This expression can also be re-expressed in term of ``plane wave''
functions (writing $\alpha_{\bm k}\equiv \alpha _{\bm k}^{\rm
  R}+i\alpha_{\bm k}^{\rm I}$)
\begin{eqnarray}
\label{eq:superHcslplanewave}
\lim_{\lambda_{\bm k}/\ell_{\rm H}\rightarrow 0}f_{\bm{k}}(\eta)&=&
\frac{A_{\bm k}}{\sqrt{2\pi}}{\rm e}^{\alpha^{\rm R}_{\bm k}\vert \eta \vert 
-i\alpha^{\rm I}_{\bm k}\eta -i\pi/4}
\nonumber \\ & & 
+\frac{B_{\bm k}}{\sqrt{2\pi}}{\rm e}^{-\alpha^{\rm R}_{\bm k}\vert \eta \vert 
+i\alpha^{\rm I}_{\bm k}\eta +i\pi/4},
\end{eqnarray}
where the coefficients $A_{\bm k}$ and $B_{\bm k}$ can be expressed as
linear combinations of $C_{\bm k}$ and $D_{\bm k}$, namely
\begin{eqnarray}
\label{eq:linkacd:1}
A_{\bm k} &=& C_{\bm k}\, {\rm e}^{-i\pi(\beta+1/2)/2}
+D_{\bm k}\, {\rm e}^{i\pi(\beta+1/2)/2} \\
B_{\bm k} &=& C_{\bm k}\, {\rm e}^{i\pi(\beta+1/2)/2}
+D_{\bm k}\, {\rm e}^{-i\pi(\beta+1/2)/2}.
\label{eq:linkbcd:2}
\end{eqnarray}
The solution~(\ref{eq:superHcslplanewave}) is nothing but the
Wentzel-Kramers-Brillouin (WKB) mode function $\exp( \pm i\int \omega
{\rm d}\tau)/\sqrt{2\omega}$. The reason for this result is that, in
the sub-Hubble regime, the WKB approximation is still valid even in
presence of the CSL term. As is well known, this approximation is
satisfied when the quantity $\vert Q/\omega^2\vert \ll 1$, where $Q$
is given by
\begin{eqnarray}
Q&\equiv&
\frac{3}{4}\frac{1}{\omega^2}\left(
\frac{\dd\omega}{\dd\eta}\right)^2
-\frac{1}{2\omega}\frac{\dd^2\omega}{\dd\eta^2}.
\end{eqnarray}
Since, in the limit under consideration, $\omega^2 $ tends toward a
constant, namely $\omega^2=k^2-2i\gamma $, and since $Q$ is given in
terms of derivatives of $\omega$, it is obvious that the criterion is
satisfied. As already mentioned, the only effect of the CSL theory is
to add the constant term $-2i\gamma $ to $\omega^2$. Although this
modifies the solution for the mode function, clearly, this cannot
change the fact that WKB is valid at the beginning of inflation.

\par

Let us now comment on Eq.~(\ref{eq:superHcslplanewave}). When $\vert
\eta \vert $ goes to infinity, the second branch of the above solution
is going to die away since $\alpha^{\rm R}_{\bm k}>0$. As a
consequence, only the first branch remains and, since $\Omega_{\bm
  k}$ is given in terms of a ratio, \ie $-if'_{\bm k}/(2f_{\bm k})$,
the remaining constant $A_{\bm k}$ disappears from the final
expression. Therefore, $\Omega_{\bm k}$ becomes independent of the
initial conditions and is given by $\Omega _{\bm k}\simeq i\alpha_{\bm
  k}/2$, which implies that $\Rea  \Omega _{\bm k} \simeq -\alpha_{\bm
  k}^{\rm I}/2\simeq -k/2$. Returning to
Eq.~(\ref{eq:SingleGaussianR}), this means that the wavefunction is
not bounded at infinity and is not normalizable. The deep reason is
that, in the CSL context, $z_{\bm k}$ (or $\alpha_{\bm k}$) is complex
and this implies that the WKB solution acquires either a growing or a
decaying exponential component which automatically kills one of the
two branches. And, of course, $z_{\bm k}$ (or $\alpha_{\bm k}$) is
complex because of the CSL term $-2i\gamma $.

\par

Based on the previous discussion, it is clear that the only meaningful
choice of initial conditions is to require that $A_{\bm k}=0$. From
Eqs.~(\ref{eq:linkacd:1}) and~(\ref{eq:linkbcd:2}), we see that this
implies
\begin{equation}
\label{eq:bunchcsl}
C_{\bm k}=-D_{\bm k}\, {\rm e}^{i\pi(\beta +1/2)}.
\end{equation}
This choice exactly coincides with the Bunch-Davies initial
conditions~(\ref{eq:bunchdaviesstandard}). From now on, we
assume Eq.~(\ref{eq:bunchcsl}) but we will come back soon to this
discussion. Then, one can re-derive the behavior of $\Omega_{\bm k}$
in the sub-Hubble regime. One obtains
\begin{equation}
\label{eq:limomegasubhubble}
\lim_{\lambda_{\bm k}/\ell_{\rm H}\rightarrow 0}
\Omega _{\bm k}(\eta)=-\frac{i}{2}\alpha_{\bm k},
\end{equation}
which is fully consistent with Eqs.~(\ref{eq:Omega:limit+})
and~(\ref{eq:Omega:limit-}). In particular, one can check that, now,
$\Rea  \Omega _{\bm k}\rightarrow k/2$ and the wavefunction becomes
normalizable (of course, it tends to the ground state
wavefunction). Therefore, we have proven that, as expected, the
cosmological perturbations behave, in the sub-Hubble regime, exactly
as the CSL harmonic oscillator.

\par

Having studied the behavior of the stochastic wavefunction in the
sub-Hubble regime, we now turn to the super-Hubble case. In the
framework of CSL, and contrary to the sub-Hubble regime studied
before, it is clear that this regime has no counter part in the case
of the harmonic oscillator. It corresponds to the limit $\ell _{\rm
  H}\ll \lambda _{\bm k}$ and, from Eq.~(\ref{eq:argbesselcsl}), we
see that this means $\vert z_{\bm k}k\eta \vert \rightarrow
0$. Let us notice that one could also consider the
  case where $k_{\rm phys}\ll M_{\gamma}$ such that $M_{\gamma}/k_{\rm
    phys}\gg 1$ compensates the ratio $\ell_{\rm H}/\lambda_{\bm k}$
  in Eq.~(\ref{eq:argbesselcsl}) resulting in a large $\vert z_{\bm
    k}k\eta \vert $, even in the super-Hubble regime. Below, we
  briefly comment on this case. Here, we assume that $M_{\gamma}$ is
  such that this does not happen. Then, upon using the asymptotic
  behavior of the Bessel functions for small values of their argument,
  one arrives at
\begin{widetext}
\begin{equation}
\label{eq:omegasuperhubblecsl}
\frac{\Omega_{\bm k}}{k}=-\frac{i(1+\beta)}{2k\eta}
-\frac{i(-k\eta)}{4(\beta +3/2)}-\frac{(-k\eta)}{2(\beta +3/2)}
\frac{\gamma}{k^2}
+i\frac{D_{\bm k}}{C_{\bm k}}
\left(1-2i\frac{\gamma}{k^2}\right)^{-\beta-1/2}
\hskip-1mm 2^{2\beta+1}
\hskip-1mm
\left(\beta+\frac12\right)
\frac{\Gamma(\beta+1/2)}{\Gamma(-\beta -1/2)}
(-k\eta)^{-2\beta -2}
+\cdots .
\end{equation}
This equation should be compared to the corresponding non-CSL
formula~(\ref{eq:omegasuperhubble}). If $\gamma =0$ and if one takes
the Bunch-Davies initial conditions, $D_{\bm k}=-C_{\bm k}{\rm
  e}^{-i\pi(\beta +1/2)}$, then the above equation exactly reduces to
Eq.~(\ref{eq:omegasuperhubble}). Here, although we argued before that
one should use the Bunch-Davies initial
conditions~(\ref{eq:bunchcsl}), we temporarily keep the coefficients
$C_{\bm k}$ and $D_{\bm k}$ arbitrary because, later on, we shall want
to comment on their influence on the shape of the CSL power
spectrum. Let us also notice that the last term of the above
expression is in fact proportional to $z_{\bm k}^{-(2\beta +1)}$. If
we write $z_{\bm k}$ in polar form, $z_{\bm k}\equiv \vert z_{\bm
  k}\vert {\rm e}^{i\theta_{\bm k}}$ (of course, $\theta_{\bm k}$
should not be confused with the squeezing angle) where the modulus and
the phase can be read off directly from Eq.~(\ref{eq:defz}), and
parametrize the initial conditions as $C_{\bm k}=\vert C_{\bm
  k}\vert{\rm e}^{i\theta_c}$ and $D_{\bm k}=\vert D_{\bm k}\vert {\rm
  e}^{i\theta_d-i\pi\beta +i\pi/2}$ (so that the Bunch-Davies limit is
simply $\theta_d-\theta_c=0$), then it is easy to determine the real
and imaginary parts of the function $\Omega_{\bm k}$. One finds
\begin{eqnarray}
\label{eq:reomegacsl}
\Rea  \Omega_{\bm k}(\eta) &=& -\frac{k}{2(\beta +3/2)}\frac{\gamma}{k^2}
(-k\eta)+\frac{\vert D_{\bm k}\vert}{\vert C_{\bm k}\vert}
\vert z_{\bm k}\vert ^{-(2\beta +1)}\cos\left[\pi \beta +\left(2\beta +1\right)
\theta_{\bm k}-\theta_d+\theta_c\right]
\nonumber \\ & & \times 
\frac{k\, \pi\, 2^{2\beta+1}}{\Gamma^2(-\beta -1/2)\cos(\pi \beta)}
(-k\eta)^{-2\beta -2}
+\cdots ,\\
\label{eq:imomegacsl}
\Ima  \Omega _{\bm k}(\eta ) &=& -\frac{k}{2k\eta}(1+\beta)
-\frac{k}{4(\beta +3/2)}(-k\eta)
-\frac{k}{\pi}
\frac{\vert D_{\bm k}\vert}{\vert C_{\bm k}\vert}
\vert z_{\bm k}\vert ^{-(2\beta +1)}2^{2\beta}
\nonumber \\ & & \times
\frac12
\sin\left[\pi \beta +\left(2\beta +1\right)
\theta_{\bm k}-\theta_d+\theta_c\right]\cos(\pi \beta)
\Gamma ^2\left(\beta +\frac32\right)
(-k\eta)^{-2\beta -2}
+\cdots . 
\end{eqnarray}
\end{widetext}
These equations are the CSL counterparts of Eqs.~(\ref{eq:reomegaeta})
and~(\ref{eq:imomegaeta}). Of course, for $\gamma =0$ and the 
Bunch-Davies initial conditions, they exactly reduce to those equations. We
see that the main effect of the CSL theory is to strongly modify $\Rea 
\Omega _{\bm k}$ since its leading term in the above expansion is a
term which cancels if $\gamma =0$. We also see that we still have
$\Rea \Omega_{\bm k}\rightarrow 0$ in the super-Hubble limit. In absence
of the CSL term, we would obtain the same limit but not with the same
power. Compared to $\Rea  \Omega _{\bm k}$, $\Ima  \Omega _{\bm k}$ is
much less modified since the first correction show up only in the
third term of the expansion. As a consequence, we still have
$\Ima \Omega _{\bm k} \rightarrow \infty$ in the super-Hubble regime.

\par

We now use the above results to discuss the collapse of the
wavefunction in more detail. Since we have assumed in
Eq.~(\ref{eq:SchroModes}) that the ``collapse operator'' is
$\hat{v}_{\bm k}$, we expect the non linear and stochastic terms in
the modified Schr\"odinger equation to drive the initial Gaussian
state to an eigenvector of $\hat{v}_{\bm{k}}$, that is to say to the
Dirac function $\delta\left(v_{\bm k}-\bar{v}_{\bm
    k}\right)$. However, in practice, as we learned from the harmonic
oscillator example in Sec.~\ref{sec:HarmOsc}, this is not what
happens. In practice, we find that the wavefunction tends towards a
Gaussian state with a constant spread in position and that the larger
the value of $\gamma $, the smaller the amplitude of this spread, \ie
$\underset{t\to \infty}{\mathrm{lim}}\sigma_x\rightarrow
[\hbar/(4m\gamma )]^{1/4}$ when $\gamma \rightarrow
\infty$. Therefore, strictly speaking, the exact localization is
obtained only in the $\gamma \to \infty$ limit. Of course, if the
spread is very small, then for all practical purposes, the collapse
has been achieved. In fact, this is the essence of the amplification
mechanism discussed in Sec.~\ref{subsec:reviewcsl}. The effective
value $\gamma_{_\mathrm{CM}}$ of $\gamma $ for a macroscopic object
(or for its center of mass) is the fundamental $\gamma $ times the
number of particles in that object which results in a huge effective
$\gamma $ and, therefore, a very efficient localization. As a
consequence, a collapse can occur for macroscopic objects while it
does not happen for microscopic particles even if their behavior is
slightly disturbed.

\par

Let us now see how the previous discussion applies to inflation. The
first difference is that the standard deviation, $1/(2\sqrt{\Rea
  \Omega _{\bm k}})$, does not go to a constant as for the harmonic
oscillator but to infinity since Eq.~(\ref{eq:reomegacsl}) implies
that $\Omega _{\bm k}\propto \eta \rightarrow 0$. We remark that the
divergence is less violent than when $\gamma \neq 0$ since, in that
case, $\Omega _{\bm k}\propto \eta^2 \rightarrow 0$, according to
Eq.~(\ref{eq:reomegaeta}). This is of course due to the influence of
the non linear and stochastic terms. However, this influence is not
sufficient to prevent the divergence of the variance and, therefore,
to ensure an efficient localization. As a matter of fact, we see that,
in the limit $\eta \rightarrow 0$, the main divergence in the
Hamiltonian comes from the term $\propto\omega^2 v_{\bm{k}}^2$ while
the CSL term goes like $\gamma v_{\bm{k}}^2$. Hence, it is because the
term $\omega^2\propto\eta^{-2}$ diverges at the end of inflation that
the Hamiltonian strongly dominates the dynamics of the system,
preventing the CSL terms $\propto \gamma v_{\bm{k}}^2$ to carry out
its job and to localize $v_{\bm{k}}$ (however, see Appendix
\ref{sec:appendix}).  This is certainly a problem for the inflationary
CSL theory. This issue can also be related to the fact that it is
unclear how an amplification mechanism could be implemented in quantum
field theory. As a consequence, the collapse mechanism is controlled
by the parameter $\gamma $ and no effective $\gamma $ can be derived
which would ensure a better localization.

\par

Finally, let us mention that one could wonder whether the localization
can be achieved during the radiation dominated era that takes place
after inflation. In this case, the scale factor behaves as $a(\eta
)\propto \eta $ and, therefore, $\epsilon_1=2$ and
$(a\sqrt{\epsilon_1})''/(a\sqrt{\epsilon_1})=0$. As consequence, the
mode equation for $f_{\bm k}$ is exactly that of an harmonic
oscillator. This means that the variance now goes to a constant, see
Sec.~\ref{sec:HarmOsc}, which seems to cure the problem discussed
above. However, one can show that the corresponding value remains
large for modes of astrophysical interest today. Therefore, this
remains an unsatisfactory solution.

\subsection{The CSL Power Spectrum}
\label{subsec:cslps}

We now turn to one of the main goal of the present paper, namely the
determination of the power spectrum predicted by the CSL theory. It
was shown in Eqs.~(\ref{eq:calculmeanv2}) and~(\ref{eq:linkzetav})
that the power spectrum of the conserved quantity $\zeta_{\bm k}$ can
be expressed as
\begin{equation}
{\cal P}_\zeta(k)=\frac{k^3}{16\pi^2\Mp^2}\frac{1}
{a^2\epsilon_1\Rea  \Omega _{\bm k}}.
\end{equation}
Since we have determined the quantity $\Rea  \Omega_{\bm k}$ in
Eq.~(\ref{eq:reomegacsl}), the calculation of ${\cal P}_\zeta$ becomes
straightforward. One obtains
\begin{eqnarray}
\label{eq:powercsl}
{\cal P}_{\zeta}(k)&=&
g_\gamma (k,\beta)\left[1-\frac{\gamma }{k^2}g_\gamma (k,\beta)f(\beta)
\frac{(-k\eta)^{2\beta +3}}{\beta +3/2}\right]^{-1}
\nonumber \\ & & \times 
{\cal P}_{\zeta}(k)\bigl \vert _{\rm stand},
\end{eqnarray}
where ${\cal P}_{\zeta}\bigl \vert _{\rm stand}$ is the standard power
spectrum given by Eq.~(\ref{eq:powerstandard}) and the function $f(\beta)$
has been defined in Eq.~(\ref{eq:deffuncf}). The function $g_{\gamma
}(k,\beta)$
\begin{equation}
\label{eq:g:def}
g_\gamma (k,\beta)\equiv \frac{\vert C_{\bm k}\vert}{\vert D_{\bm k}\vert}
\vert z_{\bm k}\vert^{2\beta+1}\frac{\cos(\pi \beta)}
{\cos\left[\pi \beta +\left(2\beta +1\right)
\theta_{\bm k}-\theta_d+\theta_c\right]}
\end{equation}
is seen to depend on the choice of the initial conditions. It has
the property that, for $\gamma =0$ and the Bunch-Davies initial
conditions, $g_{\gamma=0}(k,\beta )=1$. In this case, and as expected,
one can check that the modified power spectrum~(\ref{eq:powercsl})
reduces to the standard inflationary power spectrum. We also notice
that the power spectrum~(\ref{eq:powercsl}) is still a time-dependent
quantity, contrary to the conventional case where the time dependence
cancels out. For this reason, it is convenient to evaluate it at the
end of inflation. In that case, the quantity $-k\eta$ can be
rewritten as
\begin{equation}
-k\eta =-\frac{k}{k_0}(1+\beta){\rm e}^{\Delta N_*/(1+\beta)},
\end{equation}
where $k_0$ is the comoving wavenumber of the Fourier mode, the
wavelength of which equals the Hubble radius today, \ie
$k_0=a_0H_0$. The quantity $\Delta N_*$ denotes the number of
e-folds spent by a mode of cosmological relevance today outside the
Hubble radius during inflation; typically, one has $\Delta N_*\simeq
50-60$. As a consequence, the power spectrum~(\ref{eq:powercsl}) can
be re-expressed as
\begin{widetext}
\begin{eqnarray}
\label{eq:powercslfinal}
{\cal P}_{\zeta}(k)&=&
g_\gamma (k,\beta)\left[1-\frac{\gamma }{k_0^2}g_\gamma (k,\beta)f(\beta)
\frac{\vert 1+\beta\vert^{2\beta +3}}{(\beta +3/2)}
{\rm e}^{(2\beta+3)\Delta N_*/(1+\beta)}\left(\frac{k}{k_0}\right)^{2\beta +1}
\right]^{-1}
{\cal P}_{\zeta}(k)\bigl \vert _{\rm stand}.
\end{eqnarray}
\end{widetext}
Let us notice that, in Eq.~(\ref{eq:g:def})}, the quantities $\vert
z_{\bm k}\vert $  of Eq.~(\ref{eq:defz}) 
and $\theta_{\bm k}$ must now be written as
\begin{eqnarray}
\label{eq:defz:1}
\vert z_{\bm{k}}\vert &=&  
\left[1+4\frac{\gamma^2}{k_0^4}\left(\frac{k_0}{k}\right)^4\right]^{1/4}
\\
\theta_{\bm k} &=& 
-\frac{1}{2}\mathrm{arctan}\left[2\frac{\gamma}{k_0^2}
\left(\frac{k_0}{k}\right)^2\right], 
\label{eq:defz:2}
\end{eqnarray}
such that the amplitude of the CSL correction is controlled by the
dimensionless ratio $\gamma /k_0^2$. The
formula~(\ref{eq:powercslfinal}) is one of the main results of this
article and the corresponding power spectra for different values of
the ratio $\gamma/k_0^2$ are represented in
Fig.~\ref{fig:CSLspectrum}.

\begin{figure*}
\begin{center}
\includegraphics[width=0.85\textwidth,clip=true]{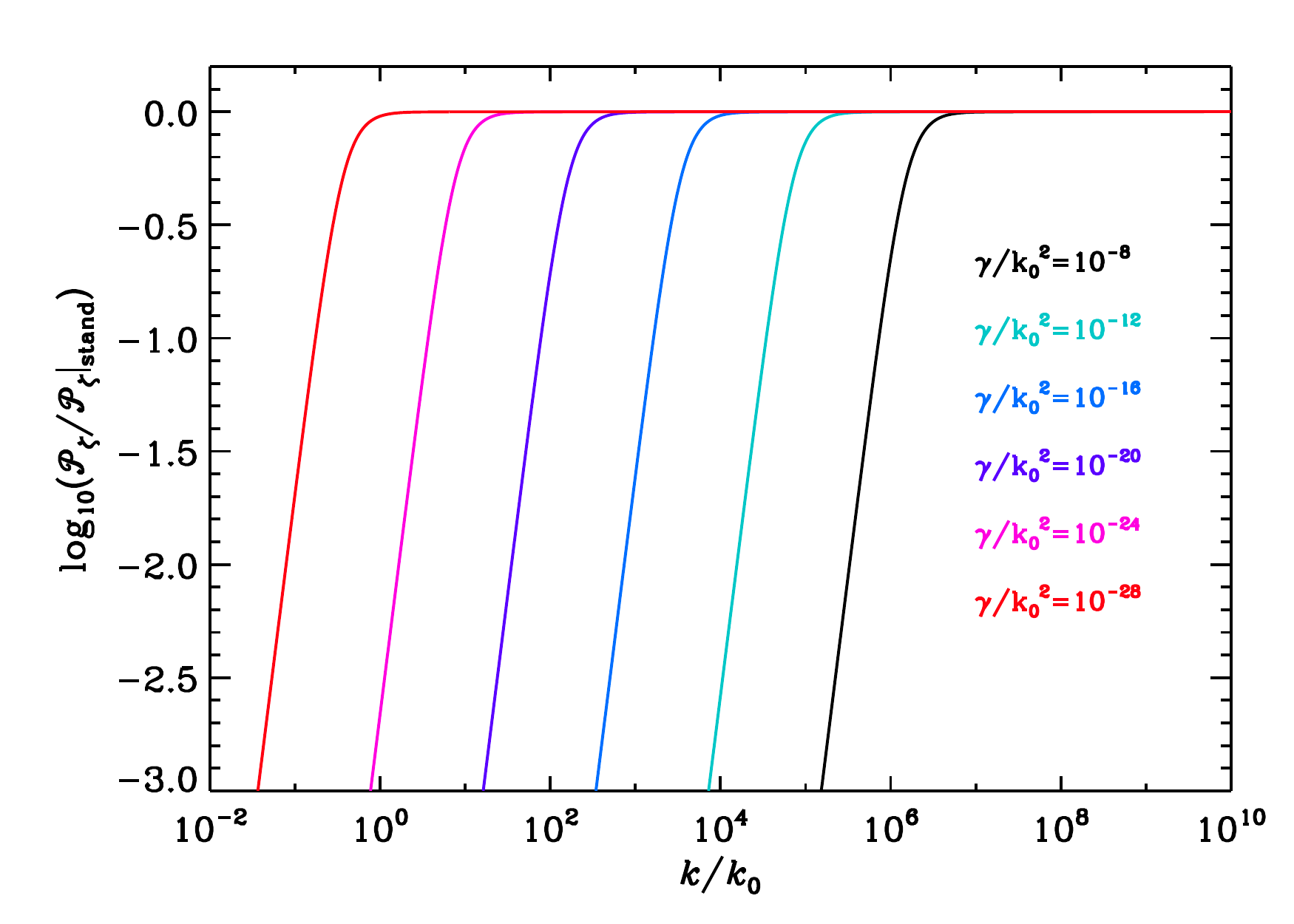}
\caption{Ratio of the power spectrum given by
  Eq.~(\ref{eq:powercslfinal}) to the standard power spectrum given by
  Eq.~(\ref{eq:powerstandard}) for different values of the parameter
  $\gamma/k_0^2$ (and for $\beta=-2.01$, a value leading to a standard
  power spectrum close to scale invariance). The number of e-folds
  between Hubble radius crossing and the end of inflation (for the
  modes of cosmological interest today) has been taken to $\Delta
  N_*=60$ and the initial conditions have been chosen to be the
  adiabatic vacuum.}
\label{fig:CSLspectrum}
\end{center}
\end{figure*}

Let us now discuss in more detail the CSL power
spectrum~(\ref{eq:powercslfinal}). Firstly, we notice that, in the
short-wavelength regime $k/k_0\rightarrow \infty$, the power spectrum
reduces to ${\cal P}_{\zeta}(k)\simeq g_\gamma(k,\beta){\cal
  P}_\zeta\vert _{\rm stand}$. Moreover, in this limit, we see that
$\vert z_{\bm k}\vert \rightarrow 1$ and $\theta_{\bm k}\rightarrow
0$. As a consequence, an almost scale invariant (namely, $n_{_{\rm
    S}}=2\beta +4$ with $\beta \lesssim -2$) power spectrum is
recovered if one assumes the Bunch-Davies initial conditions, $\vert
C_{\bm k}\vert =\vert D_{\bm k}\vert $ and $\theta _d-\theta _c=0$
since, in that case, $g_\gamma(k,\beta) =1$. This almost scale
invariant branch of the power spectrum is clearly seen in
Fig.~\ref{fig:CSLspectrum}. Secondly, there is clearly another regime
which corresponds to the case where the second term in the square
brackets in Eq.~(\ref{eq:powercslfinal}) starts playing a role. If we
neglect factors of order one, this happens at $k=k_\gamma$, where
$k_\gamma$ solves
\begin{equation}
\label{eq:conditionbreak}
\frac{\gamma }{k_0^2}g_\gamma(k_\gamma,\beta)
{\rm e}^{(2\beta+3)\Delta N_*/(1+\beta)}\left(\frac{k_\gamma}{k_0}\right)^{2\beta +1}
\simeq 1 .
\end{equation}
The value of $g_\gamma $ is mainly controlled by the value of $\vert
z_{\bm k}\vert$ which is always close to unity provided that $k\ll
k_z$ with
\begin{equation}
  \frac{k_z}{k_0}\equiv \sqrt{2}
  \left(\frac{\gamma}{k_0^2}\right)^{1/2}.
\end{equation}
Then, let us assume that $g_\gamma \simeq 1$ when the
condition~(\ref{eq:conditionbreak}) is met. In this case, the scale
$k_\gamma$ can be expressed as
\begin{equation}
\frac{k_\gamma}{k_0}\sim \left(\frac{\gamma}{k_0^2}\right)^{-1/(1+2\beta)}
\exp\left[-\frac{2\beta+3}{\left(\beta+1\right)
\left(2\beta+1\right)}\Delta N_*\right].
\end{equation}
Choosing the fiducial value $\beta\simeq -2$ leads to
$k_\gamma/k_0\sim (\gamma/k_0^2)^{1/3}\exp\left(\Delta
  N_*/3\right)$. One can check that, indeed, $k_\gamma\gg k_z$ and,
therefore, assuming $g_\gamma \simeq 1$ was, in retrospect, valid.  As
a consequence, in the range $k\ll k_\gamma$, the spectrum
approximately behaves as $\propto k^{2\beta +4}/k^{2\beta+1}=k^3$,
that is to say with a spectral index of $n_{_{\rm S}}\simeq 4$. This
second branch is also clearly visible in
Fig.~\ref{fig:CSLspectrum}. In addition, the dependence in $g_\gamma$
is canceled out which means that this prediction is actually
independent of the choice of the initial conditions, a remarkable
property indeed (this also means that, even if $k\ll k_z$, the
spectral index remains the same). Moreover, we see that this spectral
index is also independent of $\beta$ which is also remarkable. In this
sense, the CSL branch of the power spectrum can be said to be
``universal'' (unfortunately not scale invariant!).

\par

We are now in a position where we can discuss the cosmological
constraints on the parameter $\gamma $. From the high accuracy
measurements of the CMB
anisotropies~\cite{Larson:2010gs,Komatsu:2010fb,Martin:2006rs}, we
know that the power spectrum is almost scale invariant, $n_{_{\rm
    S}}\simeq 1$, and that a spectral index $n_{_{\rm S}}=4$ is
completely excluded. This means that the CSL branch must correspond to
scales much larger than the present Hubble radius, in other words
$k_\gamma/k_0\ll 1$. This condition means that, for $\beta \simeq -2$,
one has
\begin{equation}
\label{eq:cslconstraint}
\frac{\gamma }{k_0^2}\ll {\rm e}^{-\Delta N_*}\simeq 10^{-28}.
\end{equation}
To our knowledge, this is the first time that a constraint on the
parameter $\gamma$ is obtained from cosmological considerations (see,
however, Ref.~\cite{Lochan:2012di}). We see that the constraint is
expressed as a limit not on $\gamma $ itself but on the combination
$\gamma /k_0^2$ where we remind that $k_0$ is the comoving wavenumber
of the Hubble radius today. Looking at Eqs.~(\ref{eq:SchroModes})
and~(\ref{eq:u}), this was expected since the CSL modification amounts
to a redefinition of the comoving wavenumber $k^2\rightarrow
k^2-2i\gamma $. This means that, in order to characterize the
amplitude of the modification, one has to compare the comoving
wavenumber to $\gamma $, hence the ratio $\gamma/k_0^2$. The
appearance of the comoving wavenumber in the observational constraint
reflects the fact that the theory contains a built-in ``time-dependent
physical preferred scale'' $\ell _{\gamma}(\eta)$. In terms of
physical scales, the constraint~(\ref{eq:cslconstraint}) can be
rewritten as
\begin{equation}
  \left.\frac{\ell_{_{\rm H}}}{\ell _{\gamma }}\right\vert_{\mathrm{today}}\ll 10^{-13}.
\end{equation}
Clearly, the constraint is very strong and means that the scale
$\ell_{\gamma }$ is very large in comparison to the Hubble radius
today. This is another illustration of the fact that squeezed states
are fragile and easily perturbed. For the CSL theory itself, this
probably means that, in order to be compatible with cosmological
inflation, an important fine-tuning is required. Of course, this
conclusion should be toned down given the uncertainty that exists on a
CSL formulation of quantum field theory as discussed in
Sec.~\ref{subsec:modifiedforv}. One might argue for instance that the
above result could be due to the fact that our modified Schr\"odinger
equation is not necessarily the appropriate one in the context of
quantum field theory. It would also be interesting to compare the
cosmological constraint with the other constraints on $\gamma $
derived in the literature. But, as explained before, because we
assumed $\hat{v}_{\bm k}$ to be the preferred basis for the collapse,
our parameter $\gamma $ is actually different from the parameter
$\gamma $ considered elsewhere, in particular it has a different
dimension. This complicates tremendously any comparison with other
systems.

\par

Finally, before closing this section, let us discuss the following
question. In this article, we have defined the power spectrum in the
CSL theory by means of the formula $\mathbb
E\left(\mean{\hat{v}_{\bm{k}}^2}\right)-\mathbb E\left(\langle
  \hat{v}_{\bm k}\rangle^2\right)$. However, there is an issue
regarding this definition. Indeed, it is clear that it does not go to
zero when the parameter $\gamma $ vanishes. Actually, it tends towards
the standard result when the Schr\"odinger equation is recovered.
However, it was argued in Ref.~\cite{Sudarsky} that the power spectrum
should go to zero in the limit where $\gamma \rightarrow 0$ and,
therefore, cannot be given by the definition used above. The reason
advocated by Ref.~\cite{Sudarsky} is that, without a collapse, the
theory remains homogeneous and isotropic and, as consequence, there is
simply no perturbations at all. This has led Ref.~\cite{Sudarsky} to
define the CSL power spectrum by $\mathbb E
\left(\mean{\hat{v}_{\bm{k}}}^2\right)-\mathbb E^2\left(\langle
  \hat{v}_{\bm k}\rangle\right)$, a quantity which indeed vanishes
when $\gamma \rightarrow 0$ and differs from the previous one. In this
last paragraph, we explore the difference between these two
alternative definitions.

\par

At any time, the wavefunction can always be expanded as
\begin{equation}
\Psi\left(\eta, v_{\bm k}\right)=\int\Psi\left(\eta,\bar{v}_{\bm k}\right)
\delta\left(v_{\bm k}-\bar{v}_{\bm k}\right)\dd \bar{v}_{\bm k}\, ,
\end{equation} 
where the superscripts ``$\mathrm{R,I}$'' have been ignored for
convenience. If a dynamical collapse of the wavefunction takes place
then $\Psi $ is projected (collapsed) on an eigenstate of the operator
$\hat{v}_{\bm k}$, namely
\begin{equation}
\Psi \rightarrow \Psi_{\mathrm{col}}\equiv 
\delta\left(v_{\bm k}-\bar{v}_{\bm k}\right)\, ,
\end{equation} 
where $\bar{v}_{\bm k}$ depends on the specific realization under
consideration, then one obviously has
\begin{eqnarray}
\left\langle\Psi_{\mathrm{col}}\left\vert \hat{v}^2_{\bm k}\right\vert
\Psi_{\mathrm{col}}\right\rangle&=&
\left\langle\Psi_{\mathrm{col}}\left\vert \hat{v}_{\bm k}\right\vert
\psi_{\mathrm{col}}\right\rangle^2
\nonumber\\&=&
\bar{v}^2_{\bm k}\, .\nonumber\\
\end{eqnarray} 
Therefore, for each realization, one has
$\mean{\hat{v}_{\bm{k}}^2}=\mean{\hat{v}_{\bm{k}}}^2$, once the
wavefunction has collapsed. Since this is true for all realizations,
it remains the case after taking the stochastic average. Therefore,
after the collapse, one can write
\begin{equation}
\mathbb{E}\left(\mean{\hat{v}_{\bm{k}}^2}\right)=\mathbb{E}
\left(\mean{\hat{v}_{\bm{k}}}^2\right)\, ,
\end{equation}
and this remains true for any Hermitian operator. Note that this
argument strongly depends on the fact that the wavefunction has
actually collapsed to an eigenstate of the operator $\hat{v}_{\bm
  k}$. For instance, in the case of an harmonic oscillator studied in
Sec.~\ref{sec:HarmOsc}, it was shown that the asymptotic state is not
exactly a Dirac wavefunction, but a Gaussian state the spread of which
does not vanish for finite values of $\gamma$.  In that situation, the
two above expressions are not identical. 

\par

On the other hand, the second terms in both definitions of the power
spectrum differ
\begin{equation}
\mathbb{E}\left(\mean{\hat{v}_{\bm{k}}}^2\right)\neq 
\mathbb{E}^2\left(\mean{\hat{v}_{\bm{k}}}\right)\, ,
\end{equation}
so the two spectra do not coincide even after the collapse. The
difference ultimately boils down to the fact that it is built out of a
standard deviation which is not an Hermitian operator. This is a
generic question for the predictions of any theory mixing different
kinds of averages (in the case at hand, quantum and stochastic)
whenever non-linear combinations of Hermitian operators are involved.

\section{The Collapse of Cosmological Perturbations}
\label{sec:collapsepert}

In this section, we investigate the collapse mechanism and its
dynamics in more detail. In particular, we calculate
  the collapse time and compare it with the cosmological
  characteristic times. For this purpose, we now consider the
following double Gaussian quantum state~\cite{Bassi:2005fp}
\begin{widetext}
\begin{eqnarray}
\label{eq:DoubleGaussian}
\Psi_{\bm{k}}\left(\eta,v_{\bm{k}}\right)&=&
\vert N_{\bm{k}}^{(1)}\left(\eta\right)\vert
\exp\Bigl\lbrace
-\Rea  \Omega_{\bm{k}}^{(1)}\left(\eta\right)
\left[v_{\bm{k}}-\bar{v}_{\bm{k}}^{(1)}\left(\eta\right)\right]^2
+i\sigma_{\bm{k}}^{(1)}\left(\eta\right)
+i\chi_{\bm k}^{(1)}(\eta)v_{\bm{k}}
-i\Ima  \Omega_{\bm k}^{(1)}(\eta)\left(v_{\bm{k}}\right)^2\Bigr\rbrace
\nonumber \\ &+& 
\vert N_{\bm{k}}^{(2)}\left(\eta\right)\vert
\exp\Bigl\lbrace
-\Rea  \Omega_{\bm{k}}^{(2)}\left(\eta\right)
\left[v_{\bm{k}}-\bar{v}_{\bm{k}}^{(2)}\left(\eta\right)\right]^2
+i\sigma_{\bm{k}}^{(2)}\left(\eta\right)
+i\chi_{\bm k}^{(2)}(\eta)v_{\bm{k}}
-i\Ima  \Omega_{\bm k}^{(2)}(\eta)\left(v_{\bm{k}}\right)^2\Bigr\rbrace,
\end{eqnarray}
where, as before, $\vert N_{\bm{k}}^{(1,2)}\vert$, $\Rea \Omega_{\bm
  k}^{(1,2)}$, $\bar{v}_{\bm{k}}^{(1,2)}$, $\sigma_{\bm{k}}^{(1,2)}$,
$\chi_{\bm{k}}^{(1,2)}$ and $\Ima \Omega_{\bm k}^{(1,2)}$ are real,
possibly stochastic, numbers. The superscripts ``$\mathrm{R,I}$'' have
not been written for convenience but it should be remembered that they
are of course present. Inserting the above state into the modified
Schr\"odinger equation leads to the following set of formulas
\begin{eqnarray}
\label{eq:evolDoubleGaussian:N}
\frac{\left\vert N_{\bm{k}}^{(1,2)}\right\vert ^{\prime}}
{\left\vert N_{\bm{k}}^{(1,2)}\right \vert}&=&
\Ima  \Omega_{\bm{k}}^{(1,2)}+\frac{\gamma}{4\Rea  \Omega_{\bm{k}}^{(1,2)}}
-\sqrt{\gamma}\left[\left\langle \hat{v}_{\bm{k}}\right \rangle 
-\bar{v}_{\bm{k}}^{(1,2)}\right]
\frac{\dd W_\eta}{\dd \eta}
-\frac{\gamma}{2}\left[\left \langle \hat{v}_{\bm{k}}
\right \rangle -\bar{v}_{\bm{k}}^{(1,2)}\right]^2
\, ,\\
\label{eq:evolDoubleGaussian:Omega}
\left[\Rea  \Omega_{\bm{k}}^{(1,2)}\right]^\prime&=&\gamma
+4\left[\Rea  \Omega_{\bm{k}}^{(1,2)}\right]
\left[\Ima  \Omega_{\bm k}^{(1,2)}\right]\, ,\\
\label{eq:evolDoubleGaussian:h}
\left[\Ima  \Omega_{\bm{k}}^{(1,2)}\right]^\prime 
&=& -2\left[\Rea  \Omega_{\bm{k}}^{(1,2)}\right]^2+2
\left[\Ima  \Omega_{\bm{k}}^{(1,2)}\right]^2
+\frac12\omega^2\left(\eta,\bm{k}\right)\, ,\\
\label{eq:evolDoubleGaussian:barnu}
\left[\bar{v}_{\bm{k}}^{(1,2)}\right]^\prime&=&
\chi_{\bm{k}}^{(1,2)}
+\frac{\sqrt{\gamma}}{2\Rea  \Omega_{\bm{k}}^{(1,2)}}\frac{\dd W_\eta}{\dd \eta}
-2\left[\Ima  \Omega_{\bm{k}}^{(1,2)}\right]\bar{v}_{\bm{k}}^{(1,2)}
+\frac{\gamma}{\Rea  \Omega_{\bm{k}}^{(1,2)}}
\left[\left\langle \hat{v}_{\bm{k}}\right \rangle
-\bar{v}_{\bm{k}}^{(1,2)}\right]
\, ,\\
\label{eq:evolDoubleGaussian:f}
\left[\sigma_{\bm{k}}^{(1,2)}\right]^\prime&=&
-\Rea  \Omega_{\bm{k}}^{(1,2)}+2\left[\Rea  \Omega_{\bm{k}}^{(1,2)}\right]^2
\left[\bar{v}_{\bm{k}}^{(1,2)}\right]^2
-\frac12\left[\chi_{\bm{k}}^{(1,2)}\right]^2\, ,\\
\label{eq:evolDoubleGaussian:g}
\left[\chi_{\bm{k}}^{(1,2)}\right]^\prime&=&
-4\left[\Rea  \Omega_{\bm{k}}^{(1,2)}\right]^2\bar{v}_{\bm{k}}^{(1,2)}
+2\chi_{\bm{k}}^{(1,2)}\left[\Ima  \Omega_{\bm{k}}^{(1,2)}\right]\, .
\end{eqnarray}
\end{widetext}
These equations should be compared to Eqs.~(\ref{eq:evol:N}),
(\ref{eq:evol:Omega}), (\ref{eq:evol:h}), (\ref{eq:evol:barnu}),
(\ref{eq:evol:f}) and~(\ref{eq:evol:g}). They are obviously very
similar except the two last terms of
Eq.~(\ref{eq:evolDoubleGaussian:N}) and the last term of
Eq.~(\ref{eq:evolDoubleGaussian:barnu}) which are new. In the case of
a single Gaussian, one has $\langle \hat{v}_{\bm k}\rangle
=\bar{v}_{\bm k}$ and these terms disappear. In the present case, the
expression of $\langle \hat{v}_{\bm k}\rangle$ is a very complicated
function of all the parameters describing the wavefunction. Let us
also notice that, since the evolution of $\sigma_{\bm{k}}^{(1,2)}$ and
$\chi_{\bm{k}}^{(1,2)}$ depends on $\bar{v}_{\bm{k}}^{(1,2)}$, these
quantities also feel the coupling between the two Gaussian
components. However, one can see that the equations of motion for $\Rea 
\Omega_{\bm{k}}^{(1,2)}$ and $\Ima  \Omega_{\bm{k}}^{(1,2)}$ decouple
from the other equations of motion and form an independent and closed
subsystem. This means that the evolution of these two functions is
identical to that of their counterpart in the simple Gaussian case
and, moreover, that, if the initial conditions are chosen to be the
same, $\Omega_{\bm{k}}^{(1)}=\Omega_{\bm{k}}^{(2)}$ at any subsequent
time. From now on, for this reason, the superscripts ``$(1)$'' and/or
``$(2)$'' on these quantities will be dropped.

\par

It should be clear that the above system of differential equations is
rather complicated to study. However, as we shall see, the most
relevant properties of the evolution of the double Gaussian quantum
state can be analyzed in a rigorous way. In particular, it is
interesting to introduce the function $\Gamma_{\bm{k}}(\eta) \equiv
\ln\left[\left \vert N_{\bm{k}}^{(2)}\right \vert/\left \vert
    N_{\bm{k}}^{(1)}\right \vert \right]$, see
Ref.~\cite{Bassi:2005fp}. This quantity characterizes the relative
importance of one Gaussian component to the other and, therefore,
provides a criterion to decide whether the collapse has taken
place. The superposition of the two Gaussian quantum states reduces to
one of them when $\mid\Gamma_{\bm{k}}\mid$ goes to infinity. In
practice, the collapse will be said to have occurred when
$\mid\Gamma_{\bm{k}}\mid>b$ with, say, $b\sim
10$~\cite{Bassi:2005fp}. Then, by subtracting the two
equations~(\ref{eq:evolDoubleGaussian:N}), one arrives at the
following evolution equation for $\Gamma_{\bm{k}}$
\begin{eqnarray}
\label{eq:evol:Gamma}
\frac{\dd\Gamma_{\bm{k}}}{\dd\eta}&=&
\sqrt{\gamma}\left[\bar{v}_{\bm{k}}^{(2)}-\bar{v}_{\bm{k}}^{(1)}\right]
\frac{\dd W}{\dd \eta}
\nonumber\\
&-&\gamma\left[\bar{v}_{\bm{k}}^{(2)}-\bar{v}_{\bm{k}}^{(1)}\right]
\left[\bar{v}_{\bm{k}}^{(1)}+\bar{v}_{\bm{k}}^{(2)}-2\left 
\langle \hat{v}_{\bm{k}}\right \rangle \right]\, .
\end{eqnarray}
This equation remains complicated because of the presence of the term
$\left \langle \hat{v}_{\bm{k}}\right \rangle $. However, the
calculation can be simplified if one assumes that the two Gaussian
components of the wave function do not overlap, \ie have separate
supports. Technically, this means that $\Rea  \Omega_{\bm k}\left[
  \bar{v}_{\bm{k}}^{(2)}-\bar{v}_{\bm{k}}^{(1)}\right]^2 \gg 1$,
leading to the following simple formula
\begin{equation}
\left \langle \hat{v}_{\bm{k}}\right \rangle
\simeq \frac{\left\vert N_{\bm{k}}^{(1)}\right \vert ^2
\bar{v}_{\bm{k}}^{(1)}
+\left \vert N_{\bm{k}}^{(2)}\right \vert ^2\bar{v}_{\bm{k}}^{(2)}}
{\left \vert N_{\bm{k}}^{(1)}\right \vert ^2
+\left \vert N_{\bm{k}}^{(2)}\right \vert ^2}\, .
\end{equation}
Inserting this formula into Eq.~(\ref{eq:evol:Gamma}) and defining
$X_{\bm k}$ by $X_{\bm k}\equiv \bar{v}^{(2)}_{\bm
  k}-\bar{v}^{(1)}_{\bm k}$, one obtains the following expression
\begin{eqnarray}
\label{eq:eqgamma}
\frac{\dd\Gamma_{\bm{k}}}{\dd\eta}&=&
\sqrt{\gamma}X_{\bm{k}}\frac{\dd W_\eta}{\dd \eta}
+\gamma X_{\bm{k}}^2\mathrm{tanh}\left(\Gamma_{\bm{k}}\right)\, .
\end{eqnarray}
This stochastic differential equation can be further
simplified. Indeed, using the new timelike variable~\cite{Bassi:2005fp}
\begin{equation}
\label{eq:timechange}
s_{ \bm{k}}\equiv\gamma\int_{\eta_{\mathrm{ini}}}^{\eta}
X_{\bm{k}}^2\left(u\right)\dd u\, ,
\end{equation}
Eq.~(\ref{eq:eqgamma}) can be rewritten as
\begin{equation}
\label{eq:evol:Gamma:2}
\frac{\dd\Gamma_{\bm{k}}}{\dd s_{\bm k}}=\frac{\dd {W_s}}{\dd s_{\bm k}}
+\tanh\left(\Gamma_{\bm{k}}\right),
\end{equation}
where
\begin{equation} {W_s}=\sqrt{\gamma}\int_{0}^{s_{\bm k}}X_{\bm{k}}\,
  \dd W_\eta\,
\end{equation}
is another Wiener process with respect to the time variable $s_{\bm
  k}$.

\subsection{Collapse Time: Definition}
\label{subsec:timeofcollapse}

Let us now study the stochastic differential equation driving the
evolution of $\Gamma_{\bm{k}}$ in more detail. In particular, we would
like to know how much time it takes for the wavefunction to collapse
or, in technical terms, we would like to determine the value of
$s_{\bm k}$ such that $\left\vert\Gamma_{\bm{k}}\right\vert>b$. The
quantity $\Gamma _{\bm k}$ being stochastic, two complications
arise. Firstly, once it has reached a value larger than $b$, there is
no guarantee that it will stay in this region. The random behavior of
$\Gamma _{\bm k}$ could temporally brings it back to the region $\vert
\Gamma _{\bm k}\vert \leq b$. However, since the average trend is
clearly to have a collapse, this would happen for a limited amount of
time only before $\Gamma _{\bm k}$ returns in the regime where $\vert
\Gamma _{\bm k}\vert \geq b$. For this reason, we will consider that
the wavefunction has collapsed when $\Gamma _{\bm k}$ has crossed the
value $\pm b$ for the first time. Technically, this means that we are
led to define the ``collapse time'', $S_{\bm k}$, as $S_{\bm
  k}\equiv\mathrm{inf}\left(s_{\bm k}\right)$ such that
$\left\vert\Gamma_{\bm{k}}\left(s_{\bm k}\right)\right\vert>b$, see
also Ref.~\cite{Bassi:2005fp}. A second issue is that, clearly, the
value of $S_{\bm k}$ will differ from one realization to the other or,
in other words, that $S_{\bm k}$ is still a random
variable. Therefore, we will rather define the collapse time as the
ensemble average value of $S_{\bm k}$ but we will also be interested
in calculating its higher order momenta.

\par

We now seek an explicit expression for the quantity $S_{\bm k}$. It
can be obtained in the following manner. Let us consider a function
$c(\Gamma _{\bm k})$ that we do not characterize in more detail for
the moment (but see below). It can always be Taylor expanded in
$\dd\Gamma_{\bm{k}}$. At second order, the result reads
\begin{eqnarray}
c\left(\Gamma_{\bm{k}}+\dd\Gamma_{\bm{k}}\right)&=&
c\left(\Gamma_{\bm{k}}\right)
+c^{\prime}\left(\Gamma_{\bm{k}}\right)\dd\Gamma_{\bm{k}}
\nonumber\\& &
+\frac{1}{2}c^{\prime\prime}\left(\Gamma_{\bm{k}}\right)
\dd\Gamma_{\bm{k}}^2+\mathcal{O}\left(\dd \Gamma_{\bm{k}}^3\right)\, ,
\end{eqnarray}
where $\dd \Gamma_{\bm{k}}$ is given by
Eq.~(\ref{eq:evol:Gamma:2}). At first order in $\dd s_{\bm k}$, this
leads to
\begin{eqnarray}
  \dd c\left[\Gamma_{\bm{k}}\left(s_{\bm k}\right)\right]&=&
  c^{\prime}\left[\Gamma_{\bm{k}}\left(s_{\bm k}\right)\right]\dd W_s
\nonumber \\ &+&
  c^{\prime}\left[\Gamma_{\bm{k}}\left(s_{\bm k}\right)\right]\mathrm{tanh}
  \left[\Gamma_{\bm{k}}\left(s_{\bm k}\right)\right]\dd s_{\bm k}
\nonumber \\ &+&
  \frac{1}{2}c^{\prime\prime}
  \left[\Gamma_{\bm{k}}\left(s_{\bm k}\right)\right]\dd s_{\bm k}\, .
\end{eqnarray}
Then, integrating the above expression between $s_{\bm k}=0$ where
$\Gamma_{\bm{k}}\left(s_{\bm k}=0\right)=b_0$ and $s_{\bm k}=S_{\bm
  k}$ where $\Gamma_{\bm{k}}\left(s_{\bm k}=S_{\bm k}\right)=\pm b$,
one gets the following (It\^o) formula
\begin{align}
\label{eq:itoint}
&c\left(\pm b\right)-c\left(b_0\right)=
\int_0^{S_{\bm k}} c^{\prime}\left[\Gamma_{\bm{k}}\left(s_{\bm k}\right)\right]\dd W_s
\nonumber\\ &
+\int_0^{S_{\bm k}} \left\lbrace c^{\prime}\left[\Gamma_{\bm{k}}
\left(s_{\bm k}\right)\right]\mathrm{tanh}
\left[\Gamma_{\bm{k}}\left(s_{\bm k}\right)\right]
+\frac{1}{2}c^{\prime\prime}\left[\Gamma_{\bm{k}}
\left(s_{\bm k}\right)\right]\right\rbrace\dd s_{\bm k}\, .
\nonumber\\
\end{align}
At this stage, we now specify the function $c$. We require it to be
the solution of the differential ordinary equation
\begin{equation}
\label{eq:eqc}
\frac{1}{2}c^{\prime\prime}\left(x\right)+\mathrm{tanh}\left(x\right) 
c^{\prime}\left(x\right)=-1\, ,
\end{equation}
with boundary conditions $c\left(-b\right)=c\left(+b\right)=0$. It is
easy to show that $c(x)=b\tanh (b)-x\tanh (x)$. This means that the
first term on the left hand side of Eq.~(\ref{eq:itoint}) vanishes and
that the integrand of the second term on the right hand side is just
$-1$. Therefore, Eq.~(\ref{eq:itoint}) can be rewritten as
\begin{equation}
\label{eq:S:stocha}
S_{\bm k}=c\left(b_0\right)+\int_0^{S_{\bm k}} c^{\prime}\left[\Gamma_{\bm{k}}
\left(s_{\bm k}\right)\right]\dd W_s\, ,
\end{equation}
and this gives an (implicit) expression for the quantity $S_{\bm
  k}$. Finally, by averaging over all realizations, one
obtains~\cite{Bassi:2005fp}
\begin{equation}
  \mathbb{E}\left(S_{\bm k}\right)=c\left(b_0\right)=b\tanh(b)-b_0\tanh(b_0)\, .
\end{equation}
The fact that the stochastic average of the integral in
Eq.~(\ref{eq:S:stocha}) vanishes comes from the fact that
$c^{\prime}\left[\Gamma_{\bm{k}} \left(s_{\bm k}\right)\right]$
depends only on stochastic events occurring at $s^\prime_{\bm
  k}<s_{\bm k}$. As a consequence, it can be expressed as an
integration over $\dd s^\prime_{\bm k}$ and $\dd W_{s^\prime}$ where
$s^\prime_{\bm k}<s_{\bm k}$.  Since $\mathbb{E}\left( \dd
  W_{s^\prime}\dd W_{s} \right)=\delta\left(s^\prime_{\bm k}-s_{\bm
    k}\right)\dd s_{\bm k}^2$, at first order in $\dd s_{\bm k}$, the
stochastic average of the integral term in Eq.~(\ref{eq:S:stocha})
vanishes. Actually, things are slightly more complicated since the
upper bound of this integral, $S_{\bm k}$, is a stochastic quantity
itself. Therefore, the averaging process should also be carried out on
this upper bound, and a generalized demonstration which includes this
case can be found in Ref.~(\cite{Gihman:1972}) (theorem $1$ on
p.~$28$).

\par

In order to characterize better the properties of this collapse time,
it is also important to determine its variance. Interestingly enough,
the same technique described above can be used in order to calculate
iteratively higher orders of $S_{\bm k}$. Upon using
Eq.~(\ref{eq:S:stocha}) one has
\begin{equation}
\label{eq:s2}
\mathbb{E}\left(S^2_{\bm k}\right)=c^2\left(b_0\right)
+\int_0^{S_{\bm k}} c^{\prime 2}\left[\Gamma_{\bm{k}}
\left(s_{\bm k}\right)\right]\dd s_{\bm k}\, .
\end{equation}
We see that we now need to evaluate the integral in the above
expression. For this purpose, we consider a new function
$e(\Gamma_{\bm k})$. As was done before, it can be Taylor expanded and
this leads exactly to Eq.~(\ref{eq:itoint}) (with, of course, $c$
replaced by $e$). Compared with the proof that allowed us to obtain
$\mathbb{E}(S_{\bm k})$, at this point, the strategy changes. We now
require the function $e\left(x\right)$ to be the solution of the
following ordinary differential equation [compare with
Eq.~(\ref{eq:eqc})]
\begin{equation}
\label{eq:eqe}
\frac{1}{2}e^{\prime\prime}\left(x\right)
+\mathrm{tanh}\left(x\right) e^{\prime}\left(x\right)
=-e^{\prime 2}\left(x\right)\, ,
\end{equation}
with boundary conditions $e\left(-b\right)=e\left(b\right)=0$. As
before, one can use this differential equation into the It\^o formula
to simplify the second integral in Eq.~(\ref{eq:s2}) [more precisely,
the integrand is replaced by $-e^{\prime 2}\left(x\right)$]. Taking
the stochastic average of the resulting equation, one gets
\begin{equation}
e\left(b_0\right)=\int_0^{S_{\bm k}}e^{\prime 2}\left[\Gamma_{\bm{k}}
\left(s_{\bm k}\right)\right]\dd s_{\bm k}\, .
\end{equation}
As a consequence, we deduce that
\begin{equation}
\mathbb{E}\left(S^2_{\bm k}\right)=c^2\left(b_0\right)+e\left(b_0\right)\, .
\end{equation}
The only thing which remains to be done is to solve
Eq.~(\ref{eq:eqe}). In fact, it turns out to be more convenient to
solve the slightly simpler differential equation satisfied by
$e_1\left(x\right)\equiv c^2\left(x\right)+e\left(x\right)$, namely
$e_1^{\prime\prime}(x)/2+\mathrm{tanh}\left(x\right)e_1^{\prime}\left(x\right)
=-2c\left(x\right)$, with boundary conditions
$e_1\left(-b\right)=e_1\left(b\right)=0$. It is straightforward to
show that $e_1(x)=x^2-b^2+\left[1+2b\tanh\left(b\right)\right]
\left[b\tanh\left(b\right)-x\tanh\left(x\right)\right]$. Then, the
second momentum of $S$ can be simply expressed as
$\mathbb{E}\left(S^2_{\bm k}\right)=e_1\left(b_0\right)$ which,
therefore, gives an explicit expression for the variance of the
collapse time. Since $b$ is supposed to be a large number $b\gg 1$ and
if we assume that the two Gaussians have comparable initial weights
which implies that $b_0\sim 0$, then one obtains, at leading order in
$b$,
\begin{eqnarray}
  \mathbb{E}\left(S_{\bm k}\right)&\simeq &b\, ,\\
  \sqrt{\mathbb{E}\left(S^2_{\bm k}\right)-\mathbb{E}^2\left(S_{\bm k}\right)}
  &\simeq &\sqrt{b}\, .
\end{eqnarray}
These two equations tell us that the relative standard deviation
scales as $1/\sqrt{b}$ and, therefore, that the distribution of
$S_{\bm k}$ becomes more peaked as $b$ increases. For this reason, in
the following, we will simply estimate the collapse time by means of
the sloppy requirement that $s_{\bm k}=b$. Finally, let us mention
that one could also apply the technique used in this section in order
to determine the higher order correlation functions of the process
$S_{\bm k}$.

\subsection{Collapse Time in the sub-Hubble Regime}
\label{subsection:collapsesubhubble}

In the last section, we have explained how to determine the
collapse time in terms of the variable $s_{\bm k}$. In order to
translate this result in terms of a more physical time (conformal time
or, better, number of e-folds), we need to use
Eq.~(\ref{eq:timechange}) which, in turn, requires the knowledge of
the function $X_{\bm k}$. This one cannot be determined in full
generality but it is easy to characterize it in the sub- and 
super-Hubble regimes. In this section, we investigate the sub-Hubble
regime.

\par

Let us define $K_{\bm{k}}\equiv
\chi_{\bm{k}}^{(2)}-\chi_{\bm{k}}^{(1)}$. This quantity measures the
shift in momentum between the two Gaussian components of the
wavefunction~(\ref{eq:DoubleGaussian}) (we recall that $X_{\bm{k}}$
measures the shift in position). Then, taking the difference between
the versions ``$(1)$'' and $(2)$'' of
Eq.~(\ref{eq:evolDoubleGaussian:barnu}) on the one hand, and versions
``$(1)$'' and $(2)$'' of Eq.~(\ref{eq:evolDoubleGaussian:g}) on the
other hand, we arrive at a closed system which can be written in a
matrix form, namely
\begin{equation}
\label{eq:XandKsystem}
\frac{\dd}{\dd \eta}\left(
\begin{array}{c}
X_{\bm{k}}\\
K_{\bm{k}}
\end{array}
\right)=\left(
\begin{array}{cc}
\displaystyle
-2 \Ima  \Omega_{\bm{k}}-\frac{\gamma}{\Rea  \Omega_{\bm{k}}} & 1\\
-4\left(\Rea \Omega_{\bm{k}}\right)^2 & 2\Ima  \Omega _{\bm{k}}
\end{array}\right)
\left(
\begin{array}{c}
X_{\bm{k}}\\
K_{\bm{k}}
\end{array}
\right)\, .
\end{equation}
At this stage, there is no approximation and the above equation is
general. In the sub-Hubble regime, one can use
Eq.~(\ref{eq:limomegasubhubble}) to simplify the expressions of $\Rea
\Omega _{\bm k}$ and $\Ima \Omega_{\bm k}$. Moreover, we are mainly
interested in computing the collapse time for the modes that
correspond to the (almost) scale invariant part of the power spectrum
since it is clearly less interesting to compute this quantity in a
regime that is already excluded by the data. As was discussed before,
this amounts to considering that $\gamma /k^2\ll 1$. Under those
conditions, one has $\Rea \Omega _{\bm k}\rightarrow k/2$ and $\Ima
\Omega _{\bm k}\rightarrow -\gamma/(2k)$ and
Eq.~(\ref{eq:XandKsystem}) can be re-expressed as
\begin{equation}
\label{eq:systemxk}
\frac{\dd}{\dd \eta}\left(
\begin{array}{c}
X_{\bm{k}}\\
K_{\bm{k}}
\end{array}
\right)=\left(
\begin{array}{cc}
-\gamma/k & 1\\
-k^2 & -\gamma/k
\end{array}\right)
\left(
\begin{array}{c}
X_{\bm{k}}\\
K_{\bm{k}}
\end{array}
\right)\, .
\end{equation}
This system of differential equations can be integrated and the solution reads
\begin{widetext}
\begin{eqnarray}
\label{eq:solk}
K_{\bm{k}}\left(\eta\right)&=&\mathrm{e}^{-\gamma\left(\eta-\eta_{\rm ini}\right)/k}
\biggl\{K_{{\bm k},{\rm ini}}\cos \left[k\left(\eta-\eta_{\rm ini}\right)\right]
-k X_{{\bm k},{\rm ini}}\sin 
\left[k\left(\eta-\eta_{\rm ini}\right)\right]\biggr\}, \\
\nonumber \\
\label{eq:solx}
X_{\bm{{\bm{k}}}}\left(\eta\right)&=&\mathrm{e}^{-\gamma\left(\eta-\eta_{\rm ini}\right)/k}
\biggl\{X_{{\bm k},{\rm ini}}\cos \left[k\left(\eta-\eta_{\rm ini}\right)\right]
+ \frac{K_{{\bm k},{\rm ini}}}{k} 
\sin \left[k\left(\eta-\eta_{\rm ini}\right)\right]\biggr\}\, ,
\label{eq:X}
\end{eqnarray}
where $K_{{\bm k},{\rm ini}}$ and $X_{{\bm k},{\rm ini}}$ are two
integration constants conveniently chosen to be the values of $K_{\bm
  k}$ and $X_{\bm k}$ at initial time $\eta=\eta_{\rm ini}$. For
simplicity, we now consider a situation such that $K_{{\bm k},{\rm
    ini}}=0$. Upon using Eq.~(\ref{eq:timechange}), one finds that
\begin{eqnarray}
  s_{\bm k} &=& -\frac{k}{4}X_{{\bm k},{\rm ini}}^2\left[{\rm e}
    ^{-2\gamma \left(\eta -\eta_{\rm ini}\right)/k}
    -1\right]
  -\frac{\gamma^2}{k^3}X_{{\bm k},{\rm ini}}^2
  \frac{1}{1+4\gamma ^2/k^4}
  {\rm e}
  ^{-2\gamma \left(\eta -\eta_{\rm ini}\right)/k}
  \left\{\cos\left[2k\left(\eta -\eta_{\rm ini}\right)\right]
    -\sin\left[2k\left(\eta -\eta_{\rm ini}\right)\right]-1\right\}.
\nonumber \\ 
  \end{eqnarray}
\end{widetext}
If we expand the above result in $\gamma /k^2$ for the reason
discussed before then, at leading order, one obtains an approximated
expression for the mapping between the variables $\eta$ and $s_{\bm
  k}$
\begin{eqnarray}
\label{eq:timessub}
s_{\bm k} &\simeq &\frac{k X_{{\bm k},{\rm ini}}^2}{4}\left[1-
\mathrm{e}^{-2\gamma\left(\eta-\eta_{\rm ini}\right)/k}\right]\, .
\end{eqnarray}
This expression means that $s_{\bm k}$ runs from $0$ to $kX_{{\bm
    k},{\rm ini}}^2/4$ when $\eta $ runs from $\eta_{\rm ini}$ to
infinity. Therefore, the time $s_{\bm k}$ evolves in a finite
range. However, in order to be consistent, one must have $\eta <\eta
_*=-1/k$ since the equations that have been used in order to derive
$s_{\bm k}$ are valid only in the sub-Hubble regime. As a consequence,
we have in fact $s_{\bm k}\in [0,s_*]$ where $s_*\equiv kX_{{\bm
    k},{\rm ini}}^2/4\{1-\exp[(2\gamma /k^2)(1+k\eta _{\rm
  ini})]\}$. Since we have $\vert k\eta_{\rm ini}\vert\gg 1$, one can
thus write $s_*\simeq kX_{{\bm k},{\rm ini}}^2/4[1-\exp(2\gamma \eta _{\rm
  ini}/k)]$. If $s<s_*$, then Eq.~(\ref{eq:timessub}) can be inverted
in order to evaluate the (total) number of e-folds in terms of the time
variable $s_{\bm k}$. One finds
\begin{eqnarray}
\label{eq:efoldcollasub}
N_{\bm k}&=&(1+\beta)\ln \biggl[1-\frac{k^2}{2\gamma}
\frac{1}{k\eta_{\rm ini}}
\ln \left(1-\frac{4s_{\bm k}}{kX_{{\bm k},{\rm ini}}^2}\right)
\biggr],\nonumber  \\
\end{eqnarray}
and one checks that if $s_{\bm k}=0$ then $N_{\bm k}=0$, if $s_{\bm
  k}=s_*$ then $N\to \infty$, and that the condition $s<s_*$ is
sufficient to guarantee that the above expression is well defined.

\par

Let us now discuss the above results in more detail. Firstly, we
notice in Eqs.~(\ref{eq:solk}) and~(\ref{eq:solx}) that the functions
$K_{\bm k}(\eta)$ and $X_{\bm k}(\eta)$ tend to zero when $\eta-\eta
_{\rm ini}\gg 1$. When this happens, the two Gaussians have the same
mean in position and momentum; in other words the two Gaussians have
merged. This ``merging phenomenon'' seems to be a generic feature and
can also be observed for the free particle~\cite{Bassi:2005fp} and/or
the harmonic oscillator in Minkowski spacetime. Therefore, it does not
come as a surprise that it also shows up in the sub-Hubble regime
where the Fourier mode under consideration does not feel spacetime
curvature. This also means that it is not a peculiar property of
inflation.

\par

The free particle situation can be studied~\cite{Bassi:2005fp} by
returning to Eqs.~(\ref{eq:noh}), (\ref{eq:reooh}), (\ref{eq:imooh}),
(\ref{eq:xbaroh}), (\ref{eq:sigmaoh}) and~(\ref{eq:chioh}). It is
sufficient to consider that $\omega=0$ in those equations to obtain
this case. This means that the mode equation~(\ref{eq:modeoh}) now
reads $f_{\bm k}''-\alpha^2f_{\bm k}=0$, where the quantity $\alpha$,
defined in Eq.~(\ref{eq:defalphaoh}) for the harmonic oscillator, now
reads $\alpha=\sqrt{2i\gamma \hbar/m}=\sqrt{\gamma \hbar/m}(1+i)$ and
is obtained from Eq.~(\ref{eq:defalphaoh}) by taking $\omega =0$. As a
consequence, the solution for $\Omega(t)$ has exactly the same form as
in Eq.~(\ref{eq:harmSol}) but now with the new $\alpha $ given
above. This implies that $\Rea  \Omega \rightarrow \sqrt{\gamma
  m/\hbar}/2$ and $\Ima  \Omega \rightarrow \sqrt{\gamma m/\hbar}/2$
when $t\rightarrow \infty$. These formulas should be compared to
Eqs.~(\ref{eq:Omega:limit+}) and~(\ref{eq:Omega:limit-}). Then,
considering the equations of motion for a double Gaussian state, and
defining $X\equiv \bar{x}_2-\bar{x}_1$ and $K\equiv\chi_2-\chi_1$,
upon using Eq.~(\ref{eq:XandKsystem}), one obtains the following set
of equations
\begin{equation}
\label{eq:systemfp}
\frac{\dd}{\dd t}\left(
\begin{array}{c}
X\\
K
\end{array}
\right)=\left(
\begin{array}{cc}
\displaystyle
-\sqrt{\gamma \hbar/m} & \hbar/m\\
\displaystyle
-\gamma & -\sqrt{\gamma \hbar/m}
\end{array}\right)
\left(
\begin{array}{c}
X\\
K
\end{array}
\right)\, .
\end{equation}
This equation should be compared to Eq.~(\ref{eq:systemxk}). In
particular, one notices that, here, the free particle case is not
simply obtained from this equation by considering $k=\omega=0$. If we
assume that $K(0)=0$, then the solution for $X(t)$ is given by
$X(t)=X(0)\exp(-t\sqrt{\hbar \gamma/m})\cos(t\sqrt{\hbar \gamma/m})$.
We see that this solution resembles the solutions~(\ref{eq:solx})
and~(\ref{eq:solk}) obtained before. Therefore, the merging is indeed
already present for a free particle in flat spacetime and is not a
specific feature of inflation. The exponential factor is mainly
responsible for the merging and this means that the ``merging time''
of the free particle is given by
\begin{equation}
T_{\rm merge}^{\rm fp}=\sqrt{\frac{m}{\hbar \gamma}}.
\end{equation}
This expression is consistent with the merging time derived in
Ref.~\cite{Bassi:2005fp}.

\par

In order to discuss our inflationary result, one should consider the
merging time of the harmonic oscillator instead of that of the free
particle since this is the appropriate limit in the sub-Hubble regime.
Following the same logic as before, it is easy to show that, for the
harmonic oscillator, Eq.~(\ref{eq:systemfp}) is replaced by
\begin{equation}
\label{eq:systemoh}
\frac{\dd}{\dd t}\left(
\begin{array}{c}
X\\
K
\end{array}
\right)=\left(
\begin{array}{cc}
\displaystyle
-\gamma \hbar/(m\omega) & \hbar/m\\
\displaystyle
-m\omega^2/\hbar & -\gamma \hbar/(m\omega)
\end{array}\right)
\left(
\begin{array}{c}
X\\
K
\end{array}
\right)\, .
\end{equation}
We see that it is indeed similar to Eq.~(\ref{eq:systemxk}) if we take
$\omega=k $ (and $m=\hbar=1$). The solution for $X(t)$ can be
expressed as $X(t)=X(0)\exp[-\hbar \gamma/(m \omega)t]\cos(\omega t)$,
assuming as before $K(0)=0$. This solution is perfectly consistent
with (\ref{eq:solk}) and~(\ref{eq:solx}). Compared to the free
particle case, one notices that the coefficient in the exponential is
now different from the frequency of the trigonometric function. But
the most important result that one can deduce from the above
considerations is that the merging phenomenon is also present for the
harmonic oscillator and that the corresponding merging time is given
by
\begin{equation}
\label{eq:mergingtimeoh}
T_{\rm merge}^{\rm ho}=\frac{m\omega}{\hbar \gamma}
=\omega \left(T_{\rm merge}^{\rm fp}\right)^2.
\end{equation}
Let us remark that the last expression could have been guessed on
dimensional grounds.

\par

In the case of inflation, the conformal merging time is given by [see
Eqs.~(\ref{eq:solk}) and~(\ref{eq:solx})]
\begin{equation}
\label{eq:mergeinf}
k\left(\eta _{\rm merge}-\eta_{\rm ini}\right)=\frac{k^2}{\gamma}.
\end{equation}
However, there is a new twist in the discussion. It is not obvious
that the above equation admits a solution because, in some sense, we
have a limited amount of time from $\eta _{\rm ini}$ to $\eta _*$, the
time of Hubble horizon crossing (defined by $\vert k\eta _*\vert
=1$). For times such that $\vert k\eta \vert <1$, we are no longer in
the sub-Hubble regime and the above equation can no longer be
used. But, given a value of $k^2/\gamma $, and an initial time $\eta
_{\rm ini}$, it is not obvious that there exists a time $\eta _{\rm
  merge}$ such that Eq.~(\ref{eq:mergeinf}) is satisfied. In fact,
there exists a solution only if $\vert k\eta_{\rm
  ini}\vert>1+k^2/\gamma $. This condition means that, for a given
$k^2/\gamma $, one can always give more time to the system to satisfy
Eq.~(\ref{eq:mergeinf}) by starting its evolution earlier (which is
equivalent to increasing $\vert \eta _{\rm ini}\vert $). It is easy to
show that the previous inequality is in fact a condition on the total
number of e-folds during inflation ($\beta\lesssim -2$), namely
\begin{equation}
N_{\rm T}\gsim  \Delta N_*
+\ln \left(1+\frac{k^2}{\gamma}\right),
\end{equation}
where $\Delta N_*\simeq 50$ for the modes of cosmological interest
today. If this condition is met, then the merging occurs after $N_{\bm
  k}^{\rm merge}$ with
\begin{equation}
N_{\bm k}^{\rm merge}=-\ln \left(1+\frac{k^2}{\gamma k\eta _{\rm ini}}
\right).
\end{equation}
Moreover, the term $k^2/(\gamma k\eta _{\rm ini})$ is of the order
$\sim k^2{\rm e}^{-N_{\rm T}+50}/\gamma$ and it seems reasonable to
assume that it is small. Indeed, typically, the total number of
e-folds during inflation is very large and, even if $k^2/\gamma \gg
1$, the factor ${\rm e}^{-N_{\rm T}}$ will entirely compensate its
influence (to be more concrete, we know that $k^2/\gamma \gsim
10^{28}$ but $N_{\rm T}$ can easily be larger than, say, $1000$ and
can even be as large as $10^8$). Then, the merging time during
inflation can be approximated by
\begin{equation}
  N_{\bm k}^{\rm merge}\simeq -\frac{k^2}{\gamma k\eta _{\rm ini}}\ll 1.
\end{equation}
We see that this expression scales as $\propto k/\gamma $ in full 
agreement with the previous considerations  on the harmonic oscillator, 
see Eq.~(\ref{eq:mergingtimeoh}).

\par

Let us now study the collapse time. First of all, the collapse can
occur in the sub-Hubble regime only if $b<s_*$. If we use the
expression of $s_*$ and assume, as before, that $k^2/(\gamma k\eta
_{\rm ini})\ll 1$, then $s_*\simeq kX_{{\bm k},{\rm ini}}^2/4$ and the
condition for having the collapse in the sub-Hubble regime can be
simply rewritten as
\begin{equation}
\label{eq:conditioncollpasesub}
b\ll \frac{kX_{{\bm k},{\rm ini}}^2}{4}.
\end{equation}
If this condition is satisfied, then the ``e-fold collapse number'' of
the mode under consideration is obtained by putting $s_{\bm k}=b$ in
the above expression~(\ref{eq:efoldcollasub}). Upon using the same
assumptions as before, we obtain that
\begin{equation}
\label{eq:coltimeinflation}
  N_{\bm k}^{\rm col}\simeq 
%  (1+\beta )\frac{k^2}{2\gamma}\frac{1}{k\eta _{\rm ini}}
%  \frac{4b}{kX_{{\bm k},{\rm ini}}^2}\simeq
   -\frac{2b}
{\gamma X_{{\bm k},{\rm ini}}^2\eta _{\rm ini}}\ll 1.
\end{equation}
At this point, several remarks are in order. Firstly, we notice that
$N_{\bm k}^{\rm col}/N_{\bm k}^{\rm merge}=4b/(kX_{{\bm k},{\rm
    ini}}^2)\ll 1$. This means that the collapse occurs on a much
smaller time scale than the merging. This property was also noticed in
the case of a free particle in Ref.~\cite{Bassi:2005fp}. This means
that the merging cannot be viewed as a substitute for the
collapse. Secondly, we notice that $N_{\bm k}^{\rm col}$ is actually
independent of $k$. We interpret this fact as meaning that, on
sub-Hubble scales, the mode under consideration must behave as in flat
spacetime. Indeed, for a free particle or the harmonic oscillator in
Minkowski spacetime, the condition for the collapse to occur can be
written as $s=\gamma \int X^2(\tau){\rm d}\tau \simeq \gamma
X(0)^2T_{\rm col}^{\rm fp,ho}=b$, where we have used $X(t)\simeq X(0)$
since we have shown that the merging takes place on a much longer time
scale. This implies that
\begin{equation}
T_{\rm col}^{\rm fp,ho}\simeq \frac{b}{\gamma X(0)^2},
\end{equation}
and one verifies that it is similar to
Eq.~(\ref{eq:coltimeinflation}). Therefore, if the collapse occurs on
sub-Hubble scales, its properties are, as expected, similar to what
happens in flat spacetime. Finally, if one starts from an initial
state made of several well-separated Gaussian wavefunctions, the
previous calculation suggests that it will almost instantaneously turn
into a single Gaussian state. As a matter of fact, it is a general
property~\cite{Diosi:1988vj,Halliwell:1995qa} of the CSL dynamics that
it asymptotically leads to Gaussian states. A posteriori, this remark
reinforces the assumption of using a Gaussian state for the
calculation of the spectrum in Sec.~\ref{subsec:cslps}.

\par

When the condition~(\ref{eq:conditioncollpasesub}) is not satisfied,
there will be no collapse on sub-Hubble scales. However, we can still
hope it will happen on super-Hubble scales. In fact, the claim that
the collapse has occurred depends on the value chosen for $b$. Before,
we used $b\simeq 10$ and for this value, given that our working
assumption is $kX_{{\bm k},{\rm ini}}^2\gg 1$, the
condition~(\ref{eq:conditioncollpasesub}) is probably always
satisfied. Therefore, it is only if we are more demanding about the
criterion that defines the collapse that this condition can be
violated. It is clear that a more stringent criterion takes more time
to be satisfied and, in this case, the time ``at our disposal'' in the
sub-Hubble regime may not be sufficient. In this situation, we have to
consider the super-Hubble regime. In the next section, we turn to this
case and show that the collapse is less efficient on large scales than
it is on small scales.

\subsection{Collapse Time in the super-Hubble Regime}
\label{subsection:collapsesuperhubble}

In this section, we repeat the previous discussion but now in the
super-Hubble regime. Therefore, we restart from the
equations~(\ref{eq:XandKsystem}) but now use the super-Hubble
limit~(\ref{eq:reomegacsl}) and~(\ref{eq:imomegacsl}) for $\Rea  \Omega
_{\bm k}$ and $\Ima  \Omega_{\bm k}$.  For the modes of cosmological
interest today in the (almost) scale invariant branch of the CSL power
spectrum, one has $\gamma/k^2\lll 1$ and the solution for
$X_{\bm{k}}(\eta )$ can be simply written as
\begin{equation}
  X_{\bm{k}}(\eta )\simeq X_{{\bm{k}}*}\left(-k\eta\right)^{\beta+1}\, ,
\end{equation}
where $X_{{\bm{k}}*}$ is the value of $X_{{\bm{k}}}(\eta )$ when the
mode under consideration $\bm{k}$ crosses the Hubble radius. One can
see that $X_{\bm{k}}(\eta )$ increases with time contrary to what
happens in the sub-Hubble regime. From this expression, it is easy to
derive the relation between $s_{\bm k}$ and the conformal time. One
obtains
\begin{eqnarray}
s_{\bm k}
&=&-\frac{\gamma}{k^2}\frac{kX_{{\bm k}*}^2}{2\beta+3}
\left[\left(-k\eta\right)^{2\beta+3}-1\right]\, .
\end{eqnarray}

The last formula is valid only on super-Hubble time, that is to say
for $\eta>\eta_*=-1/k$. At $\eta =\eta_*$, $s_{\bm k}=0$ and then
$s_{\bm k}\rightarrow \infty $ as $\eta \rightarrow 0$. From this
expression, it is also possible to relate the time variable $s_{\bm
  k}$ and the number of e-folds. One arrives at
\begin{equation}
N_{\bm k}=N_*+\frac{1+\beta}{2\beta+3}\ln \left(1-\frac{k^2}{\gamma}
\frac{2\beta +3}{kX_{{\bm k}*}^2}s_{\bm k}\right).
\end{equation}
This expression is always well defined because $2\beta +3<0$. One
verifies that $s_{\bm k}=0$ corresponds to $N_{\bm k}=N_*$.

\par

Let us now derive the time of collapse. As usual, it is obtained by
$s_{\bm k}=b$. As a consequence, it is simply given by
\begin{equation}
\label{eq:collapsetimesuper}
N_{\bm k}^{\rm col}=N_*+\frac{1+\beta}{2\beta+3}\ln \left(1-\frac{k^2}{\gamma}
\frac{2\beta +3}{kX_{{\bm k}*}^2}b\right).
\end{equation}
As a first check of this equation, we notice that, when $\gamma
\rightarrow \infty$, $N_{\bm k}^{\rm col}\simeq N_*$. Of course, this
result is expected since a large value of $\gamma $ means that the
collapse mechanism is very efficient and, therefore, that the
wavefunction almost instantaneously collapses. On the other hand, the
formula~(\ref{eq:collapsetimesuper}) can be further
simplified. Indeed, if the collapse has not taken place on sub-Hubble
scales, it is also the case for the merging since $N_{\bm k}^{\rm
  col}/N_{\bm k}^{\rm merge}\ll 1$. As a consequence, $X_{\bm k}(\eta
)$ has not evolved much and one can replace $X_{{\bm k}*}$ by $X_{{\bm
    k},{\rm ini}}$. Moreover, for the same reason, one must have
$b\gsim kX_{{\bm k},{\rm ini}}^2/4$, see also
Eq.~(\ref{eq:conditioncollpasesub}). In addition, we know that
$k^2/\gamma \gg 1$. Therefore, the first term in the argument of the
logarithm in Eq.~(\ref{eq:collapsetimesuper}) can be neglected. For
$\beta \simeq -2$, this equation can be rewritten as
\begin{equation}
N_{\bm k}^{\rm col}-N_*\simeq \ln \left(\frac{k^2}{\gamma}\right)
+\ln \left(\frac{b}{kX_{{\bm k},{\rm ini}}^2}\right).
\end{equation}
Of course the result will depend on what we require for $b$ and what
we assume for $X_{{\bm k},{\rm ini}}$. However, it seems reasonable to
assume that the second logarithm will not lead to a dominant
contribution. If this is the case, then our result simply says that
the wavefunction collapses just $\ln(k^2/\gamma)$ e-folds after the
Hubble radius crossing. Given the constraint obtained from the
measurement of the power spectrum in Eq.~(\ref{eq:cslconstraint}), one
already knows that $N_{\bm k}^{\rm col}-N_*\gsim 28$. Smaller values
of $\gamma/k^2$ would of course lead to a larger number of e-folds. We
conclude this section by noticing that the
constraint~(\ref{eq:cslconstraint}) is compatible with a collapse
occurring during inflation. Only for values of $\gamma $ such that
$\gamma /k^2\ll 10^{-50}$ (and $b\gtrsim kX_{{\bm k},{\rm ini}}^2/4$)
would the collapse happen after inflation.

\subsection{The Born Rule Derived}
\label{subsec:born}

Finally, we conclude with a section where we calculate the probabilities
of collapsing to each of the two branches of the wavefunction. We show
that these probabilities are given by the Born rule, which is of
course expected since the CSL theory is precisely designed to
reproduce this result, as already discussed in Sec.~\ref{sec:grw}
(see also Ref.~\cite{Bassi:2005fp}).

\par

Let us denote by $p_1$ the probability that the system collapses on
the first Gaussian branch of the wavefunction. This is also the
probability that, from given initial conditions, the stochastic
quantity $\Gamma _{\bm k}$ reaches first the region $\Gamma_{\bm
  k}<-b$ (\ie before the region $\Gamma _{\bm k}>b$) and that,
therefore, one has $\Gamma_{\bm k}(S_{\bm k})=-b$. Clearly, the
probability $p_2$ that the wavefunction collapses on the second branch
is the probability that $\Gamma _{\bm k}(S_{\bm k})=b$. Now, let us
introduce a function $\psi(x)$ which is defined by
\begin{equation}
\label{eq:linkpsig}
\psi\left(x\right)\equiv\frac{g\left(x\right)-g\left(b\right)}
{g\left(-b\right)-g\left(b\right)}\, ,
\end{equation}
where $g(x)$ will be specified soon. By construction, one has
$\psi\left(-b\right)=1$ and $\psi\left(b\right)=0$. Since, by
definition, $\Gamma _{\bm k}(S_{\bm k})$ can only take two values
(namely $\pm b$), one has
\begin{equation}
\mathbb{E}\left\lbrace\psi\left[\Gamma_{\bm{k}}\left(S_{\bm k}\right)\right]
\right\rbrace=p_1
\psi\left(-b\right) +p_2 \psi\left(b\right)=p_1,
\end{equation}
and this gives us a method to calculate $p_1$. To do so, we follow
what was explained in Sec.~\ref{subsec:timeofcollapse}, see in
particular Eq.~(\ref{eq:itoint}), and we write the corresponding It\^o
formula
\begin{align}
  \label{eq:itopsi}
  &\psi\left[\Gamma_{\bm k}(S_{\bm k})\right]-\psi\left(b_0\right)=
  \int_0^{S_{\bm k}} \psi^{\prime}\left[\Gamma_{\bm{k}}\left(s_{\bm
        k}\right) \right]\dd W_s \nonumber\\ &+ \!\!\int_0^{S_{\bm k}}
  \!\!\!\left\lbrace \psi^{\prime} \left[\Gamma_{\bm{k}}\left(s_{\bm
          k}\right)\right]
    \mathrm{tanh}\left[\Gamma_{\bm{k}}\left(s_{\bm k}\right)\right]
    +\frac{1}{2}\psi^{\prime\prime}\left[\Gamma_{\bm{k}} \left(s_{\bm
          k}\right)\right]\right\rbrace\dd s_{\bm k} .
  \nonumber\\
\end{align}
Then, let us choose the function $g(x)$ such that it obeys the equation
\begin{equation}
\frac{1}{2}g^{\prime\prime}\left(x\right)+\mathrm{tanh}\left(x\right) 
g^{\prime}\left(x\right)=0\, ,
\end{equation}
or, equivalently, $g(x)=\tanh(x)$. Since Eq.~(\ref{eq:linkpsig})
implies that $\psi(x)$ and $g(x)$ are linearly related, $\psi(x)$ also
obeys the above differential equation. As a consequence, the second
integral in Eq.~(\ref{eq:itopsi}) vanishes. Taking the stochastic
average, one obtains
\begin{equation}
\mathbb{E}\left\lbrace\psi\left[\Gamma_{\bm{k}}
\left(S_{\bm k}\right)\right]\right\rbrace=p_1=\psi\left( b_0\right)\, ,
\end{equation}
which is explicitly known since $g(x)$ has been determined. 

\par

The probability $p_2$ can be deduced along the same lines, by
introducing a new function $\psi$ such that, this time, $\psi(-b)=0$
and $\psi(b)=1$. Another method, much simpler, is just to use the
condition $p_1+p_2=1$. The final result reads
\begin{eqnarray}
p_1&=&\frac{\tanh\left(b_0\right)-\tanh \left(b\right)}
{\tanh\left(-b\right)-\tanh\left(b\right)}\, ,\\
p_2&=&\frac{\tanh\left(b_0\right)-\tanh\left(-b\right)}
{\tanh\left(b\right)-\tanh\left(-b\right)}\, .
\end{eqnarray}
From the definition of $\Gamma_{\bm{k}}$, these two formula can be
rewritten as~\cite{Bassi:2005fp}
\begin{eqnarray}
p_1=\frac{\vert N_1\left(\eta_{\mathrm{ini}}\right)\vert^2}
{\vert N_1\left(\eta_{\mathrm{ini}}\right)\vert^2
+\vert N_2\left(\eta_{\mathrm{ini}}\right)\vert^2}\, , \\
p_2=\frac{\vert N_2\left(\eta_{\mathrm{ini}}\right)\vert ^2}
{\vert N_1\left(\eta_{\mathrm{ini}}\right)\vert^2
+\vert N_2\left(\eta_{\mathrm{ini}}\right)\vert ^2}\, ,
\end{eqnarray}
which are exactly the Born rules of conventional quantum mechanics.

\section{Conclusion}
\label{sec:conclusion}

Let us now summarize our main findings. In this paper, we have applied
the CSL theory to inflation. Since the CSL scenario addresses the
measurement problem in quantum mechanics, it is a priori relevant to
explain how the wave-packet reduction took place in the early
universe, in the absence of any observer. Assuming that the
wavefunction has to collapse on an eigenstate of the Mukhanov-Sasaki
operator, we have computed the scalar power spectrum of cosmological
perturbations and studied the dynamics of the wavefunction
collapse. We have found that, in order to preserve the scale
invariance of the power spectrum, it is necessary to fine-tune the
parameter $\gamma $ which controls the amplitude of the CSL
corrections. Typically, depending on which temporal gauge is chosen
(see the appendix~\ref{sec:appendix}), we have found that the
dimensionless parameter that can be constructed out of $\gamma$ must
be smaller than $\exp\left(-\mathrm{a}\ \mathrm{few}\ \Delta
  N_*\right)$, where $\Delta N_*\simeq 50-60$ is the number of e-folds
spent by the relevant modes outside the Hubble radius during
inflation. We have also found that the time available during the
inflationary phase is sufficient in order for the perturbations
wavefunction to collapse. However, due to the smallness of $\gamma $,
the spread of the final wavefunction is too important, rendering the
collapse process not sufficiently efficient.  Therefore, under the
assumptions made in this paper, it seems fair to claim that the
collapse theories cannot solve the inflationary
``macro-objectification'' question.

\par 

The conclusions drawn above may not be as drastic as they appear at
first sight, because they are subject to some assumptions, and in
particular the choice of the ``collapse operator'' as the Fourier
space Mukhanov-Sasaki $v_{\bm{k}}$ variable: all cosmological
predictions made to date are based on this variable, rendering this
choice very sensible, but it is by no means unique (see, \eg the
discussion in Sec.~\ref{subsec:modifiedforv}). Moreover, $v_{\bm{k}}$
can be understood as a quantum field living in a curved spacetime, so
it should be treated by a quantum field theory version of the CSL
mechanism. The present state-of-the-art of this subject technically
forbids such a direct treatment, hence our simplifying
hypothesis. Could it be that a full relativistic version of CSL,
reproducing the many successes of quantum field theory and of the
ensuing particle physics, is needed before we can even embark in
examining cosmological perturbations? We doubt so, because cosmology,
contrary to ordinary quantum field theory, is endowed with a preferred
timelike direction that renders the ``time-dependent Minkowski
approximation'' accurate enough for all practical purposes. It is left
for future investigations to verify that the potential problems raised
and stringent constraints obtained in this work could be naturally
solved in a more general, yet-unknown, framework.

\par

There are other questions that could be the subject of further
works. In particular, there is the issue that energy is
not conserved in the CSL theory. In the case of the harmonic
oscillator, this is expressed through
Eq.~(\ref{eq:nonconservationH}). In the case of cosmological
perturbations, it is easy to show that this leads to
\begin{equation}
  \frac{{\rm d}}{{\rm d}\eta }\langle \hat{\cal H}_{\bm k}\rangle 
  =\frac{\gamma }{2}+\omega \, \omega'\langle v_{\bm k}^2\rangle.
\end{equation}
The CSL contribution can easily be integrated and gives $\langle {\cal
  \hat{H}}_{\bm k}\rangle \vert_{_{\mathrm{CSL}}}\simeq \gamma \eta_{\rm
  ini}/2$ at the end of inflation. Expressed in terms of the
Hamiltonian rather than the Hamiltonian density, one arrives at
\begin{equation}
  \langle \hat{H}\rangle \vert _{_\mathrm{CSL}}\simeq -4\pi^2\frac{\gamma}{2}
\eta_{\rm ini}  \int k^2{\rm d}k,
\end{equation}
which is infinite. It does not come as a surprise as it is known that
the CSL Tomanaga-Schwinger equation precisely leads to this type of
divergences~\cite{Bassi:2003gd,Bassi:2012bg}. It could be regularized
by introducing an ultraviolet cutoff although we notice on the above
equation the weird property that the infinite integral is over
comoving wavenumbers rather than over physical ones. This energy
non-conservation should cause a continuous increase of energy density
during inflation. It is interesting to notice that it cannot occur at
first order in the perturbations since $\mathbb{E}\left(\langle
  \widehat{\delta \rho}_{\bm k}\rangle \right) =0$. This means that it
will be important at second order only. Then, it would be important to
quantify this effect and, in particular, to compare it to the
background energy density $\simeq H^2\Mp ^2$ in order to check whether
this leads to a backreaction problem.

\par

Another point is that we have shown that the power spectrum, contrary
to what happens in the standard case, remains a time-dependent
quantity, \ie still evolves with time on large scales during
inflation. It is therefore not obvious that ${\cal P}_\zeta$ evaluated
at the end of inflation is exactly the power spectrum that should be
used at recombination. In fact, what happens just after the end of
inflation is of great interest for the cosmological consequences of
CSL. Indeed, just after inflation, the stages of pre-heating and
re-heating begin~\cite{Traschen:1990sw,Shtanov:1994ce,Kofman:1997yn};
this is also shown in Fig.~\ref{fig:scalecsl}. During this phase of
evolution, the inflaton field oscillates at the bottom of its
potential, $\varphi(t)\propto \sin (mt+\Delta)/(mt)$ where $\Delta $
is a phase and $m$ the mass of the inflaton (in the case of power-law
inflation, the potential has no minimum and, therefore, can only be
used to describe the slow-roll regime. Here, we assume that the
potential can be approximated by $m^2\varphi^2$ in the vicinity of the
minimum). In this case, the equation of motion~(\ref{eq:eomv}) for the
Mukhanov-Sasaki variable takes the form of a Mathieu
equation~\cite{Finelli:1998bu}. As is well known, this equation
possesses unstable solutions when the parameters falls in the resonant
bands. In the case of inflation, one can show that the large-scale
perturbations are in the first instability band which makes $v_{\bm
  k}$ growing and $\zeta_{\bm k}$ staying
constant~\cite{Finelli:1998bu,Jedamzik:2010dq}. In the CSL case, the
corresponding Mathieu equation would read
\begin{equation}
\frac{{\rm d}^2v_{\bm k}}{{\rm d}z^2}
+\left[{\cal A}_{\bm k}-2q \cos(2z+2\Delta)\right]v_{\bm k}
=0,
\end{equation}
where $z\equiv mt+\pi/4$, $a_{\rm e}$, $t_{\rm e}$ denoting the scale
factor and the cosmic time at the end of inflation and with
\begin{eqnarray}
{\cal A}_{\bm k} &=& 1+\frac{k^2-2i\gamma }{m^2a^2}\, ,\\
q &=& 
\frac{2}{mt_{\rm e}}\left(\frac{a_{\rm e}}{a}
\right)^{3/2}\, .
\end{eqnarray}
Since $q\ll 1$, in the regular case when $\gamma=0$,
the condition to be in the first resonant bands, $1-q<{\cal A}_{\bm
  k}<1+q$, is equivalent to $0<k/a<\sqrt{3Hm}$. In the CSL case, the
coefficient ${\cal A}_{\bm k}$ becomes complex. Therefore, in order to
determine the corresponding Floquet index, it now becomes necessary to
study the instability chart of the Mathieu equation in the complex
domain. Although this is beyond the scope of this paper, this is
certainly a subject worth investigating. In particular, it would be
interesting to see whether the instability is enhanced in this case as
one can, maybe naively, suspect. If so, maybe the preheating stage can
put even more stringent constraints on the parameter $\gamma $.

\par

We have seen that the study of the CSL cosmological perturbations is
in fact equivalent to the study of the CSL parametric oscillator (\ie
an harmonic oscillator with a time-dependent frequency). The previous
discussion suggests that it would be interesting to investigate the
case of a parametric oscillator in the presence of a resonance in the CSL
framework. In quantum field theory, this is a common situation and
typical examples are the dynamical Schwinger
effect~\cite{Brezin:1970xf} (the analogy between cosmological
perturbations and the Schwinger effect was discussed in
Ref.~\cite{Martin:2007bw}) or the dynamical Casimir
effect~\cite{moore:1970} which was recently observed for the first
time~\cite{2011Natur.479..376W} in the laboratory. In fact, if we want
to avoid the objection that the quantum field CSL theory is not yet
ready, it would be even more interesting to find a non-relativistic
system governed by a Mathieu equation and to investigate its behavior
within the CSL theory. We believe that all the equations presented in
the present article can be straightforwardly applied to this
case. Here, we suggest that a Paul trap~\cite{Leibfried:2003zz}
could be such an example. As for the inflationary preheating, we
expect the coefficients of the Mathieu equation to become complex
because of the $-2i\gamma $ term. This will probably make the system
extremely unstable and, as a consequence, it will probably be possible
to put very tight constraints on the value of $\gamma $. We hope that
this case will be treated in details soon.

\acknowledgements

We would like to thank D.~Sudarsky for interesting and enjoyable
discussions.

\appendix

\section{``Gauge Invariance'' of the CSL Power Spectrum}
\label{sec:appendix}

In section~\ref{subsec:modifiedforv}, we discussed the choice of the
``collapse operator'', \ie the operator that appears in the non-linear
and stochastic part of the modified Schr\"odinger equation. In
principle, this operator should be determined by a more fundamental
theory. However, the CSL model is just a phenomenological approach and
the ``collapse operator'' is just put by hand in order to match what
we observe when an experiment or an observation is performed (the
position of a spot in a detector, the energy density of a field,
etc.). In the case of the cosmological primordial perturbations, we
have argued that the Mukhanov-Sasaki variables $\hat{v}_{\bm{k}}$ is
the most sensible choice. But this variable often appears factorized
by a background quantity, typically a power of the scale factor
$a(\eta )$. Therefore, instead of $\hat{v}_{\bm{k}}$, one could very
well choose the collapse operator to be $h\left(a\right)
\hat{v}_{\bm{k}}$, where $h$ is a priori an arbitrary function of the
scale factor $a$.  After all, $\hat{v}_{\bm{k}}$ and $h\left(a\right)
\hat{v}_{\bm{k}}$ share the same eigenspectrum and drive the system
towards the same target states with the same probabilities. But the
point is that, a priori and as is discussed in detail below, this does
not lead to the same solution for the mode function $f_{\bm k}(\eta)$
and, therefore, a priori, for the power spectrum.

\par

In fact, this issue is related to an even more fundamental
problem. Indeed, one could claim that the conformal time $\eta$ used
in this paper to write the modified Schr\"odinger equation is not the
physical one and that one should use instead, say, the cosmic time $t$
(of course, the discussion also applies to any other time variables
related to $\eta$ through a transformation that depends only on the
background). In fact, a choice of time is equivalent to a choice of
$h$ since it has the same effect on the modified Schr\"odinger
equation. And, of course, as already mentioned, one could worry that
different choices lead to different predictions. Therefore, the
phenomenological approach used in this article suffers from what can
be called a ``temporal gauge'' problem. This problem probably
originates from the fact that the CSL equation is not covariant under
diffeomorphisms (contrary to the standard theory of cosmological
perturbations).

\par

In this appendix, we investigate this question, showing the remarkable
property that the conclusions obtained in this paper for
$h\left(a\right)=1$ are in fact valid for any other functions $h$. It
is true that the detailed shape of the power spectrum depends on the
gauge but its global properties are independent of the choice of
$h$. This means that, a priori for any $h$ allowing meaningful initial
conditions, the power spectrum of cosmological perturbations has a
broken power-law shape, with $n_{_{\rm S}}=1$ at small wavelengths and
$n_{_{\rm S}}=4$ at large wavelengths. As a consequence, the
requirement of moving the non-scale-invariant part of the spectrum
beyond the Hubble radius today always leads to extreme constraints on
the parameter $\gamma$.

\par 

Let us now consider the modified Schr\"odinger equation of motion for
$\Psi_{\bm{k}}$ in the CSL picture, with spontaneous localization on
the $h\left(a\right) \hat{v}_{\bm{k}}$ eigenmanifolds. It reads
\begin{eqnarray}
\dd\Psi_{\bm{k}}^\mathrm{R}&=&\left[-i\hat{\mathcal{H}}_{\bm{k}}^\mathrm{R}\dd\eta
+\sqrt{\gamma}h\left(a\right)\left(\hat{v}_{\bm{k}}^\mathrm{R}
-\mean{\hat{v}_{\bm{k}}^\mathrm{R}}\right)
\dd W_\eta\right.\nonumber\\ 
& &-\left.
\frac{\gamma}{2}h^2\left(a\right)
\left(\hat{v}_{\bm{k}}^\mathrm{R}-\mean{\hat{v}_{\bm{k}}^\mathrm{R}}\right)^ 2
\dd\eta\right]
\Psi_{\bm{k}}^\mathrm{R}\, ,
\end{eqnarray} 
and a similar equation for $\Psi_{\bm{k}}^\mathrm{I}$. This equation
should be compared with Eq.~(\ref{eq:SchroModes}), the only difference
being that the operator $\hat{v}_{\bm{k}}$ is now multiplied by
$h(a)$. Parameterizing $\Psi_{\bm{k}}$ as in
Eq.~(\ref{eq:SingleGaussianR}) using again
$\Omega_{\bm{k}}=-if_{\bm{k}}^\prime/(2f_{\bm{k}})$, one is led to the 
following equation for the mode function
\begin{equation}
\label{eq:modewithh}
f_{\bm{k}}^{\prime\prime}
+\left[\omega^2\left(\eta,\bm{k}\right)
-2i\gamma h^2\left(a\right)\right]
f_{\bm{k}}=0\, .
\end{equation}
This expression should be compared with Eq.~(\ref{eq:u}): as expected,
the only difference is that an extra $h^2\left(a\right)$ appears in
front of the $\gamma$ term. For simplicity, let us choose $h$ to be a
simple power law and let us assume the inflationary dynamics to be
close to a de Sitter Universe $a(\eta )\simeq -\ell_0/\eta$. Then, the
mode function can be re-expressed as
\begin{equation}
\label{eq:u:withp}
f_{\bm{k}}^{\prime\prime}
+\left(k^2-\frac{2}{\eta^2}-2i\gamma a^p\right)
f_{\bm{k}}=0\, .
\end{equation}
If $p<0$, the Bunch-Davies vacuum state cannot be chosen at the onset
of inflation since the $k^2$ term does not dominate in the
parenthesis. This means that one must work with $p\geq 0$. In this
paper the case $p=0$ [\ie $h\left(a\right)=1$] has been studied, hence
one only needs to study the $p>0$ cases. It is interesting first to
notice that the cases $p>0$ provide a natural amplification phenomenon
depending on the physical length of the mode since the amplitude of
the term proportional to $\gamma $ now increases as the mode is
stretched by the growth of the scale factor. This is consistent with
the physical intuition which tells us that the collapse should occur
for macro extended objects only. If $p>2$, the term proportional to
$\gamma $ dominates the dynamics at the end of inflation, when $k\eta$
goes to $0$, and one can expect the power spectrum scale invariance to
be destroyed. Therefore, if $p$ is an integer, we are left with the
cases $p=1$ and $p=2$ that we now study.

\par 

\begin{figure*}
\begin{center}
\includegraphics[width=0.85\textwidth,clip=true]{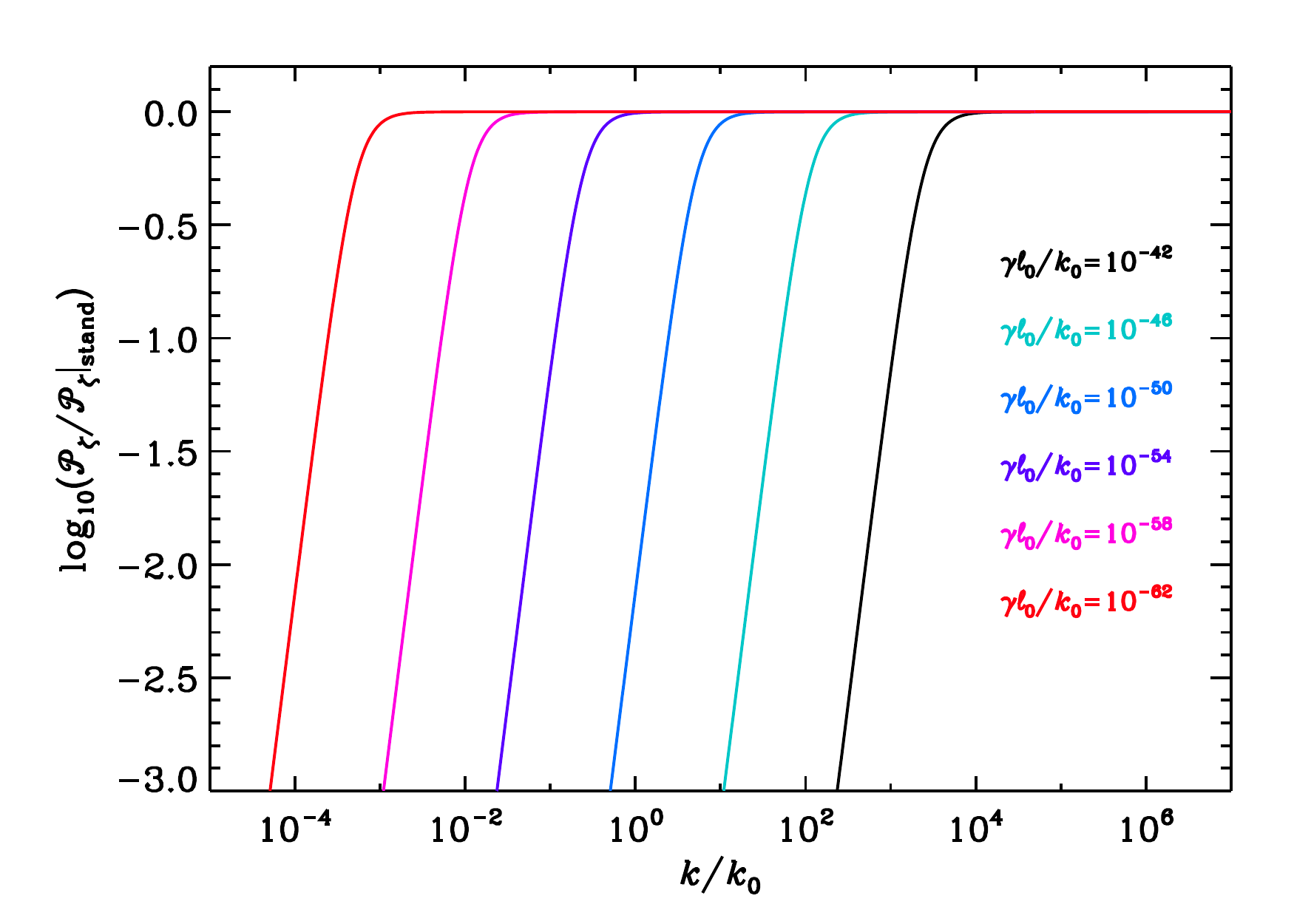}
\caption{Ratio of the power spectrum given by Eq.~
  (\ref{eq:powercslfinal:Whittaker}) ($p=1$) to the standard power
  spectrum given by Eq.~(\ref{eq:powerstandard}) for different values
  of the parameter $\gamma \ell _0/k_0$. }
\label{fig:CSLAlternativeSpectrum:1}
\end{center}
\end{figure*}

If $p=1$, the general solutions of Eq.~(\ref{eq:u:withp}) can be
expressed in terms of Whittaker functions
$W_{\mu,\kappa}\left(z\right)$~\cite{Abramovitz:1970aa,Gradshteyn:1965aa} as
\begin{equation}
f_{\bm{k}}(\eta)=C_{\bm{k}}W_{\gamma \ell_0/k,3/2}\left(2ik\eta\right)
+D_{\bm{k}}W_{-\gamma \ell_0/k,3/2}\left(-2ik\eta\right)\, ,
\end{equation}
where $C_{\bm{k}}$ and $D_{\bm{k}}$ are integration constants that can
be determined by choosing the Bunch-Davies  vacuum state for the initial
conditions. This leads to $C_{\bm{k}}=0$. Then, in the limit where
$k\eta$ goes to $0$, $\Rea {\Omega}_{\bm{k}}(\eta )$ can be Taylor
expanded, and this provides a simple expression for this quantity. In
particular, we find that $\Rea  \Omega_{\bm k}/k=\gamma
\ell_0/(2k)+{\cal O}(k\eta)$, showing that, in this case, the
spread does not divergence in the large-scale limit and that, as a
consequence, the localization of the wavefunction becomes much more
accurate. Moreover, since the inverse of $\Rea  \Omega _{\bm k}$ is
basically ${\cal P}_\zeta$, this allows us to calculate the power
spectrum, provided we push the expansion to higher orders. One obtains
\begin{widetext}
\begin{equation}
\label{eq:powercslfinal:Whittaker}
{\cal P}_{\zeta}(k)=
g\left(\frac{\ell_0\gamma}{k}\right)\left[ 1
+\frac{\ell_0\gamma}{k_0}g\left(\frac{\ell_0\gamma}{k}\right)
{\rm e}^{2\Delta N_*}\left(\frac{k_0}{k}\right)^{3}
-2\frac{\ell_0\gamma}{k}
g\left(\frac{\ell_0\gamma}{k}\right)
\left(1-\frac{\ell_0^2\gamma^2}{k^2}\right)\mathrm{log}
\left(2\frac{k}{k_0}\mathrm{e}^{-\Delta N_*}\right)
\right]^{-1}
{\cal P}_{\zeta}(k)\bigl \vert _{\rm stand},
\end{equation}
\end{widetext}
where ${\cal P}_{\zeta}\bigl \vert _{\rm stand}$ is the standard power
spectrum~(\ref{eq:powerstandard}),
and where $g\left(x\right)$ is defined by
\begin{equation}
\frac{1}{g(x)}\equiv 1+3x-3x^2-x^3
-2x\left(1-x^2\right)
\left[\psi\left(2+x\right)-2\psi\left(1\right)\right]\, ,
\end{equation}
$\psi\left(x\right)$ being the digamma Euler
function~\cite{Abramovitz:1970aa,Gradshteyn:1965aa}. Let us notice
that, in Eq.~(\ref{eq:powercslfinal:Whittaker}), we have sometimes
introduced the quantity $\ell_0\gamma/k$. Of course, the most
convenient way of dealing with this quantity is to express it as
$(\ell_0\gamma /k_0)k_0/k$ such that the dimensionless small parameter
$\ell _0\gamma /k_0$ explicitly appears. The spectrum given by
Eq.~(\ref{eq:powercslfinal:Whittaker}) should be compared with the one
obtained in Eq.~(\ref{eq:powercslfinal}) with the choice $h=1$. They
share the same broken power-law structure, with a scale-invariant part
$n_{_{\rm S}}\simeq 1$ at small scales and a branch with $n_{_{\rm
    S}}=4$ on large scales.  This spectrum is displayed in
Fig.~\ref{fig:CSLAlternativeSpectrum:1} for different values of the
parameter $\ell_0\gamma/k_0$.

The break in the power spectrum appears at $k^3/k_0^3\simeq
\ell_0\gamma/k_0\, \mathrm{e}^{2\Delta N_*}$.  Therefore, in order for
the non-scale invariant part of the power spectrum to be outside the
Hubble radius, one must have
\begin{equation}
\label{eq:conscslp1}
\frac{\gamma \ell_0}{k_0}\ll\mathrm{e}^{-2\Delta N_*}\simeq 10^{-53}\, .
\end{equation}
This equation should be compared to Eq.~(\ref{eq:cslconstraint}). We
see that, in the present case, we also obtain a constraint that can be
considered as ``extreme''. In other words, it seems that a very
important fine-tuning is necessary to maintain the consistency of the
CSL predictions with the CMB observations. We also notice that,
instead of $\gamma /k_0^2$, it is now the combination $\gamma
\ell_0/k_0$ that is constrained. Of course, this is just the
consequence of the fact that, as already discussed, changing the
collapse operator can change the dimension of the parameter $\gamma
$. In some sense, we face again the discussion of the temporal gauge
issue.

\par 

\begin{figure*}
\begin{center}
\includegraphics[width=0.85\textwidth,clip=true]{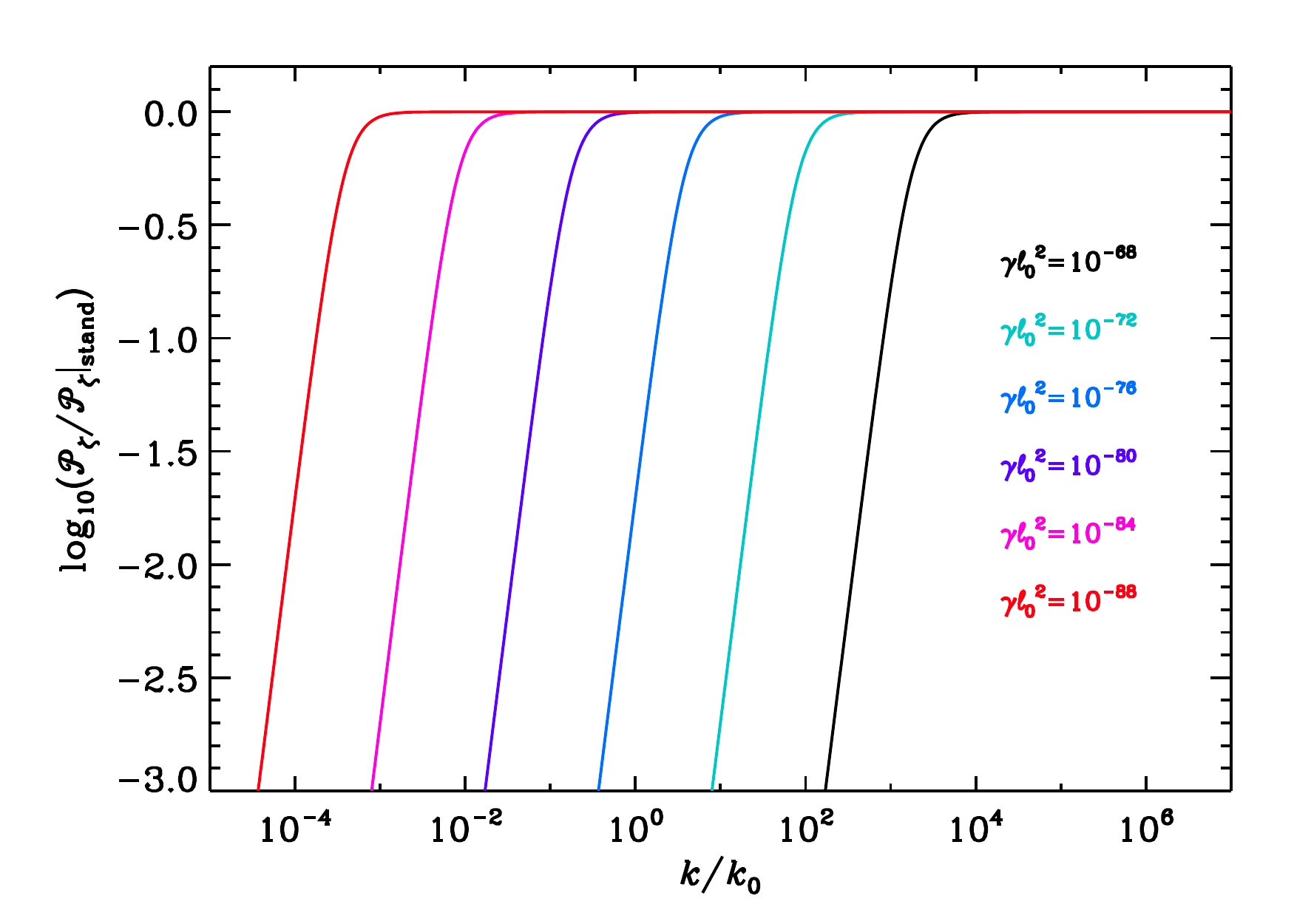}
\caption{Ratio of the power spectrum given by Eq.~
  (\ref{eq:powercslfinal:Bessel2}) ($p=2$) to the standard power
  spectrum given by Eq.~(\ref{eq:powerstandard}) for different values
  of the parameter $\gamma \ell _0^2$.}
\label{fig:CSLAlternativeSpectrum:2}
\end{center}
\end{figure*}

Let us now turn to the case $p=2$ in Eq.~(\ref{eq:u:withp}). The
general solutions of this equation can be expressed in terms of Bessel
functions with a complex
order~\cite{Abramovitz:1970aa,Gradshteyn:1965aa}, namely
\begin{eqnarray}
f_{\bm{k}}(\eta )&=&
C_{\bm{k}}\sqrt{-k\eta}J_{\frac{3}{2}\sqrt{1+\frac89 i\gamma \ell_0^2}}
\left(- k\eta\right)
\nonumber\\& &
+D_{\bm{k}}\sqrt{-k\eta}J_{-\frac{3}{2}\sqrt{1+\frac89 i\gamma \ell_0^2}}
\left(- k\eta\right)\, ,
\end{eqnarray}
where $C_{\bm{k}}$ and $D_{\bm{k}}$ are integration constants that can
be determined by requiring, as usual, the initial state to be the
Bunch-Davies  vacuum. This leads to
$C_{\bm{k}}=-D_{\bm{k}}\mathrm{e}^{3i\pi/2\sqrt{1+8/9\, i\gamma
    \ell_0^2}}$. In the limit where $k\eta$ goes to $0$,
$\Rea {\Omega}_{\bm{k}}$ can be Taylor expanded and, at first order in
the parameter $\gamma \ell_0^2$, the power spectrum reads
\begin{widetext}
\begin{equation}
\label{eq:powercslfinal:Bessel2}
{\cal P}_{\zeta}(k)\simeq
\left(1+\frac{2\pi}{3}\gamma \ell_0^2\right)
\left[ 1+\frac{2\gamma \ell_0^2}{3}
{\rm e}^{3\Delta N_*}\left(\frac{k_0}{k}\right)^{3}
+\frac{4}{3}\gamma \ell_0^2\frac{k_0}{k}\mathrm{e}^{\Delta N_*}
\right]^{-1}
{\cal P}_{\zeta}(k)\bigl \vert _{\rm stand}.
\end{equation}
\end{widetext}
The formula~(\ref{eq:powercslfinal:Bessel2}) should be compared with
Eqs.~(\ref{eq:powercslfinal}) and~(\ref{eq:powercslfinal:Whittaker}).
Again, the power spectrum has the same shape, with a scale invariant
part on small scales and a non-invariant branch with $n_{_{\rm S}}=4$
on large scales. This is clearly seen in
Fig.~\ref{fig:CSLAlternativeSpectrum:2}, where the
spectrum~(\ref{eq:powercslfinal:Bessel2}) is represented for different
values of the parameter $\gamma \ell_0^2$. The break in the power
spectrum appears at $k^3/k_0^3\simeq \gamma \ell_0^2/3\,
\mathrm{e}^{3\Delta N_*}$.  Therefore, in order for the non-scale
invariant part of the power spectrum to be outside the observational
window, one must require that
\begin{equation}
\label{eq:conscslp2}
\gamma \ell_0^2\ll\mathrm{e}^{-3\Delta N_*}\simeq 10^{-79}\, .
\end{equation}
Again, we can consider the above constraint as a fine-tuning. It is
also interesting to notice that, contrary to
Eqs.~(\ref{eq:cslconstraint}) or~(\ref{eq:conscslp1}),
Eq.~(\ref{eq:conscslp2}) involves physical quantities only. This is
because, when $p=2$, the CSL correction that should be compared to the
comoving wavenumber squared is $\propto \gamma a^2$, see
Eq.~(\ref{eq:u:withp}). In other words, $\gamma $ should
now be compared to the physical wavenumber. If we take $\ell _0\simeq
10^5 \ell_{_{\rm Pl}}$, which comes from the CMB normalization, then
one arrives at $\gamma \ll 10^{-89}$.

Let us conclude this appendix by noticing that the above results are
in fact generic and do not depend on the value of $p$. Technically,
the power spectrum is obtained by taking the super-Hubble limit of the
mode function $f_{\bm k}(\eta )$, by inserting it in the expression of
$\Rea \Omega _{\bm k}=\Rea
\left[-if^\prime_{\bm{k}}/(2f_{\bm{k}})\right]$ and by retaining only
the leading order in $k\eta $. In the standard case, the leading terms
of the mode function expansion turn out to cancel out in $\Rea \Omega
_{\bm k}$, leaving an expression which precisely gives a
scale-invariant power spectrum. This cancellation originates from the
fact that the Wronskian is conserved. In the CSL case, the fact that
$\gamma \neq 0$ implies that this symmetry no longer exists, and, as a
consequence, the nice cancellations mentioned above no longer show up
and scale invariance is immediately broken. In some sense, the fact
that the $\gamma$ term destroys the scale invariance of the power
spectrum does not come from the fact that its presence modifies the
time dependence of the effective frequency (the value of $p$ or the
choice of $h$), but is rather due to the fact that it makes the
effective frequency a complex quantity. We conclude that modifying the
definition of the ``collapse operator'' by multiplying it with a
background function, despite changing the dimension of $\gamma$,
always constrains this parameter to be extremely fine-tuned.

\bibliography{biblio}

\end{document}